\documentclass[%
 aps,
 prd,
 twocolumn,
 superscriptaddress,
 nofootinbib,
 amsmath,amssymb,
 floatfix,
]{revtex4-2}

\usepackage{graphicx}
\usepackage{dcolumn}
\usepackage{bm}
\usepackage{xcolor}
\usepackage{url}
\usepackage{hyperref}
\usepackage{orcidlink}

\newcommand{\dnnz}{{\sc DNNz}}
\newcommand{\mizuki}{{\sc Mizuki}}
\newcommand{\dempz}{{\sc DEmPz}}

\providecommand{\arcsec}{\ensuremath{^{\prime\prime}}}

\begin{document}

\title{Weak-lensing mass calibration of \emph{Planck} Sunyaev--Zel'dovich clusters with HSC-SSP Year~3}

\author{Andr\'es A. Plazas Malag\'on\,\orcidlink{0000-0002-2598-0514}}
\email{plazas@stanford.edu}
\affiliation{Kavli Institute for Particle Astrophysics and Cosmology, Stanford University, Stanford, CA, USA}
\affiliation{SLAC National Accelerator Laboratory, Menlo Park, CA, USA}
\affiliation{Department of Astrophysical Sciences, Princeton University, Peyton Hall, Princeton, NJ, USA}

\author{Hironao Miyatake\,\orcidlink{0000-0001-7964-9766}}
\affiliation{Kobayashi-Maskawa Institute for the Origin of Particles and the Universe (KMI), Nagoya University, Nagoya, 464-8602, Japan}
\affiliation{Institute for Advanced Research, Nagoya University, Nagoya 464-8601, Japan}
\affiliation{Kavli Institute for the Physics and Mathematics of the Universe (KIPMU), The University of Tokyo Institutes for Advanced Study (UTIAS), The University of Tokyo, Chiba 277-8583, Japan}

\author{Surhud More\,\orcidlink{0000-0002-2986-2371}}
\affiliation{The Inter-University Centre for Astronomy and Astrophysics, Post bag 4, Ganeshkhind, Pune 411007, India}
\affiliation{Kavli Institute for the Physics and Mathematics of the Universe (KIPMU), The University of Tokyo Institutes for Advanced Study (UTIAS), The University of Tokyo, Chiba 277-8583, Japan}

\author{Nicholas Battaglia\,\orcidlink{0000-0001-5846-0411}}
\affiliation{Department of Astronomy, Cornell University, Itaca, NY, USA}

\author{Eunseong Lee}
\affiliation{Department of Physics and Astronomy, University of Pennsylvania, Philadelphia, PA, USA}

\author{Neta Bahcall\,\orcidlink{0000-0002-8226-9825}}
\affiliation{Department of Astrophysical Sciences, Princeton University, Peyton Hall, Princeton, NJ, USA}

\date{\today}

\begin{abstract}
We present a weak gravitational lensing mass calibration of 19 \textit{Planck}
Sunyaev--Zel'dovich (SZ) selected galaxy clusters using shape measurements from
the Hyper Suprime-Cam Subaru Strategic Program (HSC-SSP) Year 3 shape catalog. We measure the stacked weak-lensing signal $\Delta\Sigma(R)$
using per-cluster lensing weights that match the measurement pipeline's
stacking scheme, and construct an analytical covariance matrix that includes
shape noise and projected large-scale structure contributions. Our primary constraint on the SZ mass bias comes from a forward-modeling
approach that integrates over the halo mass function while accounting for the
\textit{Planck} SZ selection function, Eddington bias from log-normal scatter
in the SZ mass proxy, and cluster miscentering.
Fitting four free parameters, the log mass bias $\ln(1-b)$, the miscentered
fraction $f_{\rm mis}$, the offset scale $\sigma_{\rm off}$, and the
SZ scatter $\sigma_{\ln M}$, over the
radial range $0.5$--$5.0\,h^{-1}\,\mathrm{Mpc}$, we obtain
$1-b = 0.73^{+0.10}_{-0.11}$
with $\chi^2/\mathrm{dof} = 5.2/5$
at an effective redshift $z_{\rm eff}\simeq 0.24$.
This measurement is consistent with recent weak-lensing
calibrations of SZ-selected clusters and supports the picture that significant
mass bias corrections are required to reconcile cluster abundance measurements
with primary cosmic microwave background constraints on cosmological parameters.
\end{abstract}


\maketitle

\section{Introduction}
\label{sec:intro}

Galaxy clusters are the most massive collapsed structures in the universe. Their theoretical abundances, as a function of redshift and mass, are sensitive to cosmological parameters that probe the growth of large-scale structure, such as the amplitude of the matter power spectrum, $\sigma_8$, and the matter density, $\Omega_m$ \citep{bahcall98,reiprich02,voit05,allen11,2012ARAA..50..353K}. To test these predictions, observations aim to detect ensembles of clusters and provide measurements of observables that can be related to their masses. However, systematic uncertainties in determining these mass--observable relations limit the constraining power of cluster abundance measurements. Accurate empirical calibrations of the mass--observable relation are therefore essential.

Galaxy clusters can be identified through several methods, including X-ray emission from intracluster hot gas \citep{pacaud16}, peaks in projected weak-lensing convergence maps \citep{oguri21,miyazaki18}, and overdensities or concentrations of galaxies in optical and near-infrared imaging surveys \citep{2018PASJ...70S..20O, 2014MNRAS.444..147O,rykoff14}. Another powerful method is detection via the thermal Sunyaev--Zel'dovich (tSZ) effect---the inverse Compton scattering of cosmic microwave background (CMB) photons by hot electrons in the intracluster medium \citep{sunyaev72}. Because the tSZ signal is nearly redshift-independent at fixed mass, this technique efficiently identifies clusters above a given mass threshold across a wide range of epochs.  Recent CMB experiments such as the South Pole Telescope (SPT; \citealt{Carlstrom2011}), the Atacama Cosmology Telescope (ACT; \citealt{Thornton2016_ACT}), and the \emph{Planck} satellite \citep{planck16} have produced large catalogs of tSZ-selected clusters \citep{planck16,bocquet19,hilton21,2025arXiv250721459A}. However, the tSZ effect measures only a mass proxy, the volume-integrated pressure of the intracluster medium, known as the Compton-$Y$ parameter. To use clusters for cosmology, it is necessary to establish scaling relations between this observable and the true halo mass via independent calibration methods. Each survey adopts different techniques to measure the Compton-$Y$ parameter and to calibrate the corresponding mass--observable relation. For example, the \emph{Planck} survey provides a well-characterized selection function and calibrates $Y$--$M$ using X-ray observations. These calibrations, however, typically rely on hydrostatic equilibrium and assumptions about cluster physics that may not always hold, leading to systematic uncertainties in cluster mass estimates \citep[e.g.,][]{evrard1990,2012NJPh...14e5018R}.

Weak gravitational lensing (WL)---the coherent distortion of background galaxy shapes due to foreground mass distributions---provides an unbiased probe of the total matter content of clusters, independent of its baryonic state or dynamical equilibrium. In clusters, the WL signal manifests as a small but coherent tangential shear pattern in the shapes of background galaxies. Several studies have employed WL to calibrate the masses of SZ-selected clusters, including using ACT clusters \citep{2013MNRAS.429.3627M,2016JCAP...08..058B,miyatake19,Robertson2024,Shirasaki2024,shin25}, SPT clusters \citep{2018MNRAS.474.2635S,2019MNRAS.485...69S,Schrabback2021}, {\sl Planck} clusters \citep{vonderLinden2014,Hoekstra2015,Sereno2017,medezinski18,PennaLima2017,Aymerich2025}, {\sl Planck} and SPT clusters \citep{Gruen2015}, and other massive cluster samples \citep{Hoekstra2015,Smith2016}. Mass calibration in these works is often quantified through the bias parameter
\begin{equation}
1-b = \frac{M_{\mathrm{SZ}}}{M_{\mathrm{true}}},
\end{equation}
where $M_{\mathrm{SZ}}$ is the SZ-derived mass estimate and $M_{\mathrm{true}}$ is the true cluster mass, best estimated from WL ($M_{\mathrm{true}} \equiv M_{\mathrm{WL}}$).

A long-standing tension has been reported between the values of $1-b$ obtained from weak-lensing calibrations of \emph{Planck} SZ cluster masses \citep[e.g.,][]{vonderLinden2014,Hoekstra2015} and those inferred by reconciling the \emph{Planck} primary CMB anisotropies with the \emph{Planck} cluster abundance. This discrepancy, at the $\sim 2\sigma$ level, can be partially mitigated by accounting for systematic effects such as Eddington bias \citep{battaglia16} and by incorporating updated measurements of the optical depth to reionization \citep{2016AA...594A..24P}. However, if this tension persists as the precision of cluster measurements continues to improve, it may indicate the presence of unaccounted observational systematics or point to the need for extensions to the standard $\Lambda$CDM cosmological model \citep{2016AA...594A..13P}.

The hydrostatic mass bias parameter $b$, which quantifies the ratio between the SZ-inferred mass and the true halo mass, has been extensively studied through weak-lensing mass calibration of cluster samples \citep{vonderLinden2014, Hoekstra2015, medezinski18, miyatake19, Shirasaki2024, shin25}. These analyses consistently find values of $(1-b)$ in the range $0.6$--$0.9$, indicating that assumptions of hydrostatic equilibrium tend to underpredict true cluster masses. Recent work by \citet{shin25} has further reported evidence for a redshift-dependent mass bias using the ACT~DR5$\times$DES~Y3 dataset, suggesting that the level of hydrostatic equilibrium violation may evolve with cosmic time.

In this paper, we study the WL signal around SZ-selected clusters from the \emph{Planck} mission (PSZ2 catalog) \citep{planck16} to provide mass calibrations. We employ a stacked lensing analysis to measure the average mass density distribution around 19 clusters overlapping the Year~3 (S19A) footprint of the Hyper Suprime-Cam Subaru Strategic Program (HSC-SSP; \citealt{Aihara2017b}). Stacking increases the detection significance by averaging over intrinsic shape noise and cluster-to-cluster variations. This work extends previous HSC-based \emph{Planck} calibrations \citep{medezinski18}, which focused on a smaller five-cluster sample, by leveraging the greater depth and area of the HSC-Y3 shape catalog.

This paper is organized as follows. Section~\ref{sec:data} describes the \emph{Planck} cluster catalog and the HSC-Y3 shape catalog. Section~\ref{sec:measurements} presents our WL measurements, including source selection, scale cuts, and systematic tests. Section~\ref{sec:results} describes the modeling and results of the SZ--WL mass calibration, including the analytical covariance matrix, the weighted stacking methodology, and a comparison with published measurements. Section~\ref{sec:conclusion} summarizes our findings and discusses their implications for cluster cosmology and future surveys such as the Vera C. Rubin Observatory Legacy Survey of Space and Time (LSST), which will enable high-precision cluster mass calibration with vastly larger samples. Throughout, we assume a flat $\Lambda$CDM cosmology with
$\Omega_m = 0.3$, $h = 0.7$, $\Omega_b h^2 = 0.02225$,
$\ln(10^{10}A_s)=3.094$ (corresponding to $\sigma_8\approx 0.85$),
$n_s = 0.9645$, and $w=-1.0$.
These parameters are used consistently across the \textsc{Dark Emulator}
predictions, the covariance calculation, and mass conversions.

\section{Data}
\label{sec:data}

\subsection{HSC-SSP observations}

The Hyper Suprime-Cam (HSC) \citep{2018PASJ...70S...1M} is a wide-field (1.77 deg$^{2}$) prime-focus camera mounted on the 8.2 m Subaru Telescope.
The HSC Subaru Strategic Program (HSC-SSP) is a multi-band ($grizy$), wide and deep imaging survey awarded 330 Subaru nights, designed to cover 1400 deg$^{2}$ of the northern sky to a point-source depth of $i \approx 26$ (5$\sigma$) with a median seeing of 0.6\arcsec\ in the $i$ band (measured in a 2\arcsec\ aperture). The survey consists of three layers: Wide, Deep, and UltraDeep. The Wide layer, in particular, is optimized for weak-lensing cosmology, spanning $\sim$1100 deg$^{2}$. The $i$-band observations used for weak lensing measurements are obtained under excellent seeing conditions.

In this work we use the third-year dataset (hereafter HSC-Y3), which includes galaxy shape and photometric-redshift catalogs \citep{li22}. The HSC-Y3 data are based on the S19A internal release (September 2019), incorporating observations collected between March 2014 and April 2019.

\subsection{Cluster overlap between HSC S19A and Planck}
We use the \emph{Planck} 2015 \emph{Union} catalog\footnote{\url{https://wiki.cosmos.esa.int/planckpla2015/index.php/Catalogues\#Union_catalogue}; data products available at \url{https://irsa.ipac.caltech.edu/data/Planck/release_2/ancillary-data/HFI_Products.html}.} for \emph{Planck} PDR2 \citep{planck16}. This catalog is a nearly full-sky list of 1653 SZ detections built from three independent cluster-finding pipelines: two that implement the matched multi-filter algorithm (MMF1 and MMF3) and one based on a Bayesian algorithm (PowellSnakes, PwS). All three pipelines assume a circularly symmetric pressure profile based on the universal pressure profile of \citet{arnaud10} for the SZ signal template in the detection. The Union catalog combines detections from all three pipelines and provides, for each cluster, an estimate of the integrated Compton-$Y$ parameter and an SZ-derived mass proxy $M_{\rm SZ}$ obtained through the $Y$--$M$ scaling relation calibrated with X-ray observations under the assumption of hydrostatic equilibrium \citep{2016AA...594A..24P}.

\begin{figure*}[t]
\centering
\includegraphics[width=\textwidth]{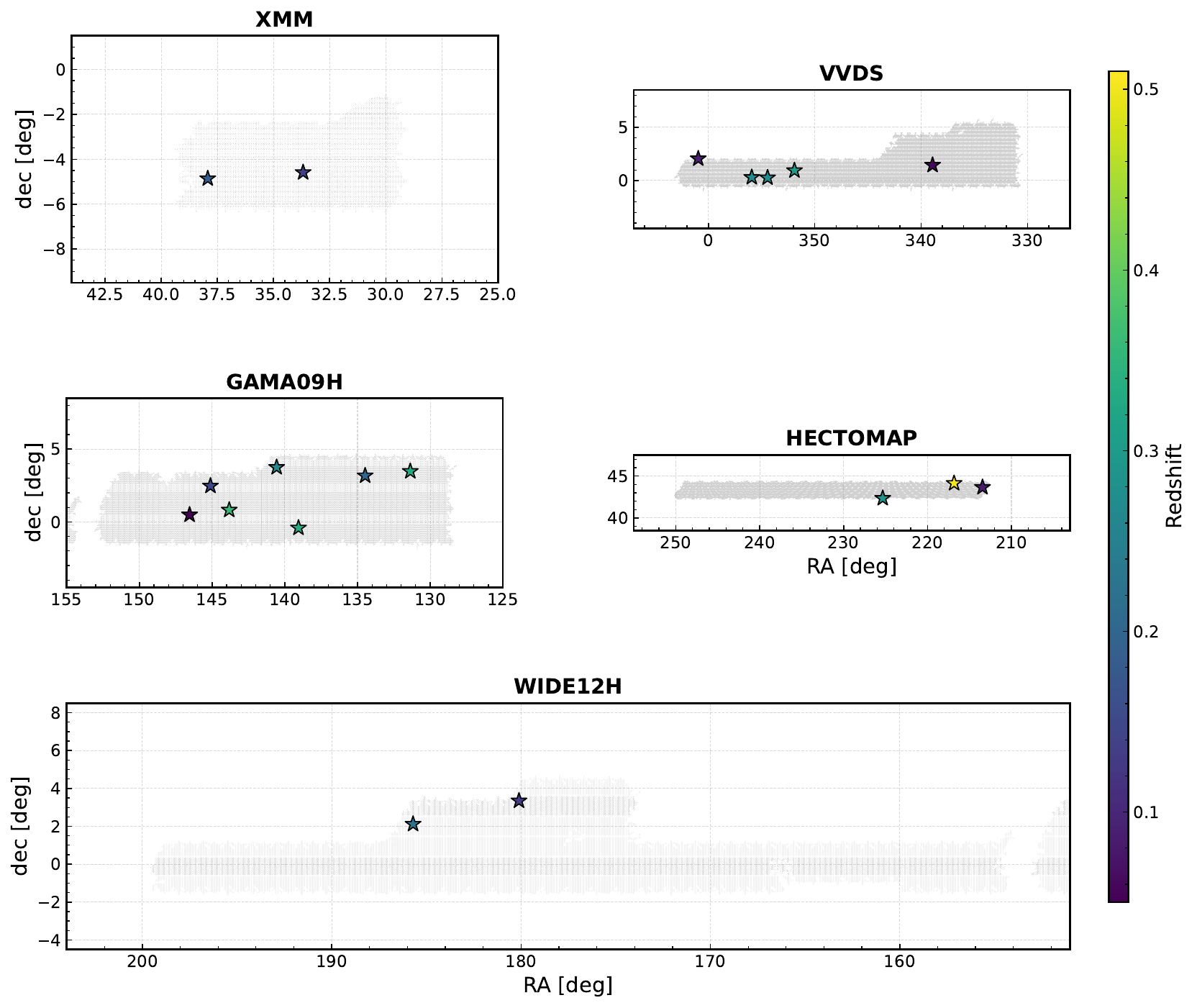}
\caption{Sky distribution of the 19 \emph{Planck} PSZ2 clusters (stars,
color-coded by redshift) overlapping the HSC-SSP S19A Wide-layer
full-depth full-color (FDFC) footprint (shaded regions). Each panel
shows one HSC field: XMM, VVDS, GAMA09H, HECTOMAP, and WIDE12H.
The GAMA15H field has no overlap with the cluster sample and is therefore not shown.}
\label{fig:psz2}
\end{figure*}

After cross-matching the \emph{Planck} catalog with the HSC-S19A Full-Depth-Full-color footprint (resulting in $\approx$ 450 deg$^2$ overlap), we find a total of 19 overlapping clusters (Fig.~\ref{fig:psz2} and Table~\ref{table1:clusters}). Five of the 19 clusters correspond to the sample analyzed in a previous and similar study by \citet{medezinski18}.

The 19-cluster subsample is defined solely by the geometric overlap
between the \emph{Planck} Union catalog and the HSC~S19A full-depth
full-color footprint: every Union-catalog cluster falling within the
HSC mask is retained, with no additional quality, mass, or
signal-to-noise cuts beyond the Union catalog's own $q>4.5$ threshold.
The one cluster without a reported \emph{Planck} mass
(PSZ2~G235.96+38.21) is included in the stack with its full
weak-lensing weight. Because the HSC footprint is contiguous within
each of the six Wide fields and the \emph{Planck} beam spans several
arcminutes, the probability that a given \emph{Planck} cluster lands
in the HSC mask is determined almost entirely by the ratio of HSC area
to the full sky ($\sim$430~deg$^2$/41\,253~deg$^2\sim 1$\%), which is
independent of cluster mass and redshift. We therefore treat this
secondary, footprint-driven selection as effectively random with
respect to the mass and redshift distributions modeled by the Planck
completeness function, and no additional selection correction is
applied.

\begin{table*}[t]
\caption{%
\emph{Planck} SZ clusters overlapping the HSC S19A footprint.
Five of the 19 clusters (shown in bold) correspond to the sample analyzed by
\citet{medezinski18}: PSZ2 G068.61-46.60 (Abell~2457),
PSZ2 G167.98-59.95 (Abell~0329),
PSZ2 G174.40-57.33 (Abell~0362),
PSZ2 G228.50+34.95 (MaxBCGJ140.53188+03.76632), and
PSZ2 G231.79+31.48 (MACSJ0916.1$-$0023/Abell~0776).
Right ascension and declination are taken from the \textsc{CAMIRA} catalog
\citep{oguri14,oguri18}, except for PSZ2 G068.61-46.60 (center from the
\emph{Planck} SZ2 catalog; \citealt{planck16}), and PSZ2 G083.85-55.43 and
PSZ2 G099.57-58.64 (centers from the SDSS \texttt{redMaPPer} DR8 catalog;
\citealt{rykoff16}).
Redshifts are taken from the \emph{Planck} SZ2 catalog, except for
PSZ2 G235.96+38.21, whose redshift is adopted from the \textsc{CAMIRA} catalog.
SZ masses are the uncorrected \emph{Planck} catalog values; Eddington bias
from scatter in the SZ mass proxy is accounted for self-consistently in the
forward model (Section~\ref{sec:eddington_model}).
PSZ2~G235.96+38.21 has no reported SZ mass in the \emph{Planck} catalog (indicated
by ``--''); its weak-lensing signal is included in the stacked measurement. The forward model does not use individual cluster SZ masses as it integrates over the halo mass function weighted by the Planck selection function at each cluster's redshift (Section~\ref{sec:eddington_model}).}
\label{table1:clusters}
\begin{ruledtabular}
\begin{tabular}{lcccc}
Name & SZ R.A. & SZ Decl. & Redshift Planck &
$M_{\mathrm{SZ}}\,[10^{14}\,h^{-1}M_{\odot}]$ \\
\hline
\textbf{PSZ2 G068.61-46.60} & 22h 35m 42.44s & 1$^{\circ}$ 26$'$ 58.0$''$ & 0.059 & 1.93 \\
PSZ2 G071.39+59.54 & 15h 1m 20.63s & 42$^{\circ}$ 20$'$ 47.1$''$ & 0.291 & 5.87 \\
PSZ2 G080.64+64.31 & 14h 27m 19.51s & 44$^{\circ}$ 7$'$ 50.9$''$ & 0.502 & 5.23 \\
PSZ2 G083.14+66.57 & 14h 13m 46.68s & 43$^{\circ}$ 39$'$ 8.2$''$ & 0.089 & 2.07 \\
PSZ2 G083.85-55.43 & 23h 27m 37.50s & 0$^{\circ}$ 56$'$ 30.9$''$ & 0.302 & 4.92 \\
PSZ2 G087.03-57.37 & 23h 37m 43.65s & 0$^{\circ}$ 16$'$ 5.6$''$ & 0.278 & 7.33 \\
PSZ2 G089.46-58.09 & 23h 43m 40.14s & 0$^{\circ}$ 18$'$ 7.5$''$ & 0.270 & 5.18 \\
PSZ2 G099.57-58.64 & 0h 3m 46.81s & 2$^{\circ}$ 3$'$ 13.1$''$ & 0.092 & 2.24 \\
\textbf{PSZ2 G167.98-59.95} & 2h 14m 44.21s & -4$^{\circ}$ 35$'$ 8.2$''$ & 0.139 & 4.31 \\
\textbf{PSZ2 G174.40-57.33} & 2h 31m 43.22s & -4$^{\circ}$ 51$'$ 40.1$''$ & 0.184 & 3.96 \\
PSZ2 G223.47+26.85 & 8h 45m 28.94s & 3$^{\circ}$ 28$'$ 31.4$''$ & 0.327 & 5.27 \\
PSZ2 G225.48+29.41 & 8h 57m 54.07s & 3$^{\circ}$ 10$'$ 27.9$''$ & 0.203 & 4.60 \\
\textbf{PSZ2 G228.50+34.95} & 9h 22m 14.54s & 3$^{\circ}$ 45$'$ 10.4$''$ & 0.270 & 5.78 \\
\textbf{PSZ2 G231.79+31.48} & 9h 16m 13.91s & -0$^{\circ}$ 24$'$ 42.5$''$ & 0.332 & 4.87 \\
PSZ2 G232.84+38.13 & 9h 40m 26.64s & 2$^{\circ}$ 28$'$ 18.9$''$ & 0.151 & 3.43 \\
PSZ2 G233.68+36.14 & 9h 35m 16.71s & 0$^{\circ}$ 49$'$ 6.4$''$ & 0.357 & 5.48 \\
PSZ2 G235.96+38.21 & 9h 46m 11.10s & 0$^{\circ}$ 29$'$ 10.0$''$ & 0.499$^{*}$ & -- \\
PSZ2 G273.59+63.27 & 12h 0m 26.18s & 3$^{\circ}$ 20$'$ 54.5$''$ & 0.134 & 5.54 \\
PSZ2 G286.39+64.06 & 12h 22m 48.77s & 2$^{\circ}$ 6$'$ 56.8$''$ & 0.229 & 4.10 \\
\end{tabular}
\end{ruledtabular}
\end{table*}

\subsection{Sources: HSC Y3 shape catalog}
In this paper, we use the HSC third-year (HSC-Y3) galaxy shape and photo-z catalogs based on the S19A data release acquired between March 2014 and April 2019 \citep{li22}. The S19 images are processed with the HSC Pipeline \citep{bosch2018} version {\tt{hscPipe v7}}, which in turn is a fork of the Vera C. Rubin Legacy Survey of Space and Time (LSST) Science Pipelines\footnote{\url{https://pipelines.lsst.io/}} \citep{2025rubn.rept...32N,bosch2019,2017ASPC..512..279J,2019ApJ...873..111I}. The HSC Y3 catalog represents a substantial improvement over the first-year HSC shape catalog \citep{mandelbaum18} based on the S16A HSC internal data release from August 2016 and processed with a previous version of the HSC Pipeline ({\tt{hscPipe v4}}, \citet{bosch2018}). A more detailed description of the improvements of the HSC Y3 shape catalog and the tests performed to ensure that it satisfies the HSC science requirements for Y3 science can be found in \citep{li22}. Here, we provide a summary of the improvements over the HSC-Y1 catalog, as well as the selection criteria and shape measurement algorithm. Key improvements in the HSC-Y3 catalog include an expanded survey area (from $\sim$137~deg$^{2}$ to $\sim$430~deg$^{2}$), updated point-spread function (PSF) modeling, improved star--galaxy separation, and more realistic image simulations for calibrating the multiplicative shear bias. The catalog comprises shapes from galaxies chosen from the ``full-depth full-color region'' in all five filters. Besides fundamental quality criteria about pixel-level data, our selection criteria encompass extended objects meeting the following conditions: extinction-corrected cmodel magnitude $i<24.5$, $i$-band signal-to-noise ratio (SNR) $\geq 10$, resolution $>0.3$, $>5\sigma$ detection in a minimum of two bands other than $i$, a 1 arcsec diameter aperture magnitude cutoff of $i<25.5$, and a blendedness cutoff in the $i$-band of $10^{-3.8}$.
The shape catalog contains 35.7 million galaxies, covering about 430 square degrees, with an effective number density of 19.9 arcmin$^{-2}$. It is divided into six regions: {\tt{XMM}}, {\tt{VVDS}}, {\tt{GAMA09H}}, {\tt{WIDE12H}}, {\tt{GAMA15H}}, and {\tt{HECTOMAP}} (see Figure 2 in \citet{li22}). Of these six fields, only five contain \emph{Planck} PSZ2 clusters from our sample; {\tt{GAMA15H}} has no overlap with the cluster sample and is therefore omitted from the per-field display in Fig.~\ref{fig:psz2}. Shape measurements in the catalog were calibrated using detailed image simulations to ensure that the uncertainty in the multiplicative shear bias is less than approximately $10^{-2}$. In addition to the HSC-Y3 shape catalog, a photometric redshift catalog is produced, using three methods, as outlined in \citep{nishizawa20}. The \mizuki\ software uses template fitting to estimate photometric redshifts, while \dempz\ and \dnnz\ use machine learning techniques. Each method provides a posterior distribution function $P(z_{\rm s})$ for the redshift of individual galaxies.

\section{Measurements}
\label{sec:measurements}

\subsection{Weak lensing}
\label{sec:lensing}

Gravitational lensing induces coherent distortions in the observed shapes of
background galaxies due to the gravitational potential of foreground matter.
Because galaxies have intrinsically nonzero ellipticities, the weak-lensing
signal must be extracted statistically by averaging over a large ensemble of
sources. The HSC-Y3 shape catalog provides, for each galaxy, the ellipticity
components $(e_1,e_2)$, inverse-variance weights $w_{\rm s}$, estimates of
multiplicative and additive shear biases $(m,c_1,c_2)$, and the ellipticity
dispersion $e_{\rm RMS}$. Galaxy shapes are measured on coadded \emph{i}-band
images using the re-Gaussianization method \citep{hirata2003}.

For a given lens--source pair, the tangential component of the galaxy ellipticity
relative to the lens position is defined as
\begin{equation}
e_{\rm t} = -e_1 \cos(2\phi) - e_2 \sin(2\phi),
\end{equation}
where $\phi$ is the angle between the lens--source separation vector and the
$x$-axis of the coordinate system in which the ellipticity components are measured. The ensemble average of $e_{\rm t}$ provides an estimator of the
tangential shear $\gamma_{\rm t}(R)$ as a function of projected comoving
separation $R$. The radial profile is measured in 20 logarithmically
spaced bins spanning $R\in[0.1,\,80]\,h^{-1}\,\mathrm{Mpc}$.

The tangential shear is related to the excess surface mass density,
\begin{equation}
\gamma_{\rm t}(R) =
\frac{\bar{\Sigma}(<R) - \Sigma(R)}{\Sigma_{\rm cr}(z_l,z_s)}
\equiv
\frac{\Delta\Sigma(R)}{\Sigma_{\rm cr}(z_l,z_s)},
\end{equation}
where $\Sigma(R)$ is the azimuthally averaged projected surface mass density at
projected radius $R$, and $\bar{\Sigma}(<R)$ is its mean within $R$.

The critical surface density is given by
\begin{equation}
\Sigma_{\rm cr}(z_l,z_s) =
\frac{c^2}{4\pi G}
\frac{D_A(z_s)}{(1+z_l)^2\,D_A(z_l)\,D_A(z_l,z_s)},
\end{equation}
where $D_A$ denotes angular diameter distances and the $(1+z_l)^2$ factor
accounts for the use of comoving coordinates.

The weak-lensing signal is measured using the pipeline described in
\citet{more23}, which implements the estimator below.
Lens and source positions are matched using a kd-tree, and per-galaxy
shape corrections (multiplicative bias, additive bias, selection bias,
and PSF leakage) are applied consistently.
Prior to the lensing measurement, we apply the standard HSC bright-star
masks and pixel-level quality flags in all bands to remove contaminated
regions \citep{li22}.

The observable lensing signal is the excess surface density $\Delta\Sigma(R)$,
which we estimate using a weighted average over lens--source pairs:
\begin{equation}
\Delta\Sigma(R) =
\frac{1}{1+\hat{m}}
\frac{
\sum_{\rm ls}
w_{\rm ls}\,
e_{{\rm t},\,{\rm ls}}\,
\langle \Sigma_{\rm cr}^{-1} \rangle_{{\rm ls}}^{-1}
}{
2\,\mathcal{R}\,
\sum_{\rm ls} w_{\rm ls}
},
\label{eq:dsigma_estimator}
\end{equation}
where $w_{\rm ls} = w_{\rm l}\,w_{\rm s}$ is the product of lens and source
weights. The factor $\mathcal{R}$ is the shear responsivity, which accounts for
the response of the measured ellipticity to a small applied shear. For the HSC
re-Gaussianization catalog, the responsivity is given by
\begin{equation}
\mathcal{R} = 1 - e_{\rm RMS}^2,
\end{equation}
and the factor of $2\mathcal{R}$ converts ellipticity to shear. The term
$\hat{m}$ denotes the weighted mean multiplicative shear bias correction,
computed from the per-galaxy $m$ values provided by the shape catalog.

The inverse critical surface density averaged over the source photometric
redshift posterior is
\begin{equation}
\left\langle \Sigma_{{\rm cr},\,{\rm ls}}^{-1} \right\rangle =
\frac{
\int_0^\infty P_s(z)\,\Sigma_{\rm cr}^{-1}(z_l,z)\,dz
}{
\int_0^\infty P_s(z)\,dz
},
\end{equation}
where $P_s(z)$ is the photometric-redshift posterior of the source galaxy.

The re-Gaussianization shear estimates in the S19A catalog are calibrated using
image simulations, which are used to determine the multiplicative and additive
shear biases as functions of galaxy properties. These calibration corrections
are applied consistently in the estimator above.

As part of the HSC-Y3 blinding protocol \citep{li22}, the multiplicative
shear bias values in the shape catalog are blinded by adding a small
unknown offset, preventing confirmation bias during the analysis (``user-level" blinding, as described in \citep{li22}).

\begin{figure*}[t]
\centering
\includegraphics[width=0.45\textwidth, page=1]{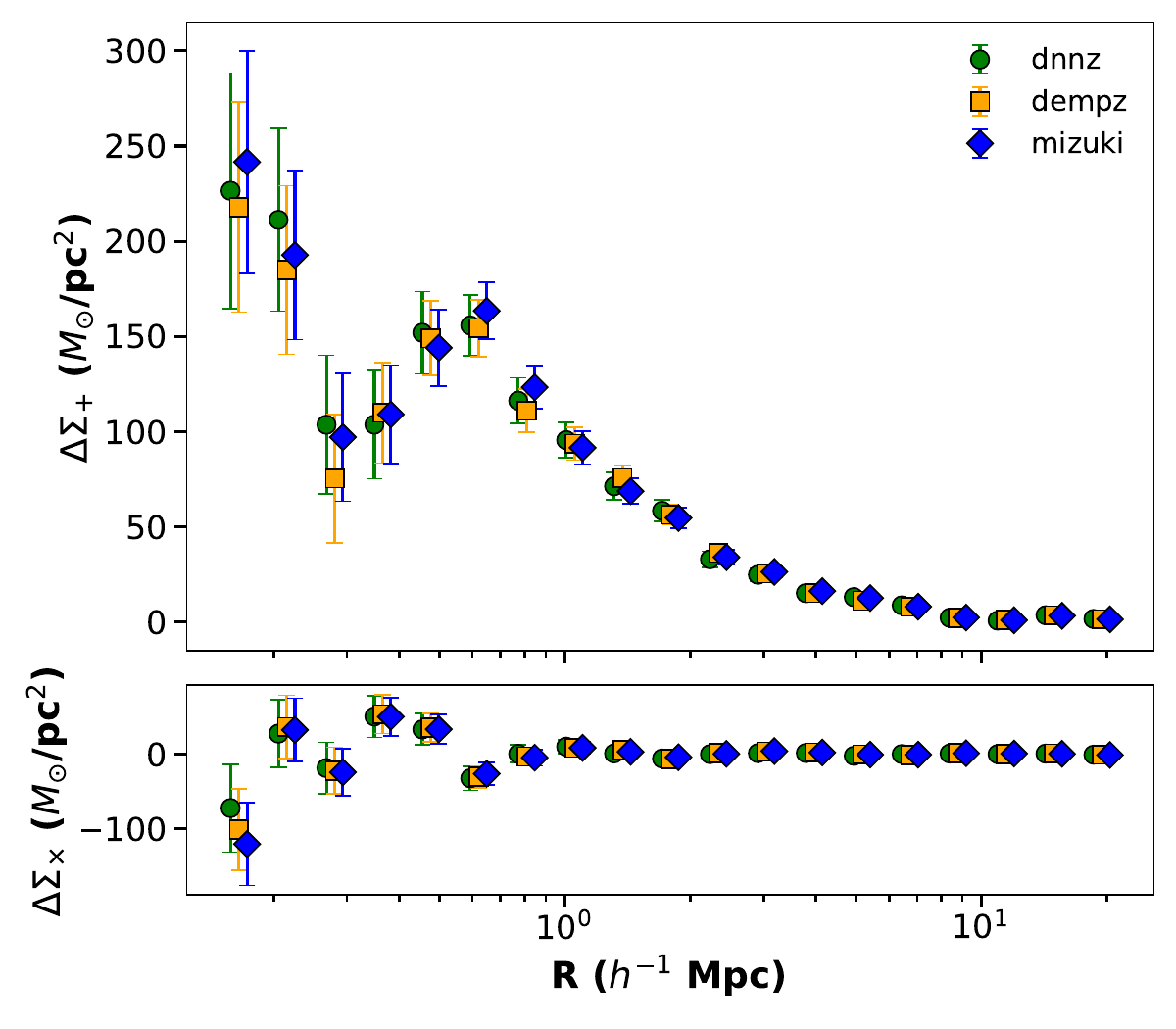}\hfill
\includegraphics[width=0.45\textwidth, page=2]{figures/delta_sigma_photo_zs.pdf}
\caption{Stacked weak-lensing signal from the 19 clusters in the sample (Table \ref{table1:clusters}) using the ``$P(z)$-cut'' source-galaxy selection, shown for the three HSC-SSP photometric-redshift estimators \mizuki, \dnnz, and \dempz. \emph{Left}: $\Delta\Sigma(R)$. \emph{Right}: the same measurements multiplied by $R$, i.e.\ $R\,\Delta\Sigma(R)$. The measurements are shifted in the horizontal axis in both panels for visualization.}
\label{fig:delta_r_cuts_photo_z_comparison}
\end{figure*}

\subsection{Source galaxy selection}

Contamination of the background source sample by galaxies physically associated with the cluster (cluster members) or in the foreground dilutes the measured weak-lensing signal, biasing the inferred mass low. We tested two source-selection methods following \citet{medezinski18} and \citet{miyatake19}. The first method is based on color--color cuts (the ``CC-cut'' method; \citealt{2011MNRAS.414.1840M,medezinski18}), defined in the $g-i$ versus $r-z$ space to minimize contamination from cluster members and foreground galaxies. The specific boundaries of the cuts differ for clusters at $z_l < 0.4$ and $z_l > 0.4$, ensuring that galaxies behind low-redshift clusters are not excessively removed.

The second approach, the ``$P(z)$-cut'' method \citep{oguri14,miyatake19}, uses the full photometric-redshift probability distribution function, $P(z)$, to identify secure background galaxies that satisfy
\begin{equation}
    p_{\mathrm{cut}} < \int_{z_{\mathrm{min}}}^{\infty} P(z)\,dz ,
\end{equation}
where $p_{\mathrm{cut}} = 0.98$ and $z_{\mathrm{min}} = z_l + \Delta z$, with $z_l$ denoting the cluster redshift and $\Delta z = 0.2$ in our analysis \citep{miyatake19,medezinski18}.

We verified that both the CC-cut and the $P(z)$-cut methods, when applied with the \mizuki\ photometric-redshift estimates, yield consistent stacked lensing signals with no statistically significant differences in amplitude or shape.

The HSC photometric-redshift (photo-$z$) catalog includes estimates from several independent algorithms that adopt complementary approaches \citep{nishizawa20}.
\textsc{Mizuki} is a Bayesian template-fitting method that matches observed multi-band photometry to spectral energy distribution (SED) models while incorporating physical priors on galaxy properties such as stellar mass and star-formation history \citep{Tanaka2015}.
\textsc{DEmPz} (Direct Empirical Photometric redshifts) is an empirical algorithm that uses local polynomial fitting trained on a spectroscopic reference sample to predict redshifts from observed colors and magnitudes \citep{HSC_PZ}.
\textsc{DNNz} employs a deep neural network architecture that maps multi-band fluxes and shape parameters to redshift probability distributions, capturing complex non-linear relationships between observables and redshift \citep{nishizawa20}.

Figure~\ref{fig:delta_r_cuts_photo_z_comparison} shows the measured $\Delta\Sigma$ and $\Delta\Sigma\times R$ profiles obtained with the $P(z)$-cut method for three different photometric-redshift algorithms: \dnnz, \dempz, and \mizuki. The results from these three photo-$z$ estimators are mutually consistent within the statistical uncertainties across all radial bins. This insensitivity to the photo-$z$ method demonstrates that the stacked lensing signal is robust against the choice of photometric-redshift algorithm.
As an additional diagnostic, the measurement pipeline records the
stacked source photo-$z$ probability distributions $P(z_s)$ in radial
bins around each cluster. We verified that these distributions do not
show evidence of significant cluster-member contamination at any radius
within our fitting range, confirming the effectiveness of the $P(z)$-cut
selection.

Given this agreement, and for consistency with other HSC Year~3 weak-lensing analyses \citep[e.g.,][]{more23,Shirasaki2024}, we adopt the $P(z)$-cut method with \dempz\ photometric redshifts as our fiducial source selection in the remainder of this work.

\subsection{Cluster centers}
\label{sec:centers}
The choice of cluster center directly affects the measured lensing profile, particularly on small scales where miscentering suppresses the signal \citep{miyatake19,shin25}. We examined the effect of different choices of cluster centers on the measured lensing signal by comparing profiles computed using centers from \emph{Planck} SZ positions and from the \textsc{Camira} optical cluster catalog \citep{oguri14,oguri18}. The \textsc{Camira} algorithm identifies the brightest cluster galaxy (BCG) as the cluster center, which is typically closer to the true halo center than the SZ centroid determined from the low-angular-resolution \emph{Planck} maps. The comparisons indicate that (1) the lensing signal on small scales ($R \lesssim 1\,h^{-1}\,\mathrm{Mpc}$) is slightly higher when using \textsc{Camira} centers, consistent with reduced miscentering, and (2) the $P(z)$-cut selection yields a marginally higher signal than the CC-cut method, as expected from the more conservative background selection. Most of the 19 clusters in our sample have an identified \textsc{Camira} center; however, for three systems no \textsc{Camira} match was available, in which case we adopted the center from the SDSS \texttt{redMaPPer} DR8 catalog \citep{rykoff16} or from the \emph{Planck} catalog itself (see Table~\ref{table1:clusters}). The residual effect of miscentering is modeled explicitly in our forward-model analysis (Section~\ref{sec:miscenter_model}).

\begin{figure}[t]
  \centering
  \includegraphics[width=0.95\columnwidth]{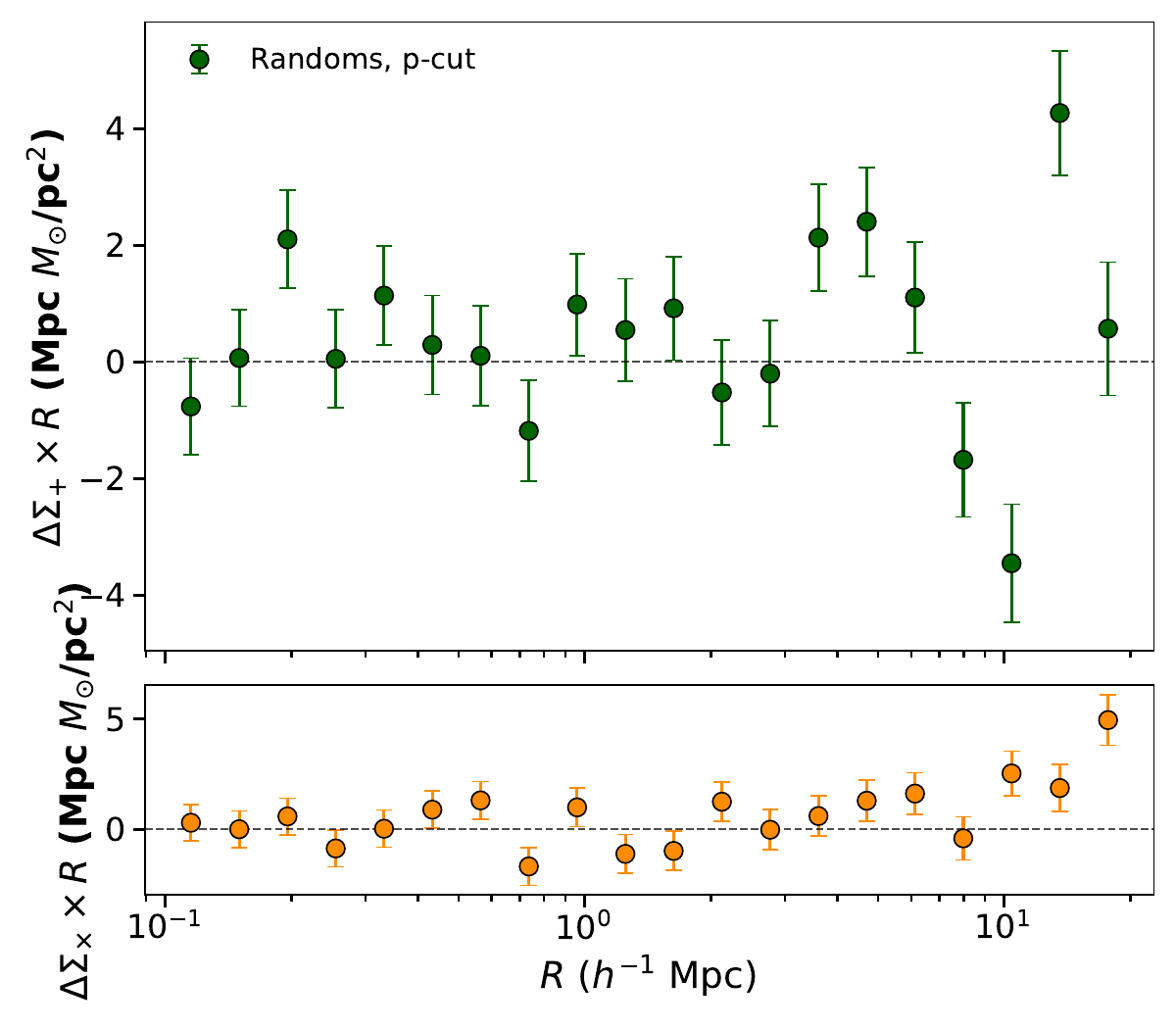}
  \caption{
Stacked lensing signal measured around random points using the $P(z)$-cut source selection. The dashed line in each panel marks $\Delta\Sigma=0$.
\emph{Top}: Tangential (E-mode) component, $\Delta\Sigma_{+} \times R$.
Over the full radial range the signal is inconsistent with zero
($\chi^2/\mathrm{dof}=58.1/20$, $p=1.4\times10^{-5}$),
indicating large-scale survey-related systematics.
On the extended test range
$0.5 < R < 10\,h^{-1}\,\mathrm{Mpc}$ the tension is mild
($\chi^2/\mathrm{dof}=21.3/11$, $p=0.030$), and on the fiducial fitting range
$0.5 < R < 5\,h^{-1}\,\mathrm{Mpc}$ the E-mode is marginally consistent with zero
($\chi^2/\mathrm{dof}=17.0/9$, $p=0.048$).
\emph{Bottom}: Cross (B-mode) component, $\Delta\Sigma_{\times} \times R$, consistent with zero on the extended test range
($\chi^2/\mathrm{dof}=17.6/11$, $p=0.092$) and on the fiducial fitting range
($\chi^2/\mathrm{dof}=14.5/9$, $p=0.11$).
These results motivate our conservative choice of
$R_{\rm max}=5\,h^{-1}\,\mathrm{Mpc}$ for the fiducial analysis
(Section~\ref{sec:fit_range}).
}
  \label{fig:signal_randoms}
\end{figure}

\subsection{Signal around random points}
\label{sec:randoms}

A standard null test in stacked weak-lensing analyses is to measure the tangential shear signal around random points that follow the same angular footprint and selection criteria as the lens sample \citep{mandelbaum2006,miyatake19,Shirasaki2024,shin25}. A non-zero signal around random points would indicate the presence of additive systematic biases (e.g., residual PSF leakage, survey geometry effects, or depth variations) that could contaminate the cluster lensing measurement. We measured the lensing signal around 1900 random points (100 per cluster, to match the statistical weight of the cluster sample) distributed across all 5 HSC fields used in this work. The random positions are drawn from the HSC random catalog, which is downsampled by a factor of $10^5$ to produce a manageable point set that uniformly samples the survey footprint while respecting the bright-star masks and depth variations. Each random point is assigned a redshift drawn (with replacement) from the set of cluster redshifts, so that the lens-redshift distribution of the randoms matches that of the cluster stack and the critical surface density $\Sigma_{\rm cr}$ is computed on the same redshift mix.

We assess the random-point signal using a $\chi^2$ null test,
\begin{equation}
\chi^2_{\rm rand} \equiv
\mathbf{d}_{\rm rand}^{\mathsf T}\,
\mathbf{C}_{\rm rand}^{-1}\,
\mathbf{d}_{\rm rand},
\end{equation}
where $\mathbf{d}_{\rm rand}$ is the vector of random-point $\Delta\Sigma$ measurements and
$\mathbf{C}_{\rm rand}$ is the corresponding covariance matrix.  We adopt the diagonal approximation constructed from the weak-lensing pipeline shape-noise uncertainties.
We quote $\chi^2$, the number of degrees of freedom ($\mathrm{dof}=N_{\rm bin}$), and the associated $p$-value.

Over the full radial range, both the E-mode and B-mode random-point signals are inconsistent with zero, with
$\chi^2_{\rm rand,E}/\mathrm{dof}=58.1/20$ ($p=1.4\times10^{-5}$) and
$\chi^2_{\rm rand,B}/\mathrm{dof}=48.4/20$ ($p=3.8\times10^{-4}$),
indicating the presence of large-scale survey-related systematics.
These large-scale residuals are consistent with the additive PSF-related
systematics flagged in the HSC Y3 shape catalog null tests
\citep{li22}, where the star--galaxy correlation $\xi_{\rm sys}$ reaches
the level of the cosmic shear signal on degree scales and varies among
fields. Their imprint is enhanced in $\Delta\Sigma$ relative to $\gamma_t$
by the $\Sigma_{\rm cr}$ weighting, which motivates our maximum fitting
radius and the random-point subtraction applied to the cluster signal.

Restricting the analysis to the fitting range
$0.5 < R < 10\,h^{-1}\,\mathrm{Mpc}$ substantially reduces these residuals.
On this range, the B-mode random-point signal is statistically consistent with zero,
$\chi^2_{\rm rand,B}/\mathrm{dof}=17.6/11$ ($p=0.092$),
while the E-mode shows only mild tension,
$\chi^2_{\rm rand,E}/\mathrm{dof}=21.3/11$ ($p=0.030$).
Further restricting to the fiducial fit range
$0.5 < R < 5\,h^{-1}\,\mathrm{Mpc}$ yields
$\chi^2_{\rm rand,E}/\mathrm{dof}=17.0/9$ ($p=0.048$) and
$\chi^2_{\rm rand,B}/\mathrm{dof}=14.5/9$ ($p=0.11$),
indicating that the random-point signals are statistically consistent with zero on the scales used for the cluster mass fit, with only marginal residual tension in the E-mode.
We therefore adopt a conservative maximum fitting radius of
$R_{\rm max}=5\,h^{-1}\,\mathrm{Mpc}$ for our fiducial stacked analysis and verify that
extending to $R_{\rm max}=10\,h^{-1}\,\mathrm{Mpc}$ yields consistent results
(Section~\ref{sec:robustness}).  We subtract the random-point signal from the measured stacked signal in the single-mass bin and stacked modeling fits in Sections \ref{sec:single-mass-bin-fit} and \ref{sec:stacked_model}.

\subsection{Boost factor}
\label{sec:boost}

The boost factor is defined as the ratio of source--cluster pairs to source--random pairs as a function of projected separation $R$. A non-unity boost factor indicates that some source galaxies are physically associated with the lens clusters, which can bias the weak-lensing signal by diluting the measured shear. This quantity therefore serves as a diagnostic of the effectiveness of the background-galaxy selection.

Following \citet{miyatake19}, we define the boost factor as
\begin{equation}
    B(R_i) =
    \frac{
        \displaystyle \sum_{ls \in R_i} w_{ls}\,\langle \Sigma_{\mathrm{crit}}^{-1} \rangle^{-1} / \sum_{l} w_{l}
    }{
        \displaystyle \sum_{rs \in R_i} w_{rs}\,\langle \Sigma_{\mathrm{crit}}^{-1} \rangle^{-1} / \sum_{r} w_{r}
    },
\end{equation}
where the numerator represents the weighted number density of lens--source pairs and the denominator that of random--source pairs, both averaged within a given radial bin $R_i$. The boost factor has been widely used to diagnose and correct for signal dilution due to physically associated sources mistakenly included in the background sample \citep{mandelbaum2006,hirata2004,miyatake19,Shirasaki2024}. In practice, this correction involves comparing weighted source densities around random points that follow the same angular mask and selection criteria as the lens sample \citep{sheldon2004}.

To measure $B(R)$, we generated multiple random realizations, each containing the same number of random points as the cluster sample (1900 in our case). We computed the boost factor for each realization and used the ensemble to estimate the mean $B(R)$ and its covariance matrix, including correlations among radial bins.

\begin{figure}[t]
\centering
\includegraphics[width=0.95\columnwidth]{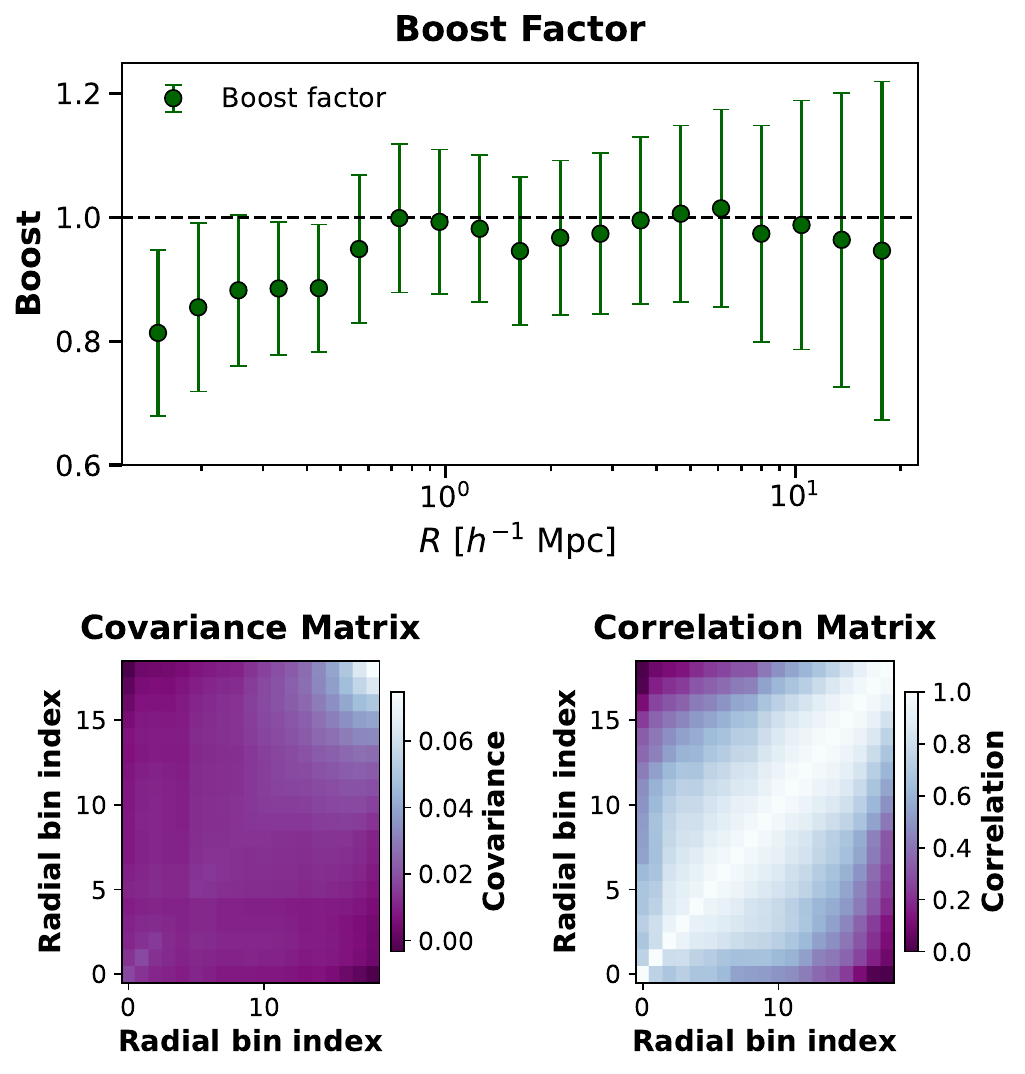}
\caption{
\emph{Top}: Measured boost factor as a function of projected separation $R$.
\emph{Bottom left}: Covariance matrix of the boost factor, estimated from 50 random realizations.
\emph{Bottom right}: Corresponding correlation matrix.
The boost factor remains consistent with unity across most scales, indicating negligible contamination by physically associated source galaxies.
A mild suppression at $R < 0.5\,h^{-1}\,\mathrm{Mpc}$ is likely caused by obscuration from cluster member galaxies, and we exclude these scales from our analysis.
}
\label{fig:boost_factor}
\end{figure}

The measured boost factor, shown in Fig.~\ref{fig:boost_factor}, is consistent with unity across most scales. A $\chi^2$ test of $B(R)=1$ over all 19 radial bins yields $\chi^2/\mathrm{dof}=18.0/19$, indicating no statistically significant contamination by physically associated sources. At small scales ($R \lesssim 0.5\,h^{-1}\,\mathrm{Mpc}$), we observe a mild suppression of the boost factor ($B \sim 0.8$--$0.9$), likely due to obscuration by bright cluster galaxies; these scales are excluded from the fiducial analysis. At intermediate scales ($R \sim 1$--$3\,h^{-1}\,\mathrm{Mpc}$), the boost factor is slightly above unity but remains consistent within the $1\sigma$ uncertainties. The absence of significant deviations from unity in the boost factor indicates that contamination by physically associated sources is not responsible for the large-scale systematics observed in the random-point shear signal.

\section{Modeling and results}
\label{sec:results}

\subsection{Individual cluster weak-lensing fits}
\label{sec:indiv}

As a diagnostic of the data quality and modeling assumptions, we fit the
measured excess surface density profiles $\Delta\Sigma(R)$ of individual
clusters with a spherical Navarro--Frenk--White (NFW) halo profile
\citep{navarro1995simulations}. These per-cluster fits are not intended to
provide high-precision mass measurements, but rather to assess the consistency
of the measured profiles with standard halo models on intermediate scales and
to identify potential outliers or systematic issues. Our fiducial mass
calibration is based on the stacked analysis described in
Section~\ref{sec:stacked_model}.

The three-dimensional density profile is given by
\begin{equation}
  \rho_{\rm NFW}(r) =
  \frac{\rho_s}{(r/r_s)\,\left(1+r/r_s\right)^2},
\end{equation}
and is parameterized by the halo mass $M_{200\mathrm{m}}$ and concentration
$c_{200\mathrm{m}}\equiv R_{200\mathrm{m}}/r_s$, where $M_{200\mathrm{m}}$ is the
mass enclosed within the radius $R_{200\mathrm{m}}$ for which the mean enclosed
density equals $200$ times the mean matter density at the cluster redshift $z$.
For a given $(M_{200\mathrm{m}},c_{200\mathrm{m}},z)$, we compute the predicted
$\Delta\Sigma(R)$ by projecting the NFW profile.

We evaluate the model using the public \textsc{Colossus} package
\citep{diemer18}, which self-consistently adopts the mean matter density
$\bar{\rho}_{\mathrm{m}}(z)$ for the mass definition and provides the projected
surface density for NFW halos. The measured profiles use comoving radii in
$h^{-1}\,\mathrm{Mpc}$ and comoving surface density in
$h\,M_\odot\,\mathrm{pc}^{-2}$. When evaluating the model, we convert radii to
physical $h^{-1}\,\mathrm{kpc}$ for \textsc{Colossus} and transform the output
surface density (physical $h\,M_\odot\,\mathrm{kpc}^{-2}$, reflecting the
$M_\odot/h$ mass and $\mathrm{kpc}/h$ distance units used internally by
\textsc{Colossus}) back to our comoving convention via
\begin{equation}
  \Delta\Sigma_{\rm com} =
  \frac{\Delta\Sigma_{\rm phys}}{10^{6}\,(1+z)^2}.
\end{equation}

Following \citet{miyatake19}, we restrict the fits to the radial range $0.3~h^{-1}\,\mathrm{Mpc} < R < 3~h^{-1}\,\mathrm{Mpc}$, which reduces sensitivity to blending, obscuration, and potential miscentering at small radii, as well as to the increasing contribution of the two-halo term and large-scale systematics at larger separations. We note that this radial range differs from that adopted in the stacked analysis (Section~\ref{sec:stacked_model}), where the scale cuts are optimized based on null tests of the full sample. In contrast, the per-cluster fits presented here are intended as a qualitative diagnostic of profile shapes and are therefore restricted to intermediate scales where the single-halo NFW model provides an adequate description and the signal-to-noise is highest.

We estimate $(M_{200\mathrm{m}},c_{200\mathrm{m}})$ using Bayesian inference with
a Gaussian likelihood,
\begin{equation}
  -2\ln\mathcal{L} =
  \sum_i
  \left[
  \frac{\Delta\Sigma_i -
  \Delta\Sigma_{\rm model}(R_i\,|\,M_{200\mathrm{m}},c_{200\mathrm{m}})}
  {\sigma_i}
  \right]^2,
\end{equation}
where $\Delta\Sigma_i$ and $\sigma_i$ are the measured surface density contrast
and its statistical uncertainty in radial bin $i$. For these individual-cluster
fits, we adopt a diagonal covariance constructed from the measurement
uncertainties (shape noise and measurement noise) provided by the weak-lensing pipeline.

The posterior distribution is sampled using the affine-invariant ensemble
sampler \texttt{emcee} \citep{Foreman-Mackey13}. We adopt broad, flat priors
$10^{13}\,h^{-1}M_\odot < M_{200\mathrm{m}} < 10^{16}\,h^{-1}M_\odot$ and
$0.5 < c_{200\mathrm{m}} < 8$. For each cluster, we report the marginalized
median and 68\% credible intervals of $M_{200\mathrm{m}}$ and $c_{200\mathrm{m}}$.
All halo masses are quoted in units of $h^{-1}M_\odot$, consistent with the
\textsc{Colossus} convention.

\begin{figure*}[t]
\centering
\includegraphics[width=\textwidth]{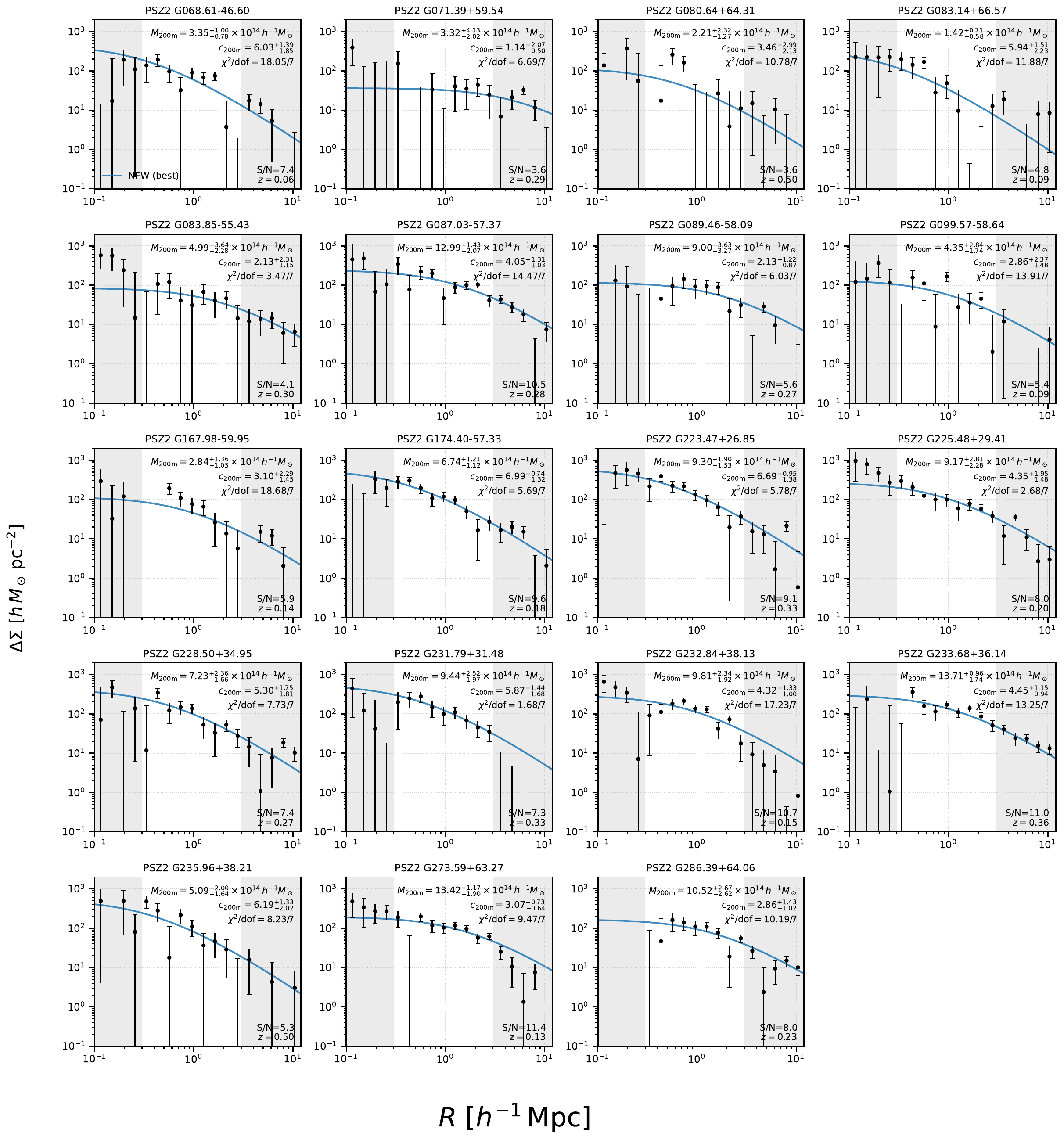}
\caption{
Excess surface density profiles $\Delta\Sigma(R)$ for individual
\textit{Planck}-selected clusters, shown with their best-fitting NFW models.
Fits use the radial range $R\in[0.3,\,3.0]\,h^{-1}\,\mathrm{Mpc}$ (shaded
band), the median $M_{200\mathrm{m}}$ and $c_{200\mathrm{m}}$ with 68\%
credible intervals, $\chi^2/\mathrm{dof}$, signal-to-noise ratio, and
cluster redshift are indicated in each panel.
These fits are used as diagnostics of the weak-lensing measurements and
modeling assumptions; the fiducial mass calibration relies on the stacked
analysis in Section~\ref{sec:stacked_model}.
}
\label{fig:indiv_nfw_fits}
\end{figure*}

Figure~\ref{fig:indiv_nfw_fits} shows the measured $\Delta\Sigma(R)$ profiles for
individual clusters together with their best-fitting NFW models. The
signal-to-noise ratio is computed over the fitting range as
\begin{equation}
  (\mathrm{S/N})^2 =
  \sum_i \left(\frac{\Delta\Sigma_i}{\sigma_i}\right)^2,
\end{equation}
and the cluster redshift is indicated in each panel.

Because these per-cluster fits neglect an explicit two-halo term and do not
model miscentering, the inferred concentrations should be interpreted with
caution and are treated primarily as diagnostic parameters. Both effects can
bias $c_{200\mathrm{m}}$ or increase $\chi^2$ at fixed mass. Residual shear- and
photometric-redshift-calibration uncertainties enter as an overall
multiplicative factor on $\Delta\Sigma$; at the signal-to-noise of individual
clusters, we find that such uncertainties have a negligible impact on
$M_{200\mathrm{m}}$. We also verify that the cross-component
$\Delta\Sigma_\times$ is consistent with zero for all clusters.
The parameter uncertainties reported in
Fig.~\ref{fig:indiv_nfw_fits} are the 68\% credible intervals from the
marginalized posteriors; for clusters with very high signal-to-noise, the
uncertainties on $c_{200\mathrm{m}}$ can appear small, but these should be
interpreted with the above caveats in mind (omission of the two-halo term and
miscentering can bias the inferred concentration).

\subsection{Single mass bin fit for stacked signal}
\label{sec:single-mass-bin-fit}

\begin{figure}[t]
\centering
\includegraphics[width=0.95\columnwidth]{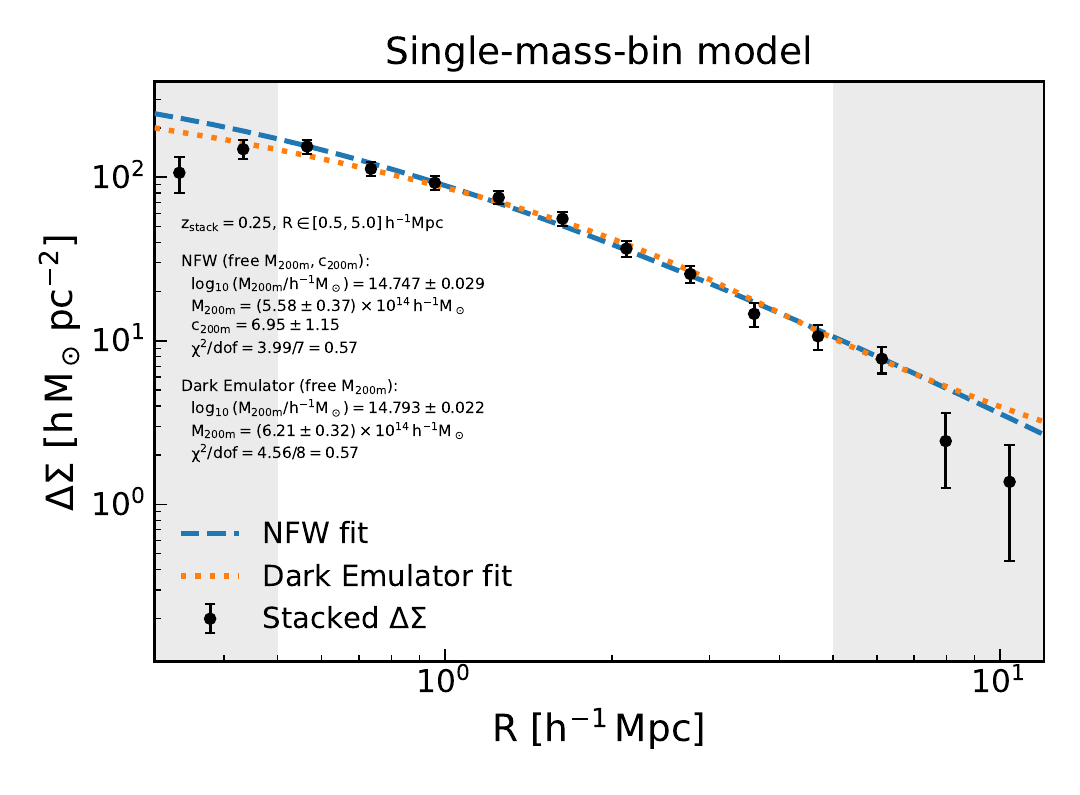}
\caption{
Single-mass-bin fits to the stacked weak-lensing excess surface density
profile $\Delta\Sigma(R)$, assuming all clusters are described by a single
halo mass at the effective stack redshift ($z_{\rm stack}=0.25$).
The random-subtracted stacked signal (black points with error bars) is fit
using an NFW profile (blue dashed line) with free
$(M_{200\mathrm{m}},c_{200\mathrm{m}})$ and the Dark Emulator prediction
(orange dotted line) with free $M_{200\mathrm{m}}$.
The fit uses scales $R\in[0.5,\,5.0]\,h^{-1}\,\mathrm{Mpc}$, matching the
headline stacked-model analysis; the shaded region indicates excluded scales.
The two models yield consistent masses
($M_{200\mathrm{m}}\approx 5.6$ and $6.2\times10^{14}\,h^{-1}M_\odot$,
respectively), with the $\sim$10\% residual offset attributable to the
two-halo term included in the Dark Emulator but absent in the analytic NFW
profile.
These single-mass-bin fits are diagnostic only; the primary mass calibration
comes from the stacked forward model (Fig.~\ref{fig:stacked_model_fit}).
}
\label{fig:models_fits}
\end{figure}

As an initial characterization of the stacked lensing
measurement, we fit the observed excess surface density profile
$\Delta\Sigma(R)$ with single-halo models assuming that the entire cluster
sample can be represented by a single characteristic halo at an effective lens
redshift \citep{miyatake19}.
This ``single-mass-bin'' description is intended only as a diagnostic and visual
summary of the signal (Fig.~\ref{fig:models_fits}); our
constraint on the hydrostatic mass bias parameter $b$ is obtained from the
stacked-model analysis described in Section~\ref{sec:stacked-model-fit}, which
accounts for selection effects and cluster miscentering.

The stacked signal used in Fig.~\ref{fig:models_fits} is the
\emph{random-subtracted} measurement,
$\Delta\Sigma_{\rm corr}(R)\equiv \Delta\Sigma_{\rm data}(R)-\Delta\Sigma_{\rm rand}(R)$,
where $\Delta\Sigma_{\rm rand}$ is measured around random points that follow the
survey footprint.
This subtraction mitigates additive systematics on large scales (e.g., residual
shear systematics, survey geometry, and depth variations) that can otherwise
bias the stacked tangential shear and hence $\Delta\Sigma$.

For the single-mass-bin fits, we fix the lens redshift to the mean redshift of
the cluster sample, $z_{\rm stack}=0.25$, and fit two model families:
(i) an NFW profile with free $(M_{200\mathrm{m}},c_{200\mathrm{m}})$
computed with \textsc{Colossus}, and
(ii) the \textsc{Dark Emulator} prediction \citep{Nishimichi2019} with free
$M_{200\mathrm{m}}$.
Fits are performed over $R\in[0.5,5.0]\,h^{-1}\,\mathrm{Mpc}$, matching
the radial range of the headline stacked-model analysis.
The minimum scale is motivated
by the observed suppression of the boost factor at $R<0.5\,h^{-1}\,\mathrm{Mpc}$,
likely due to obscuration by cluster member galaxies, and the maximum scale
is chosen to reduce sensitivity to the two-halo term and large-scale systematics.
Because these fits neglect the mass/redshift distribution of the sample, the
\textit{Planck} selection function, and explicit miscentering, the inferred
single-halo parameters should be interpreted as effective values rather than a
mass calibration; they are therefore not used to derive our final constraint on
$1-b$.

Similar single-halo modeling approaches have been applied in earlier stacked weak-lensing studies of \textit{Planck}-selected clusters, yielding comparable effective mass constraints and emphasizing the limitations of such models for precise cosmological interpretation \citep[e.g.,][]{PennaLima2017, Sereno2017, Schrabback2021}. While useful for consistency checks, these simplified fits do not account for the intrinsic cluster mass distribution, selection effects, or miscentering, and hence are not used here to constrain the hydrostatic mass bias.

\subsection{Stacked model fit}
\label{sec:stacked_model}

Our primary goal is to constrain the hydrostatic mass bias parameter $b$,
relating the \textit{Planck} SZ mass proxy $M_{\rm SZ}$ to the true halo mass $M$ via $M_{\rm Planck} = (1-b)\,M_{\rm true}$,
using the stacked weak-lensing measurement of the \textit{Planck}-selected
cluster sample. Values of $(1-b)<1$ indicate that \textit{Planck} underestimates
the true mass (hydrostatic mass bias), and $(1-b)=1$ means no bias.

We follow the stacked-model methodology of
\citet[][Section~4.2.2]{miyatake19}, in which the model prediction for each
cluster $i$ at redshift $z_i$ is computed by integrating the halo--matter
$\Delta\Sigma$ profile over the halo mass function weighted by the effective
selection function:
\begin{equation}
\label{eq:single_cluster_model}
\Delta\Sigma_{\rm model}^{(i)}(R) =
\frac{
  \int d\ln M\; W(M,z_i;b)\,
  \Delta\Sigma_{\rm tot}(R|M,z_i)
}{
  \int d\ln M\; W(M,z_i;b)
}\,,
\end{equation}
where
\begin{equation}
W(M,z_i;b) \equiv
M\,\frac{dn}{d\ln M}(M,z_i)\,
\langle S\rangle(M,z_i;b).
\end{equation}
and where $dn/d\ln M$ is the halo mass function \citep{tinker08} computed via the
\textsc{Dark Emulator} \citep{Nishimichi2019}, $\langle S\rangle$ is the
effective selection function convolved with log-normal scatter
(Section~\ref{sec:eddington_model}), and $\Delta\Sigma_{\rm tot}$ includes
miscentering (Section~\ref{sec:miscenter_model}).

This forward-modeling framework accounts self-consistently for Eddington bias \citep{battaglia16} (the preferential up-scattering of low-mass halos into an SZ-selected sample due to the steeply falling halo mass function) and has been successfully applied in recent cluster mass calibration studies \citep{miyatake19,Shirasaki2024,shin25}. We adopt a similar approach here, incorporating emulator-based predictions for the halo--matter correlation function \citep{Nishimichi2019} and the \citet{Diemer2015} concentration--mass relation. This approach is preferred over a simple ratio of stacked WL mass to stacked SZ mass because it properly accounts for the non-trivial relationship between the observed and true mass distributions induced by selection effects and scatter in the mass--observable relation \citep{miyatake19}.

\subsubsection{Weighted stacking}
\label{sec:weighted_stacking}
The measurement pipeline computes the stacked $\Delta\Sigma$ using per-cluster
weak-lensing weights $w_i(R)$ that reflect the effective number of useful
source--lens pairs at each projected radius $R$ for each cluster $i$.
Specifically, $w_i(R) = \sum_{\rm ls\in R} w_{\rm ls}\,(1+m_s)$,
where $w_{\rm ls} = w_l\,w_s\,\langle\Sigma_{\rm cr}^{-1}\rangle^2$ is
the per-pair lensing weight and $(1+m_s)$ is the per-source multiplicative
bias correction, as output by the measurement pipeline \citep{more23}. The stacked data are therefore
\begin{equation}
\label{eq:weighted_stack}
\Delta\Sigma_{\rm stack}(R) =
\frac{\sum_{i=1}^{N_{\rm cl}} w_i(R)\,\Delta\Sigma_i(R)}
     {\sum_{i=1}^{N_{\rm cl}} w_i(R)}\,.
\end{equation}
To ensure consistency between the model prediction and the data, we apply the
same per-cluster weights when computing the theoretical stacked signal,
\begin{equation}
\label{eq:weighted_model}
\Delta\Sigma_{\rm stack}^{\rm (model)}(R) =
\frac{\sum_{i=1}^{N_{\rm cl}} w_i(R)\,\Delta\Sigma_{\rm model}^{(i)}(R)}
     {\sum_{i=1}^{N_{\rm cl}} w_i(R)}\,,
\end{equation}
where $\Delta\Sigma_{\rm model}^{(i)}$ is given by
Eq.~(\ref{eq:single_cluster_model}).

The effective redshift of the weighted stack is
\begin{equation}
z_{\rm eff} = \frac{\sum_i \bar{w}_i\,z_i}{\sum_i \bar{w}_i}\,,
\end{equation}
where $\bar{w}_i=\langle w_i(R)\rangle_R$ is the mean weight of cluster $i$
across radial bins. For our sample, $z_{\rm eff}\simeq 0.24$, compared with a
simple unweighted mean of $\bar{z}=0.25$.

\subsubsection{Planck selection function and Eddington bias}
\label{sec:eddington_model}
We adopt the official \textit{Planck} selection function\footnote{\texttt{HFI\_PCCS\_SZ-selfunc-union-cosmolog\_R2\_08.fits}, \url{https://irsa.ipac.caltech.edu/data/Planck/release_2/ancillary-data/HFI_Products.html}},
which provides the completeness as a function of $(Y_{500},\theta_{500})$
for the Union catalog detection pipeline.
We apply a signal-to-noise threshold of $\mathrm{SNR}>4.5$, consistent
with the \emph{Planck} SZ2 catalog selection \citep{planck16}.
The completeness grid is converted from $(Y_{500},\theta_{500})$ to
$(M_{500\mathrm{c}},z)$ using the $Y$--$M$ scaling relation of
\citet{2016AA...594A..24P}:
\begin{equation}
\label{eq:YM_scaling}
\begin{split}
D_A^2 Y_{500} &=
10^{-0.19}\times 10^{-4}
\left(\frac{M_{500\mathrm{c}}}{6\times10^{14}\,M_\odot}\right)^{1.79} \\
&\quad \times E(z)^{0.66}\;[\mathrm{Mpc}^2] .
\end{split}
\end{equation}
where $D_A$ is the angular diameter distance and $E(z)=H(z)/H_0$.

The SZ observable $Y_{500}$ scales with the \textit{Planck} mass
$M_{\rm Planck}=(1-b)\,M_{\rm true}$, so the selection function is evaluated at
$M_{\rm eff}=(1-b)\,M_{\rm true}$. To account for Eddington bias from the
steeply falling mass function, we convolve with a log-normal scatter:
\begin{equation}
\label{eq:eff_selection}
\begin{split}
\langle S\rangle(M_{\rm true},z)
&=
\int_{-\infty}^{\infty}
\frac{dt}{\sqrt{2\pi}}\,
e^{-t^2/2} \\
&\quad \times
S\!\left(
(1-b)M_{\rm true}\,e^{\sigma_{\ln M}t},z
\right).
\end{split}
\end{equation}

where $\sigma_{\ln M}$ is the log-normal SZ scatter, treated as a free
parameter in the MCMC fit with a Gaussian prior centered at $0.25$
(Table~\ref{tab:priors}).

We note that the angular size $\theta_{500}$ entering the Planck
selection function is recomputed self-consistently at each trial mass and
redshift: the precomputed completeness grid maps each $(M_{500\mathrm{c}},z)$
point to $(Y_{500},\theta_{500})$ via the $Y$--$M$ scaling relation and
$\theta_{500}=R_{500\mathrm{c}}(M,z)/D_A(z)$. When the MCMC varies $(1-b)$
and $\sigma_{\ln M}$, the effective mass
$M_\mathrm{eff}=(1-b)\,M_\mathrm{true}$ (or its scatter-shifted value) is
used to look up completeness on this grid, so both $Y_{500}$ and
$\theta_{500}$ change in lockstep with the sampled mass.

Masses at the grid edges of the completeness map are clipped to the boundary
rather than zeroed, ensuring that the high-mass tail is not artificially
truncated. Only masses genuinely below the minimum of the selection-function grid
receive $S=0$.

\subsubsection{Lensing signal model}
For $\Delta\Sigma_{\rm model}(R|M,z)$ we use the \textsc{Dark Emulator}, which provides accurate predictions for the halo--matter
correlation function including both 1-halo and 2-halo contributions.
Halo masses are defined as $M_{200\mathrm{m}}$ (the native definition used in
the emulator and mass function) and are converted to $M_{500\mathrm{c}}$ using
\textsc{Colossus} \citep{diemer18} when evaluating the selection function.
The mass conversion uses the \citet{Diemer2015} concentration--mass relation
$c_{200\mathrm{m}}(M,z)$ to determine the NFW profile for each halo mass
and redshift, from which $M_{500\mathrm{c}}$ is computed self-consistently.
The mass integration is performed on a logarithmic grid of 65~bins
spanning $M_{200\mathrm{m}}\in[10^{13},\,3\times10^{15}]\,M_\odot$.
To avoid redundant emulator calls during MCMC sampling, the halo mass
function, concentrations, and $\Delta\Sigma(R|M,z)$ profiles are
precomputed for each unique cluster redshift and cached.

\subsubsection{Miscentering model}
\label{sec:miscenter_model}
A fraction $f_{\rm mis}$ of clusters is assumed to have centers offset from the
true halo center by a distance $r_{\rm off}$ following a Rayleigh distribution
with scale $\sigma_{\rm off}$:
\begin{equation}
P(r_{\rm off})=\frac{r_{\rm off}}{\sigma_{\rm off}^2}
\exp\!\left(-\frac{r_{\rm off}^2}{2\sigma_{\rm off}^2}\right).
\end{equation}
The miscentered profile is computed by azimuthally averaging the centered profile
over all possible offset positions:
\begin{equation}
\begin{split}
\Delta\Sigma_{\rm mis}(R)
&=
\int_0^\infty dr_{\rm off}\,P(r_{\rm off})
\int_0^{2\pi}\frac{d\theta}{2\pi} \\
&\quad \times
\Delta\Sigma_{\rm cen}\!\left[
\left(
R^2+r_{\rm off}^2
+2Rr_{\rm off}\cos\theta
\right)^{1/2}
\right].
\end{split}
\end{equation}

The total model profile is a mixture:
$\Delta\Sigma_{\rm tot}=(1-f_{\rm mis})\,\Delta\Sigma_{\rm cen}+
f_{\rm mis}\,\Delta\Sigma_{\rm mis}$.

Of the 19~clusters, 16 use \textsc{Camira} BCG centers, 2 use SDSS
\texttt{redMaPPer} DR8 centers, and 1 uses the \emph{Planck}
SZ centroid. Optically identified BCG centers (from \textsc{Camira}
and \texttt{redMaPPer}) are typically accurate to $\lesssim 0.1$--$0.2$
$h^{-1}\,\mathrm{Mpc}$, while the \emph{Planck} SZ centroid has
larger uncertainties ($\sim$1--2~arcmin, corresponding to
$\sim$0.2--0.5~$h^{-1}\,\mathrm{Mpc}$ at our redshifts) due to the
low angular resolution of the \emph{Planck} beam.
Ideally, one would fit separate $f_{\rm mis}$ values for the
optically centered and SZ-centered subsamples; however, with only
1~cluster using the \emph{Planck} centroid, such a split is not
statistically meaningful. Our single-population miscentering model
with Gaussian priors $f_{\rm mis}\sim\mathcal{N}(0.30,\,0.10^2)$ and
$\sigma_{\rm off}\sim\mathcal{N}(\ln 0.30,\,0.35^2)$ is motivated by
\citet{miyatake19} and effectively represents a population-averaged
centering accuracy appropriate for a sample dominated by optical BCG
centers. The conservative minimum scale of
$R_{\min}=0.5\,h^{-1}\,\mathrm{Mpc}$ further reduces sensitivity to
centering offsets.

\subsubsection{Fitting range}
\label{sec:fit_range}
The data vector is the random-subtracted stacked profile
$\Delta\Sigma_{\rm corr}(R)$ (Section~\ref{sec:single-mass-bin-fit}). We adopt
a fiducial (conservative) fitting range of
$R\in[0.5,5.0]\,h^{-1}\,\mathrm{Mpc}$. The minimum scale is motivated by the
boost factor (Section~\ref{sec:boost}), and the maximum scale is chosen to reduce
sensitivity to two-halo modeling systematics and large-scale random-point
residuals (Section~\ref{sec:randoms}). As a robustness check, we also fit over the extended range
$R\in[0.5,10.0]\,h^{-1}\,\mathrm{Mpc}$ and find consistent results
(Table~\ref{tab:robustness_combined}).

\subsubsection{Analytical covariance matrix}
\label{sec:covariance}
We construct an analytical covariance matrix for the stacked
$\Delta\Sigma$ profile following the approach of \citet{miyatake19} and \citet{wu19}, comprising three contributions:
\begin{equation}
\label{eq:cov_total}
\begin{split}
C(R_a,R_b)
&=
C_{\rm shape}(R_a,R_b)
+ C_{\rm LSS}(R_a,R_b) \\
&\quad
+ C_{\rm rand}(R_a,R_b).
\end{split}
\end{equation}

\paragraph{Shape noise.} The diagonal contribution from intrinsic ellipticity
dispersion and the finite number of source--lens pairs, read directly from the
pipeline output. This term dominates at small radii
($R\lesssim 1\,h^{-1}\,\mathrm{Mpc}$) and already reflects the weighted stacking
of the measurement.

\paragraph{Uncorrelated LSS.} The off-diagonal covariance from projected
large-scale structure along the line of sight, computed via a full Limber
integral:
\begin{equation}
\label{eq:Pkappa}
P_\kappa(\ell) = \int_0^{\chi_{\rm max}} d\chi\;
\frac{W^2(\chi)}{\chi^2}\;
P_{mm}^{\rm NL}\!\left(k=\frac{\ell+0.5}{\chi},z(\chi)\right),
\end{equation}
where $W(\chi)$ is the lensing kernel averaged over the mean source
redshift distribution $n(z_s)$ (obtained by averaging the per-source
$P(z)$ posteriors over all radial bins), and
$P_{mm}^{\rm NL}$ is the non-linear matter power spectrum from Halofit via the Core Cosmology Library
\citep{CCL2019}.
To convert from the convergence power spectrum $P_\kappa(\ell)$ to the
covariance of the binned $\Delta\Sigma$ in physical (comoving) units,
we define a bin-averaged Bessel filter for each radial bin
$[R_a^{\rm in},\,R_a^{\rm out}]$ and cluster $i$ at redshift $z_i$
\citep{Schneider1998,Hoekstra2003}:
\begin{equation}
\label{eq:bessel_filter}
F_a^{(i)}(\ell) = \frac{\Sigma_{{\rm cr},i}}{A_a}
\int_{\theta_a^{\rm in}}^{\theta_a^{\rm out}} \theta\,
J_2(\ell\,\theta)\,d\theta\,,
\end{equation}
where $\theta = R/(h\,\chi(z_i))$ converts comoving projected radius
($h^{-1}\,\mathrm{Mpc}$) to angular separation,
$A_a = \tfrac{1}{2}[(\theta_a^{\rm out})^2 - (\theta_a^{\rm in})^2]$
is the bin area, $\Sigma_{{\rm cr},i}$ is the effective critical surface
density for cluster $i$ averaged over $n(z_s)$, and $J_2$ is the
second-order Bessel function of the first kind. The $J_2$ kernel arises
because $\Delta\Sigma$ is a tangential-shear quantity, not a convergence.
The per-cluster LSS covariance between radial bins $a$ and $b$ is then
\begin{equation}
\label{eq:cov_lss_percl}
C_{{\rm LSS},i}(R_a,R_b) = \int \frac{\ell\,d\ell}{2\pi}\;
P_\kappa(\ell)\;F_a^{(i)}(\ell)\;F_b^{(i)}(\ell)\,,
\end{equation}
where $P_\kappa(\ell)$ is the convergence power spectrum from
Eq.~(\ref{eq:Pkappa}). The $\ell$ integration is performed over
500~logarithmically spaced multipoles from $\ell=10$ to $10^5$.
This term dominates at large radii
($R\gtrsim 3\,h^{-1}\,\mathrm{Mpc}$).

The LSS covariance term is
weighted by the per-cluster radial weights to match the weighted stacking:
\begin{equation}
\label{eq:weighted_cov}
C_{\rm LSS}(R_a,R_b)=\sum_{i=1}^{N_{\rm cl}} \alpha_i(R_a)\,\alpha_i(R_b)\,
C_{{\rm LSS},i}(R_a,R_b),
\end{equation}
where $\alpha_i(R_a)=w_i(R_a)/\sum_j w_j(R_a)$ are the normalized per-cluster
weights.

\paragraph{Randoms subtraction noise.} Because the corrected data vector is
$\Delta\Sigma_{\rm corr}=\Delta\Sigma_{\rm data}-\Delta\Sigma_{\rm rand}$
and the cluster and random-point measurements are statistically independent
(different lens positions and independent shape-noise realizations), the
variance of the corrected signal is
$\mathrm{Var}(\Delta\Sigma_{\rm corr})=\mathrm{Var}(\Delta\Sigma_{\rm data})
+\mathrm{Var}(\Delta\Sigma_{\rm rand})$. We add the diagonal
$C_{\rm rand}(R_a,R_b)=\delta_{ab}\,\sigma_{\rm rand}^2(R_a)$,
where $\sigma_{\rm rand}$ is the shape-noise uncertainty of the
randoms measurement (1900 random points). This term contributes a
fractional increase of $\lesssim 0.5\%$ to the diagonal errors but is
included for consistency.

\begin{figure*}[t]
\centering
\includegraphics[width=0.48\textwidth]{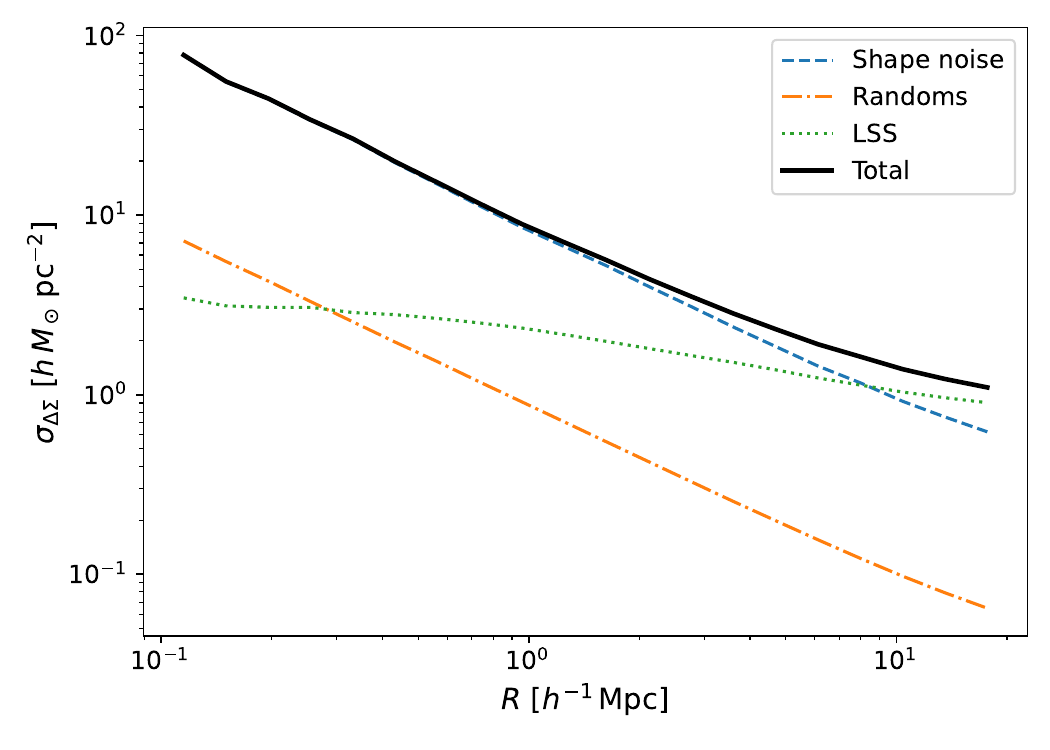}\hfill
\includegraphics[width=0.48\textwidth]{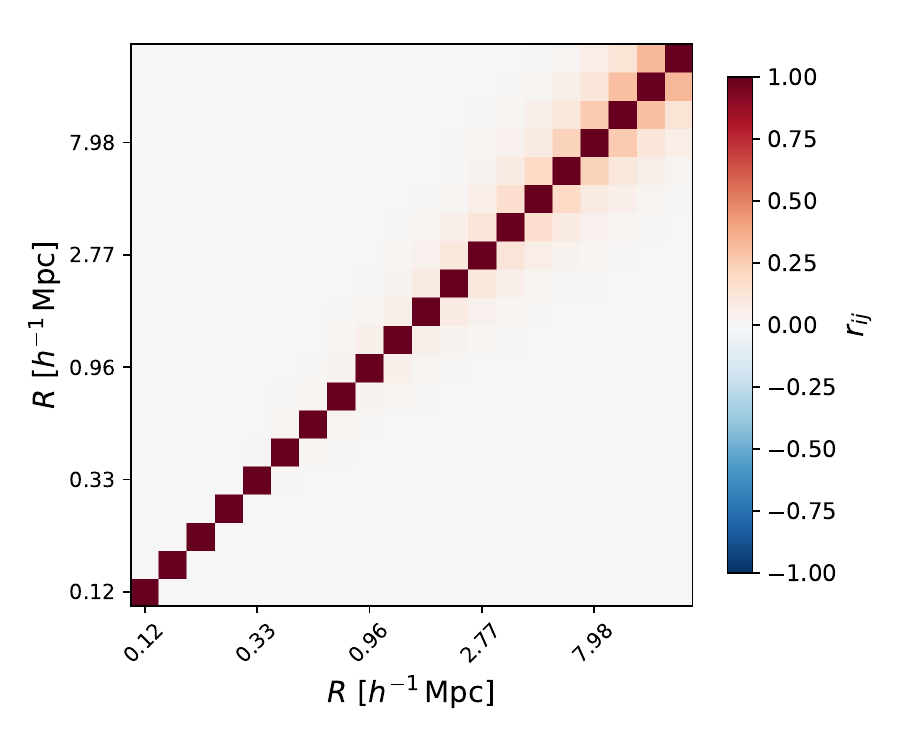}
\caption{
\emph{Left}: Diagonal of the baseline covariance matrix (shape noise
$+$ uncorrelated LSS $+$ randoms subtraction noise), decomposed into its
individual contributions. Shape noise dominates at small projected radii
($R\lesssim 1\,h^{-1}\,\mathrm{Mpc}$), while the projected LSS
contribution becomes comparable at larger separations.
\emph{Right}: Correlation matrix $r_{ab}=C_{ab}/\sqrt{C_{aa}\,C_{bb}}$
of the baseline covariance matrix. The off-diagonal structure is driven
by the projected large-scale structure term, which introduces positive
correlations between adjacent radial bins at large separations.
}
\label{fig:covariance}
\end{figure*}

Figure~\ref{fig:covariance} shows the diagonal of the baseline covariance
matrix, decomposed into its shape-noise, LSS, and randoms contributions
(left panel), along with the correlation matrix
$r_{ab}=C_{ab}/\sqrt{C_{aa}\,C_{bb}}$ (right panel).
Shape noise dominates at small radii while the LSS term becomes
comparable at $R\gtrsim 3\,h^{-1}\,\mathrm{Mpc}$; the off-diagonal
structure in the correlation matrix is driven by the projected LSS term.

\subsubsection{Likelihood and inference}
\label{sec:stacked-model-fit}
We sample the posterior distribution of four parameters,
$\boldsymbol{\theta}=(\ln(1-b),\,f_{\rm mis},\,\ln\sigma_{\rm off},\,\sigma_{\ln M})$,
using the affine-invariant ensemble sampler \texttt{emcee}
\citep{Foreman-Mackey13} with 48~walkers, 1500~burn-in steps, and
8000~production steps. The log-posterior is
\begin{equation}
\label{eq:logpost}
\ln\mathcal{P}(\boldsymbol{\theta})=
-\tfrac{1}{2}\,
[\boldsymbol{d}-\boldsymbol{m}(\boldsymbol{\theta})]^{\rm T}\,
\mathbf{C}^{-1}\,
[\boldsymbol{d}-\boldsymbol{m}(\boldsymbol{\theta})]
+\ln\pi(\boldsymbol{\theta}),
\end{equation}
where $\boldsymbol{d}=\{\Delta\Sigma_{\rm corr}(R_a)\}$,
$\boldsymbol{m}$ is the weighted stacked model
(Eq.~\ref{eq:weighted_model}), and $\mathbf{C}$ is the covariance
matrix (Eq.~\ref{eq:cov_total}). The priors on all four sampled parameters
are listed in Table~\ref{tab:priors}.

\begin{table}[t]
\caption{Fiducial priors on the four sampled parameters in the stacked model fit. $\mathcal{N}(\mu,\sigma^2)$
denotes a Gaussian prior with mean $\mu$ and standard deviation $\sigma$,
truncated to the listed bounds. See Table~\ref{tab:robustness_combined} for results involving changes in the $\sigma_{\ln M}$ priors.}
\label{tab:priors}
\begin{ruledtabular}
\begin{tabular}{lcc}
Parameter & Prior & Bounds \\
\hline
$\ln(1-b)$ & Flat & $[\ln 0.3,\;\ln 1.5]$ \\
$f_{\rm mis}$ & $\mathcal{N}(0.30,\,0.10^2)$ & $[0,\,1]$ \\
$\ln\sigma_{\rm off}$ & $\mathcal{N}(\ln 0.30,\,0.35^2)$ &
$\sigma_{\rm off}\in[10^{-3},5]\,h^{-1}\,\mathrm{Mpc}$ \\
$\sigma_{\ln M}$ & $\mathcal{N}(0.25,\,0.10^2)$ & $[0.05,\,0.80]$ \\
\end{tabular}
\end{ruledtabular}
\end{table}

The prior on $(1-b)$ is deliberately broad, extending beyond
$(1-b) = 1$, to verify that the posterior is not truncated by the
prior boundary. The prior on $\sigma_{\ln M}$, by contrast, is
informative \emph{with respect to the weak-lensing data}.
$\sigma_{\ln M}$ enters the forward model only through the
Eddington-bias convolution of the \textit{Planck} completeness
(Eq.~\ref{eq:eff_selection}) and is therefore degenerate with
$(1-b)$: increasing $\sigma_{\ln M}$ at fixed completeness
up-scatters lower-mass halos into the selection and pulls
$(1-b)$ downward. At the signal-to-noise of an $N=19$ stack the
data alone do not break this degeneracy: under a uniform prior
$\sigma_{\ln M} \sim \mathcal{U}[0.05, 0.80]$ the posterior is
$\sigma_{\ln M} = 0.37^{+0.20}_{-0.21}$, consistent with
essentially the full prior range, and the corresponding
$\chi^2/\mathrm{dof} = 5.3/5$ is statistically indistinguishable
from the fiducial value $5.2/5$
(Table~\ref{tab:robustness_combined}). The width and centroid of
the $\sigma_{\ln M}$ prior must therefore be set from external
information, namely the X-ray, SZ, and hydrodynamic-simulation
literature on the $Y_{500}$--$M_{500}$ relation.
Table~\ref{tab:sigma_lnM_priors} summarises these external
constraints. The \textit{Planck}~2015 cluster-cosmology analysis
reports an intrinsic scatter
$\sigma_{\ln Y|M} = 0.173 \pm 0.023$ in the $Y$--$M$ relation
calibrated against X-ray hydrostatic masses
\citep{2016AA...594A..24P}; propagated through the power-law slope
$\alpha = 1.79$ (Eq.~\ref{eq:YM_scaling}) this corresponds to
$\sigma_{\ln M|Y} \simeq 0.10$, but this is a strict lower bound
because it isolates the intrinsic gas-physics scatter and
excludes line-of-sight projection, triaxiality, substructure,
and non-thermal pressure support. Hydrodynamic simulations of
the integrated $Y_{500}$ observable that include AGN feedback
recover a comparable intrinsic $Y$--$M$ scatter of
$\sim 11$--$13\%$, equivalent to
$\sigma_{\ln M} \simeq 0.06$--$0.08$ from gas physics alone
\citep{Battaglia2012}. The additional scatter from line-of-sight
projection and halo triaxiality, which is relevant for any
column-integrated observable, inflates the effective scatter to
$\sigma_{\ln M} \simeq 0.20$--$0.25$ at the mass scale of our
sample \citep{BeckerKravtsov2011}. Consistent with this
decomposition, the \textit{Planck} cluster number-count
cosmology analysis adopts $\sigma_{\ln M} = 0.20$ as a baseline
and tests robustness across $[0.1, 0.4]$ \citep{salvati18};
ACT-DR5 cluster cosmology and recent weak-lensing calibration
analyses fix $\sigma_{\ln M} = 0.20$
\citep{hilton21,shin25}. The ACT$\times$HSC weak-lensing
calibration of \citet{Shirasaki2024} parameterizes the scatter
as $\sigma_{\log M}(M,z) = C_0\,\sigma_{\rm ref}(M,z)$, where
$\sigma_{\rm ref}$ is calibrated from IllustrisTNG simulations
and $C_0$ is a free nuisance parameter with a flat prior
$[0.2,\,5.0]$; at the mass scale of our sample their reference
scatter evaluates to
$\sigma_{\ln M} \simeq 0.25$--$0.28$ for $C_0 = 1$, consistent
with the range adopted here. Our fiducial choice
$\sigma_{\ln M} \sim \mathcal{N}(0.25, 0.10^2)$ spans the
$1\sigma$ interval $[0.15, 0.35]$ and therefore brackets every
entry in Table~\ref{tab:sigma_lnM_priors} that represents the
total (not merely intrinsic) scatter relevant for the
Eddington-bias convolution, while remaining wider than any
single external determination. As shown by the explicit
comparison in Table~\ref{tab:robustness_combined}, broadening
this prior to $\mathcal{N}(0.25, 0.25^2)$ or replacing it with a
uniform prior $\mathcal{U}[0.05, 0.80]$ shifts the inferred
$(1-b)$ downward by $\lesssim 1\sigma$ (from $0.73$ to $0.68$
and $0.62$ respectively), with $\chi^2/\mathrm{dof}$ unchanged.
We adopt the fiducial prior as our headline configuration
because it encodes the genuine external prior knowledge on the
\textit{Planck} SZ mass-proxy scatter; the broader-prior columns
in Table~\ref{tab:robustness_combined} should be read as upper
bounds on the sensitivity of $(1-b)$ to that external information.

\begin{table*}
\caption{External (non-weak-lensing) constraints on the log-normal
scatter of the \textit{Planck} SZ mass proxy at fixed true halo
mass, $\sigma_{\ln M}$, that motivate the fiducial prior adopted
in this work. Where the original reference reports
$\sigma_{\ln Y|M}$ on the $Y_{500}$--$M_{500}$ relation, the
equivalent mass scatter is obtained via
$\sigma_{\ln M|Y} = \sigma_{\ln Y|M} / \alpha$ with
$\alpha = 1.79$ (Eq.~\ref{eq:YM_scaling}). The third column
indicates whether the value is intrinsic to gas physics only or
already incorporates line-of-sight projection and triaxiality,
i.e.\ the total scatter relevant for the Eddington-bias
convolution in our forward model (Eq.~\ref{eq:eff_selection}).}
\label{tab:sigma_lnM_priors}
\begin{ruledtabular}
\begin{tabular}{lcl}
Reference & $\sigma_{\ln M}$ & Source / scope \\
\hline
\citet{2016AA...594A..24P}      & $\simeq 0.10$            & $Y$--$M$ vs.\ X-ray HSE; intrinsic (lower bound) \\
\citet{Battaglia2012}       & $\simeq 0.06$--$0.08$    & SPH hydro $+$ AGN feedback; intrinsic, gas physics \\
\citet{BeckerKravtsov2011}  & $0.20$--$0.25$           & $N$-body $+$ LOS projection; total \\
\citet{salvati18}         & $0.20$ (tested $[0.1,0.4]$) & \textit{Planck} cluster counts; total \\
\citet{hilton21,shin25} & $0.20$ (fixed)           & ACT-DR5 cosmology; total \\
\citet{Shirasaki2024}       & $\simeq 0.25$--$0.28$     & ACT$\times$HSC WL; sim-calibrated, $C_0=1$; total \\
\hline
This work (prior) & $\mathcal{N}(0.25,0.10^2)$ & spans $[0.15,0.35]$ ($1\sigma$); total \\
\end{tabular}
\end{ruledtabular}
\end{table*}

The chains show good convergence with a mean acceptance fraction of $\sim 0.59$
and autocorrelation lengths of $\sim$42--46~steps, yielding $\sim$4100
effective independent samples.

\subsubsection{Results}
\label{sec:headline_result}
Our headline constraint on the mass bias parameter is
\begin{equation}
\label{eq:1mb_result}
1-b = 0.73^{+0.10}_{-0.11}\,,
\end{equation}
with $\chi^2/\mathrm{dof}=5.2/5$ for the covariance
(Eq.~\ref{eq:cov_total}) and the
conservative fit range $R\in[0.5,5.0]\,h^{-1}\,\mathrm{Mpc}$.
The SZ scatter is constrained to
$\sigma_{\ln M}=0.25^{+0.10}_{-0.10}$, consistent with the prior
(Table~\ref{tab:priors}).
The best-fit miscentering parameters are
$f_{\rm mis}=0.28^{+0.10}_{-0.10}$ and
$\sigma_{\rm off}=0.28^{+0.11}_{-0.08}\,h^{-1}\,\mathrm{Mpc}$,
consistent with priors motivated by \citet{miyatake19}.
The effective weak-lensing mean mass, computed self-consistently from the
selection-weighted integral, is
$\langle M_{500\mathrm{c}}\rangle_{\rm WL}\simeq 3.5\times10^{14}\,h^{-1}M_\odot$.

\begin{figure}[t]
\centering
\includegraphics[width=0.95\columnwidth]{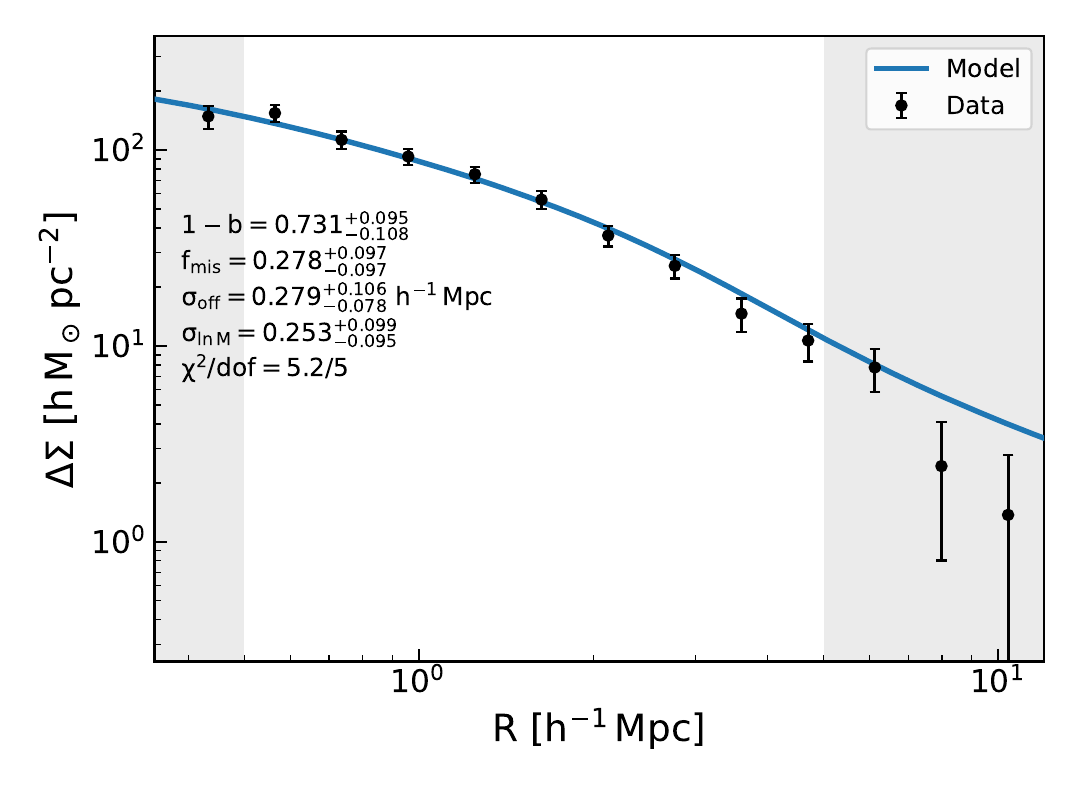}
\caption{
Stacked weak-lensing $\Delta\Sigma$ profile and best-fit forward model,
evaluated on a fine radial grid.
Best-fit parameters with 68\% credible intervals and
$\chi^2/\mathrm{dof}$ are indicated.
The shaded regions mark scales excluded from the fit
($R < 0.5$ and $R > 5.0\,h^{-1}\,\mathrm{Mpc}$).
The fit uses per-cluster weak-lensing weights
(Section~\ref{sec:weighted_stacking}) and the covariance matrix
(Section~\ref{sec:covariance}).
}
\label{fig:stacked_model_fit}
\end{figure}

\begin{figure}[t]
\centering
\includegraphics[width=0.95\columnwidth]{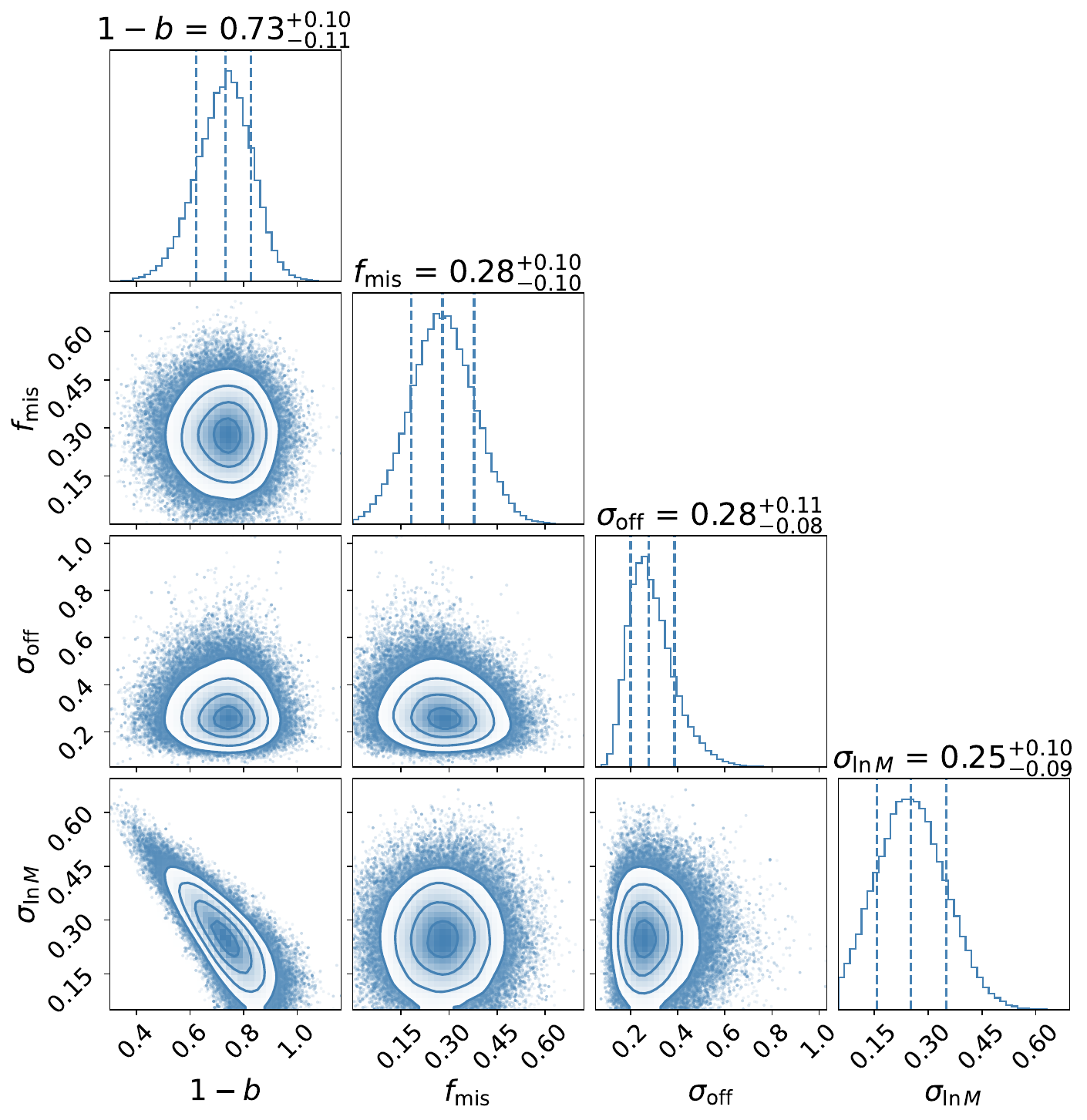}
\caption{
Joint and marginal posterior distributions from the MCMC fit of the four
sampled parameters: $(1-b)$, $f_{\rm mis}$, $\sigma_{\rm off}$, and
$\sigma_{\ln M}$.
Dashed lines indicate the 16th, 50th, and 84th percentiles. The posterior for
$(1-b)$ is well-contained within the prior boundaries, the miscentering
parameters are consistent with the Gaussian priors, and $\sigma_{\ln M}$ is
constrained near its prior center.
}
\label{fig:corner}
\end{figure}

Figure~\ref{fig:stacked_model_fit} shows the best-fit stacked model compared
with the data, and Fig.~\ref{fig:corner} shows the joint and marginal
posterior distributions of the four sampled parameters.

\subsubsection{Robustness}
\label{sec:robustness}
We systematically vary analysis choices to assess the sensitivity of the
$(1-b)$ constraint (Table~\ref{tab:robustness_combined}).

\begin{table*}[t]
\caption{Robustness of the mass bias constraint to variations in analysis
choices. All configurations sample four free parameters
($\ln(1-b)$, $f_{\rm mis}$, $\ln\sigma_{\rm off}$, $\sigma_{\ln M}$)
using the \texttt{emcee} affine-invariant MCMC sampler with 48 walkers,
1500 burn-in steps, and 8000 production steps.
The upper block uses a Gaussian prior
$\sigma_{\ln M}\sim\mathcal{N}(0.25,\,0.10^2)$ (fiducial); the
second block uses a broader prior $\sigma_{\ln M}\sim\mathcal{N}(0.25,\,0.20^2)$;
the third block uses an even broader Gaussian prior
$\sigma_{\ln M}\sim\mathcal{N}(0.25,\,0.25^2)$; and the bottom block
adopts a uniform prior $\sigma_{\ln M}\sim\mathcal{U}[0.05,\,0.80]$.
All results are statistically consistent with the headline value
$1-b=0.73^{+0.10}_{-0.11}$.}
\label{tab:robustness_combined}
\begin{ruledtabular}
\begin{tabular}{lccccc}
Configuration
& $1-b$
& $\sigma_{\ln M}$
& $f_{\rm mis}$
& $\sigma_{\rm off}$
& $\chi^2/\mathrm{dof}$ \\
\hline
\multicolumn{6}{c}{\textit{Fiducial prior: $\sigma_{\ln M}\sim\mathcal{N}(0.25,\,0.10^2)$}} \\
\hline
Headline (baseline cov, $R\in[0.5,5]$)
& $0.73^{+0.10}_{-0.11}$
& $0.25^{+0.10}_{-0.10}$
& $0.28^{+0.10}_{-0.10}$
& $0.28^{+0.11}_{-0.08}$
& 5.2/5 \\

Shape-noise-only cov
& $0.75^{+0.09}_{-0.11}$
& $0.26^{+0.10}_{-0.10}$
& $0.28^{+0.10}_{-0.10}$
& $0.28^{+0.10}_{-0.08}$
& 6.8/5 \\

Full range $R\in[0.5,10]$
& $0.74^{+0.10}_{-0.11}$
& $0.25^{+0.10}_{-0.10}$
& $0.28^{+0.10}_{-0.10}$
& $0.28^{+0.11}_{-0.08}$
& 8.7/7 \\

Conservative range $R\in[1,5]$
& $0.77^{+0.11}_{-0.12}$
& $0.25^{+0.10}_{-0.10}$
& $0.29^{+0.10}_{-0.10}$
& $0.28^{+0.11}_{-0.08}$
& 3.0/2 \\

\hline
\multicolumn{6}{c}{\textit{Broader prior: $\sigma_{\ln M}\sim\mathcal{N}(0.25,\,0.20^2)$}} \\
\hline
Baseline cov, $R\in[0.5,5]$
& $0.70^{+0.13}_{-0.18}$
& $0.29^{+0.17}_{-0.15}$
& $0.28^{+0.10}_{-0.10}$
& $0.28^{+0.11}_{-0.08}$
& 5.2/5 \\

Shape-noise-only cov ($R\in[0.5,5]$)
& $0.72^{+0.12}_{-0.19}$
& $0.28^{+0.17}_{-0.15}$
& $0.28^{+0.10}_{-0.10}$
& $0.27^{+0.10}_{-0.08}$
& 6.8/5 \\

Full range (baseline cov, $R\in[0.5,10]$)
& $0.71^{+0.13}_{-0.19}$
& $0.29^{+0.17}_{-0.15}$
& $0.28^{+0.10}_{-0.10}$
& $0.28^{+0.11}_{-0.08}$
& 8.7/7 \\

Conservative range (baseline cov, $R\in[1,5]$)
& $0.73^{+0.15}_{-0.19}$
& $0.30^{+0.17}_{-0.15}$
& $0.29^{+0.10}_{-0.10}$
& $0.28^{+0.12}_{-0.08}$
& 3.1/2 \\

\hline
\multicolumn{6}{c}{\textit{Even broader prior: $\sigma_{\ln M}\sim\mathcal{N}(0.25,\,0.25^2)$}} \\
\hline
Baseline cov, $R\in[0.5,5]$
& $0.68^{+0.14}_{-0.21}$
& $0.31^{+0.19}_{-0.16}$
& $0.28^{+0.10}_{-0.10}$
& $0.28^{+0.11}_{-0.08}$
& 5.2/5 \\

Shape-noise-only cov ($R\in[0.5,5]$)
& $0.70^{+0.14}_{-0.21}$
& $0.31^{+0.19}_{-0.16}$
& $0.28^{+0.10}_{-0.10}$
& $0.27^{+0.10}_{-0.08}$
& 6.8/5 \\

Full range (baseline cov, $R\in[0.5,10]$)
& $0.69^{+0.14}_{-0.21}$
& $0.30^{+0.19}_{-0.16}$
& $0.28^{+0.10}_{-0.10}$
& $0.28^{+0.10}_{-0.08}$
& 8.7/7 \\

Conservative range (baseline cov, $R\in[1,5]$)
& $0.71^{+0.16}_{-0.23}$
& $0.32^{+0.20}_{-0.17}$
& $0.29^{+0.10}_{-0.10}$
& $0.29^{+0.12}_{-0.08}$
& 3.1/2 \\

\hline
\multicolumn{6}{c}{\textit{Uniform prior: $\sigma_{\ln M}\sim\mathcal{U}[0.05,\,0.80]$}} \\
\hline
Baseline cov, $R\in[0.5,5]$
& $0.62^{+0.19}_{-0.22}$
& $0.37^{+0.20}_{-0.21}$
& $0.28^{+0.10}_{-0.10}$
& $0.28^{+0.11}_{-0.08}$
& 5.3/5 \\

Shape-noise-only cov ($R\in[0.5,5]$)
& $0.64^{+0.19}_{-0.24}$
& $0.37^{+0.21}_{-0.22}$
& $0.27^{+0.10}_{-0.10}$
& $0.27^{+0.10}_{-0.08}$
& 6.9/5 \\

Full range (baseline cov, $R\in[0.5,10]$)
& $0.62^{+0.19}_{-0.23}$
& $0.37^{+0.20}_{-0.21}$
& $0.28^{+0.10}_{-0.10}$
& $0.28^{+0.11}_{-0.08}$
& 8.8/7 \\

Conservative range (baseline cov, $R\in[1,5]$)
& $0.63^{+0.22}_{-0.24}$
& $0.39^{+0.21}_{-0.23}$
& $0.29^{+0.10}_{-0.10}$
& $0.29^{+0.12}_{-0.08}$
& 3.1/2 \\
\end{tabular}
\end{ruledtabular}
\end{table*}

The inferred $(1-b)$ is stable to the choice of maximum fitting radius
($R_{\rm max}=5$ versus $10\,h^{-1}\,\mathrm{Mpc}$) and to the covariance
model (shape-noise-only versus baseline).
Table~\ref{tab:robustness_combined} shows that the results are also
robust to the width of the Gaussian prior on $\sigma_{\ln M}$: broadening
the prior from $\sigma=0.10$ to $\sigma=0.20$ shifts $(1-b)$ from $0.73$
to $0.70$, within the statistical uncertainties, while the inferred
$\sigma_{\ln M}$ remains consistent across all configurations.
The residual sensitivity to the prior width arises because the data alone
provide only a modest constraint on $\sigma_{\ln M}$, and a broader prior
allows the posterior to explore higher scatter values, which shift the
effective mass through the selection-weighted mass integrand.

To further test this sensitivity, we repeat the full suite of
robustness tests with an even broader Gaussian prior,
$\sigma_{\ln M}\sim\mathcal{N}(0.25,\,0.25^2)$, and with a
uniform (uninformative) prior, $\sigma_{\ln M}\sim\mathcal{U}[0.05,\,0.80]$
(Table~\ref{tab:robustness_combined}).
With the $\mathcal{N}(0.25,\,0.25^2)$ prior, the baseline configuration
yields $(1-b)=0.68^{+0.14}_{-0.21}$, consistent with the headline value
at the $<\!0.5\sigma$ level. The inferred scatter is
$\sigma_{\ln M}=0.31^{+0.19}_{-0.16}$, and the other fitted parameters
($f_{\rm mis}$, $\sigma_{\rm off}$) and goodness-of-fit statistics are
virtually unchanged. The uniform prior, which places no informative weight
on $\sigma_{\ln M}$, returns $(1-b)=0.62^{+0.19}_{-0.22}$ for the
baseline configuration. Although the central value shifts downward
relative to the fiducial result, the two constraints overlap within
$1\sigma$, and the shift is driven by the posterior
for $\sigma_{\ln M}$ extending to higher values
($\sigma_{\ln M}=0.37^{+0.20}_{-0.21}$), which increases the
Eddington-bias correction and pulls $(1-b)$ lower.
Across all twelve configurations in the bottom two blocks,
the $\chi^2/\mathrm{dof}$ values remain comparable to those in the
fiducial and broader-prior blocks, confirming that the goodness of fit
does not prefer a particular prior choice.
These tests demonstrate that, while the data alone do not strongly
constrain $\sigma_{\ln M}$, the inferred $(1-b)$ is robust at the
$\lesssim\!1\sigma$ level to prior assumptions spanning from tightly
informative to fully uninformative. As justified in Table~\ref{tab:sigma_lnM_priors},
the fiducial Gaussian prior encodes external constraints on the
total \textit{Planck} SZ mass-proxy scatter; the broader and
uniform priors in this block deliberately weaken that external
information to bracket its impact.

Our fiducial Gaussian prior
$\sigma_{\ln M}\sim\mathcal{N}(0.25,\,0.10^2)$ is motivated by
$N$-body and hydrodynamical simulation results.
\citet{BeckerKravtsov2011} quantified the scatter in weak-lensing
cluster mass estimates arising from halo triaxiality, correlated
large-scale structure, and uncorrelated line-of-sight projections,
finding $\sigma_{\ln M}\approx 0.20$ for the most massive clusters
($M \gtrsim 10^{14.5}\,h^{-1}\,M_\odot$) and up to $\sim\!0.30$ for
lower-mass systems.
\citet{Grandis2021} calibrated the full weak-lensing-to-halo-mass
relation using hydrodynamical simulations that include observational
systematics (shear bias, photometric redshift uncertainties,
miscentering, member contamination, and projection effects), obtaining
intrinsic scatter values in the range $\sigma_{\ln M}\approx 0.20$--$0.30$
depending on the minimum fitting radius.
Our prior, centered at $0.25$ with a width of $0.10$, spans the
$1\sigma$ interval $[0.15,\,0.35]$ and thus encompasses the full
range of simulation-based estimates for clusters at the mass scale
of our sample
($\langle M_{500\mathrm{c}}\rangle\sim 3.5\times10^{14}\,h^{-1}\,M_\odot$).
Similar or consistent values have been adopted or inferred in recent
SZ-selected cluster weak-lensing analyses
\citep{medezinski18,miyatake19,Herbonnet2020,shin25,Grandis2024}.

Additional systematic uncertainties affect the inferred mass bias at a level below our current statistical precision.

\paragraph{Baryonic effects.}
Neglecting baryonic corrections to the density profile may bias cluster
masses high by $\sim$5--7\%, as estimated from hydrodynamical simulations
\citep{shirasaki_lau17,cromer21,shin25}. With our minimum fitting scale of
$R_{\min}=0.5\,h^{-1}\,\mathrm{Mpc}$, residual baryonic suppression of
$\Delta\Sigma$ is expected at the few-percent level
\citep{shirasaki_lau17,giri21}. At the mass scale of our sample
($\langle M_{500\mathrm{c}}\rangle\sim 3.5\times10^{14}\,h^{-1}M_\odot$),
AGN feedback redistributes gas outward, reducing the inner density profile
relative to dark-matter-only predictions. The effect on $\Delta\Sigma$ at
$R=0.5\,h^{-1}\,\mathrm{Mpc}$ is $\lesssim 3$--$5$\% for massive
clusters \citep{giri21}, well below our $\sim$14\% statistical
uncertainty but worth noting as a systematic floor for future analyses.

\paragraph{Cluster triaxiality and orientation bias.}
SZ-selected clusters are preferentially elongated along the line
of sight, since this orientation boosts the integrated Compton-$Y$
signal through the increased path length of hot gas
\citep{Corless2007,Sunayama2020}. This preferential orientation
can bias the stacked weak-lensing mass low by $\sim$5--10\%
relative to the true spherically averaged mass, because the projected
surface density is lower for prolate halos viewed along the major axis
\citep{Corless2007,BeckerKravtsov2011,shin25}. The magnitude of this
effect depends on the cluster mass, redshift, and the selection
observable, and is partially degenerate with the hydrostatic mass bias.

These systematic effects are individually subdominant
relative to the $\sim$14\% statistical uncertainty on $(1-b)$ in our
analysis, but will become increasingly important as sample sizes grow
with upcoming surveys such as LSST.

\subsection{Comparison with the literature}
\label{sec:comparison}

\begin{figure*}[t]
\centering
\includegraphics[width=0.85\textwidth]{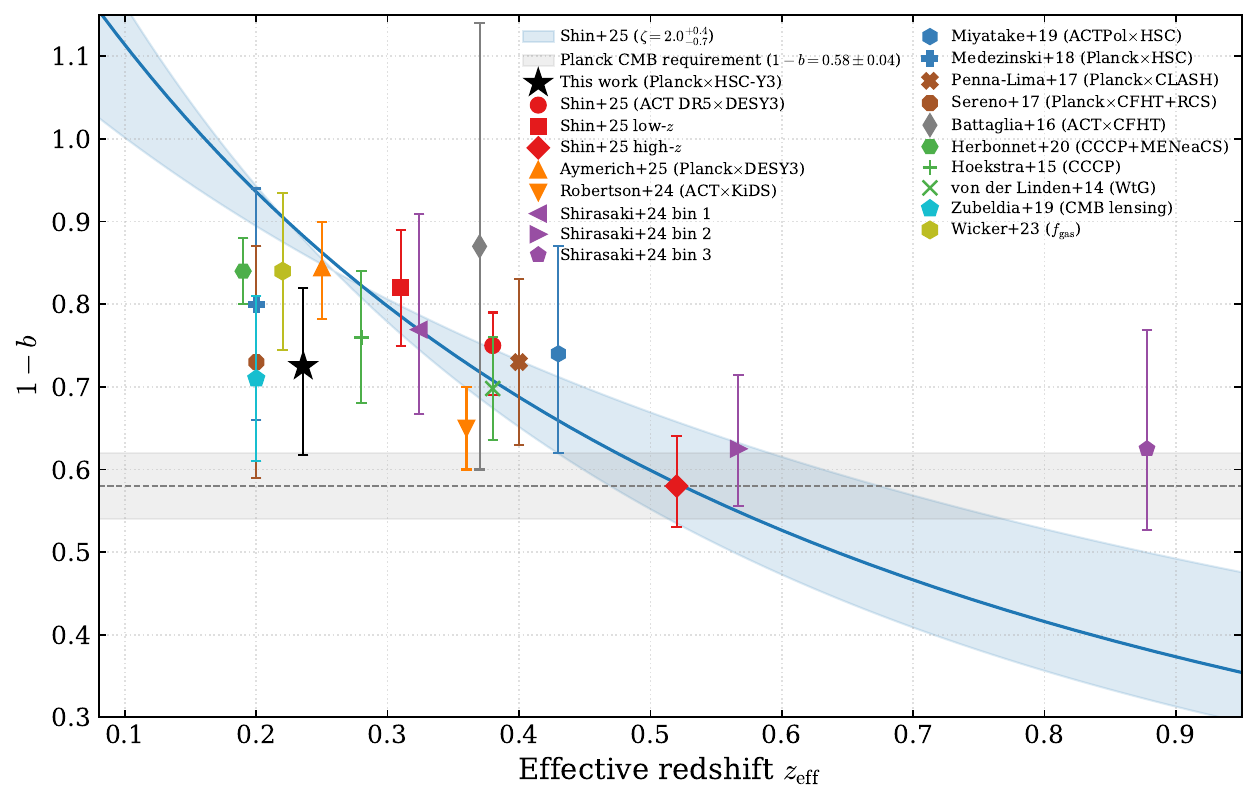}
\caption{
Comparison of $(1-b)$ measurements as a function of effective redshift.
This work (black star) is compared with published results from
SZ$\times$WL calibrations
(SZ-cluster masses calibrated against weak-lensing masses, i.e.\
$(1-b)\equiv M_\mathrm{SZ}/M_\mathrm{WL}$):
\citet{battaglia16},
\citet{vonderLinden2014},
\citet{Hoekstra2015},
\citet{PennaLima2017},
\citet{Sereno2017},
\citet{medezinski18},
\citet{miyatake19},
\citet{Herbonnet2020},
\citet{Shirasaki2024},
\citet{Robertson2024},
\citet{Aymerich2025},
and \citet{shin25};
from CMB lensing of the SZ-cluster sample:
\citet{Zubeldia2019};
from cluster gas-mass fractions:
\citet{Wicker2023};
and the \textit{Planck} CMB requirement, i.e.\ the value of $(1-b)$
needed to reconcile the \textit{Planck} primary-CMB anisotropies with
the PSZ cluster counts, shown as a gray horizontal band
(\citealt{2016AA...594A..24P}) because it represents a cosmological
average and is not attached to any specific cluster redshift.
The blue band shows the \citet{shin25} power-law redshift evolution
$(1-b)=A_{\rm mass}^{-1}[(1+z)/(1+0.45)]^{-\zeta}$.
}
\label{fig:comparison_z}
\end{figure*}

Figure~\ref{fig:comparison_z} compares our constraint on $(1-b)$ with published
results from weak-lensing calibrations of SZ-selected clusters, X-ray cluster
samples, CMB lensing, and the gas mass fraction method. We also show the
power-law redshift evolution reported by \citet{shin25} from the ACT~DR5$\times$DES~Y3
analysis, $(1-b)=A_{\rm mass}^{-1}[(1+z)/(1+0.45)]^{-\zeta}$ with
$A_{\rm mass}=1.56^{+0.11}_{-0.13}$ and $\zeta=2.0^{+0.4}_{-0.7}$, and the
value $1-b\simeq 0.58\pm0.04$ required to reconcile \textit{Planck} primary CMB
anisotropies with the SZ cluster abundance \citep{2016AA...594A..24P}.

Our measurement of $1-b=0.73^{+0.10}_{-0.11}$ at $z_{\rm eff}\simeq 0.24$ is
consistent within $1\sigma$ with earlier \textit{Planck}$\times$HSC results
from \citet{medezinski18} ($1-b=0.80\pm0.14$, 5~clusters) and
\citet{miyatake19} ($1-b=0.74^{+0.13}_{-0.12}$, ACTPol$\times$HSC), as well
as with other recent calibrations including \citet{Robertson2024}
($1-b=0.65\pm0.05$, ACT$\times$KiDS), \citet{Aymerich2025}
($1-b=0.84^{+0.06}_{-0.06}$, \textit{Planck}$\times$DES~Y3),
\citet{shin25} ($1-b=0.75^{+0.04}_{-0.06}$, ACT~DR5$\times$DES~Y3), and
\citet{Shirasaki2024} ($1/(1-b)=1.3\pm0.2$ at $z\sim 0.27$,
ACT$\times$HSC three-bin analysis).
The measurement is $\sim\!1.5\sigma$ above the \textit{Planck} CMB requirement,
consistent with the picture that the hydrostatic mass bias alone is
insufficient to fully reconcile cluster counts with the primary CMB.

The recent analysis by \citet{shin25} obtains
$1-b\simeq 0.75$ with $\sim$7\% precision from $\sim$7000
ACT~DR5 clusters calibrated with DES~Y3 weak lensing, and finds strong
evidence for redshift evolution in $(1-b)$ with
$\zeta=2.0^{+0.4}_{-0.7}$. Our result at $z_{\rm eff}\simeq 0.24$ is
consistent with their low-redshift bin. Our sample size ($N=19$) does
not permit a meaningful test of redshift evolution, but the consistency
of the two results, obtained with independent cluster catalogs, lensing
surveys, and analysis pipelines, supports the robustness of the
inferred mass bias at low redshift. Compared to \citet{shin25}, our
analysis uses a different selection function (Planck Union rather than
ACT matched-filter), does not explicitly model the boost factor (our
conservative P($z$)-cut and radial cuts render it negligible), and
employs the Dark Emulator rather than an analytic halo model for the
lensing signal; these methodological differences make our result a
complementary cross-check.

In addition, recent eROSITA cluster mass calibrations using DES~Y3
\citep{Grandis2024} and HSC~Y3 \citep{2504.01076,2022AA...661A..11C} weak lensing
adopt fully Bayesian population models with explicit
contamination/boost modeling and WL-bias calibrations. While our
P($z$)-cut source selection with $\Delta z=0.2$ and $p_{\rm cut}=0.98$
makes cluster-member contamination negligible (as confirmed by the
near-unity boost factor; Section~\ref{sec:boost}), the end-to-end
forward-modeling approach of the eROSITA analyses provides a more
complete treatment of systematics at scale. These complementary
approaches bracket the systematic uncertainty landscape and will
converge as sample sizes grow.

Our mass-bias result, with its current $\sim$14\%
uncertainty, is consistent with the emerging consensus from
multi-probe cluster analyses \citep{Ghirardini2024,2602.12238} that
$(1-b)\sim 0.7$--$0.8$ at low redshift. This range of mass bias
partially alleviates, but does not fully resolve, the tension between
\textit{Planck} cluster counts and primary CMB constraints on $S_8$.
The remaining $\sim$1--2$\sigma$ discrepancy may arise from
a combination of residual mass-calibration systematics (including
baryonic effects, triaxiality, and selection biases) and possible
extensions to $\Lambda$CDM.

\section{Conclusion}
\label{sec:conclusion}

In this work, we have presented a weak-lensing mass calibration of 19
\textit{Planck} SZ-selected galaxy clusters using the HSC-SSP Year~3 (S19A)
data release. Using a forward-modeling approach that integrates over the halo
mass function and the \textit{Planck} selection function with per-cluster
weak-lensing weights and an analytical covariance matrix, we constrain the
mass bias parameter to be
$1-b = 0.73^{+0.10}_{-0.11}$
($\chi^2/\mathrm{dof}=5.2/5$) for a conservative fit range of
$R\in[0.5,5.0]\,h^{-1}\,\mathrm{Mpc}$, with the SZ scatter
simultaneously constrained to $\sigma_{\ln M}=0.25^{+0.10}_{-0.10}$.
The corresponding effective weak-lensing mass is
$\langle M_{500\mathrm{c}}\rangle_{\rm WL}\simeq 3.5\times10^{14}\,
h^{-1}M_\odot$ at an effective redshift $z_{\rm eff}\simeq 0.24$.

This result extends previous HSC-based \textit{Planck} calibrations
\citep{medezinski18}, which analyzed a smaller five-cluster sample, by
leveraging the improved depth and expanded area coverage of the HSC-Y3 shape
catalog and by introducing a consistent weighted-stacking methodology for
both the model prediction and the analytical covariance. Our findings are
consistent with recent weak-lensing mass calibrations of SZ-selected clusters
from various surveys
\citep{miyatake19,Robertson2024,Shirasaki2024,Aymerich2025,shin25},
supporting a picture where the SZ observable--mass relation requires a
significant mass bias correction to reconcile cluster abundance measurements
with primary CMB constraints on cosmological parameters
(Fig.~\ref{fig:comparison_z}).

While our stacked analysis provides valuable constraints on the average mass
bias, several limitations warrant discussion.

First, our sample size of 19~clusters, though larger than previous HSC--Planck
studies, remains modest and is subject to sample variance. Although we now
fit $\sigma_{\ln M}$ jointly with $(1-b)$, the data provide only a modest
constraint on the scatter (Table~\ref{tab:robustness_combined}), and the
inferred $(1-b)$ retains some sensitivity to the width of the prior on
$\sigma_{\ln M}$. Improved external constraints from
joint SZ--X-ray--WL analyses will be important for reducing this degeneracy.

Second, our analysis models the cluster density distribution using the
\textsc{Dark Emulator} without explicit baryonic components.
\citet{cromer21} demonstrated that neglecting baryonic effects leads to a
$\sim$7.5\% systematic overestimation of cluster masses, and \citet{shin25}
found a $\sim$7\% effect in the ACT~DR5$\times$DES~Y3 analysis. Baryonic
feedback processes, such as AGN-driven gas expulsion and star formation, can
redistribute matter within halos, modifying the density profile relative to
dark-matter-only predictions \citep{shirasaki_lau17}. While this effect is
smaller than our $\sim$14\% statistical uncertainty, it will become
increasingly important with larger samples from projects such as LSST. Future analyses should
incorporate baryonic modeling through explicit generalized NFW (GNFW) profiles
for the gas and stellar components \citep{cromer21,shin25}, semi-analytic
baryonification methods \citep{Shirasaki2024}, or emulators trained on
hydrodynamical simulations \citep{giri21}.

Third, our modeling does not account for potential mass- or
redshift-dependent variations in $(1-b)$. The Shin+2025 analysis found
evidence for redshift evolution with $\zeta=2.0^{+0.4}_{-0.7}$. While our
measurement is consistent with a constant bias, larger datasets spanning a
wider mass and redshift range will be needed to test for such dependencies.

Looking ahead, the Rubin LSST will transform cluster mass calibration through weak lensing.
With its unprecedented combination of survey depth
($i\sim 27.5$ for point sources in the Wide-Fast-Deep survey), area
($\sim$18,000~deg$^2$), and multi-band photometry, LSST will enable
weak-lensing measurements of thousands of SZ-selected clusters from
current and future CMB experiments, including the Simons Observatory
\citep{SimonsObservatory2019} and CMB-S4 \citep{CMBS42016}. Deep LSST imaging will allow precise weak-lensing mass
measurements of individual high-mass clusters out to $z\sim 1.5$, moving
beyond the stacking analyses necessary with current datasets, and samples
of thousands of clusters will enable robust tests of potential mass- and
redshift-dependent variations in the SZ observable--mass relation. The
Early Science Program planned by Rubin Observatory \citep{guy2024rubin}
will provide opportunities to validate shape measurement algorithms and
systematic error budgets using Data Previews based on commissioning and science verification data,
before the first annual data release. Joint analyses combining LSST weak
lensing with multi-wavelength cluster observations (X-ray, SZ, optical
richness) will provide comprehensive cross-checks of mass calibration
systematics. Achieving the full scientific potential of LSST for cluster
cosmology will require addressing the theoretical modeling challenges
identified in this work, in particular developing accurate models that
incorporate baryonic effects while maintaining computational tractability
for large samples. The combination of improved observational constraints
from LSST with advances in hydrodynamical simulations and emulator-based
modeling approaches promises to push cluster mass calibration systematics
below the percent level, unlocking the power of cluster abundance as a
precision cosmological probe competitive with other Stage~IV dark energy
experiments.

\begin{acknowledgments}
This work is dedicated to the memory of R\'egulo Plazas.

We thank Masamune Oguri for providing the \textsc{Camira} cluster catalog and for useful discussions, and Xiangchong Li for assistance with the HSCY3 S19 catalog and helpful discussions, as well as the HSC WL working group. This paper underwent HSC collaboration internal review prior to submission. The work of AAPM was supported by the U.S. Department of Energy under contract number DE-AC02-76SF00515. AAPM thanks the Department of Physics of Harvard University and the Laboratory of Particle Astrophysics and Cosmology, the Cosmology Group at Boston University, and the Department of Physics at Washington University in St. Louis for their hospitality during the preparation of this paper.

The Hyper Suprime-Cam (HSC) collaboration includes the astronomical communities of Japan and Taiwan, and Princeton University.  The HSC instrumentation and software were developed by the National Astronomical Observatory of Japan (NAOJ), the Kavli Institute for the Physics and Mathematics of the Universe (Kavli IPMU), the University of Tokyo, the High Energy Accelerator Research Organization (KEK), the Academia Sinica Institute for Astronomy and Astrophysics in Taiwan (ASIAA), and Princeton University.  Funding was contributed by the FIRST program from the Japanese Cabinet Office, the Ministry of Education, Culture, Sports, Science and Technology (MEXT), the Japan Society for the Promotion of Science (JSPS), Japan Science and Technology Agency  (JST), the Toray Science  Foundation, NAOJ, Kavli IPMU, KEK, ASIAA, and Princeton University.

This paper is based [in part] on data collected at the Subaru Telescope and retrieved from the HSC data archive system, which is operated by Subaru Telescope and Astronomy Data Center (ADC) at NAOJ. Data analysis was in part carried out with the cooperation of Center for Computational Astrophysics (CfCA) at NAOJ.  We are honored and grateful for the opportunity of observing the Universe from Maunakea, which has the cultural, historical and natural significance in Hawai'i.

This paper makes use of software developed for Vera C. Rubin Observatory. We thank the Rubin Observatory for making their code available as free software at \url{http://pipelines.lsst.io/}.

The Pan-STARRS1 Surveys (PS1; \citealt{2016arXiv161205560C,2012ApJ...756..158S,2013ApJS..205...20M,2012ApJ...750...99T}) and the PS1 public science archive have been made possible through contributions by the Institute for Astronomy, the University of Hawaii, the Pan-STARRS Project Office, the Max Planck Society and its participating institutes, the Max Planck Institute for Astronomy, Heidelberg, and the Max Planck Institute for Extraterrestrial Physics, Garching, The Johns Hopkins University, Durham University, the University of Edinburgh, the Queen's University Belfast, the Harvard-Smithsonian Center for Astrophysics, the Las Cumbres Observatory Global Telescope Network Incorporated, the National Central University of Taiwan, the Space Telescope Science Institute, the National Aeronautics and Space Administration under grant No. NNX08AR22G issued through the Planetary Science Division of the NASA Science Mission Directorate, the National Science Foundation grant No. AST-1238877, the University of Maryland, Eotvos Lorand University (ELTE), the Los Alamos National Laboratory, and the Gordon and Betty Moore Foundation.

During the preparation of this work, Claude Code (Opus 4.6 and Sonnet 4.6, Anthropic) and ChatGPT (5.5, OpenAI) were used as coding assistants and to help overcome language barriers associated with the use of English as an additional language \citep{Giglio2023, alonso_pavon2025} by improving grammar and writing. After using these tools/services, the authors reviewed and edited the content as needed and take full responsibility for the final publication.
\end{acknowledgments}

\section*{Data Availability}

The HSC-SSP S19A shape and photometric-redshift catalogs are available
through the HSC-SSP data release
site.\footnote{\url{https://hsc-release.mtk.nao.ac.jp/}}
The \emph{Planck} SZ2 catalog and selection function products are available
from the Planck Legacy
Archive.\footnote{\url{https://irsa.ipac.caltech.edu/data/Planck/release_2/}}
The weak-lensing measurement pipeline is described in \citet{more23}.
The analysis code used to produce the results in this paper, including
the stacked model fitting, covariance computation, and figure generation,
will be made available upon request.

\bibliography{main}

@ARTICLE{2012ARAA..50..353K,
       author = {{Kravtsov}, Andrey V. and {Borgani}, Stefano},
        title = "{Formation of Galaxy Clusters}",
      journal = {\araa},
         year = 2012,
        month = sep,
       volume = {50},
        pages = {353-409},
          doi = {10.1146/annurev-astro-081811-125502},
archivePrefix = {arXiv},
       eprint = {1205.5556},
 primaryClass = {astro-ph.CO},
       adsurl = {https://ui.adsabs.harvard.edu/abs/2012ARA&A..50..353K}
}

@ARTICLE{2014MNRAS.444..147O,
       author = {{Oguri}, Masamune},
        title = "{A cluster finding algorithm based on the multiband identification of red sequence galaxies}",
      journal = {\mnras},
         year = 2014,
        month = oct,
       volume = {444},
       number = {1},
        pages = {147-161},
          doi = {10.1093/mnras/stu1446},
archivePrefix = {arXiv},
       eprint = {1407.4693},
 primaryClass = {astro-ph.CO},
       adsurl = {https://ui.adsabs.harvard.edu/abs/2014MNRAS.444..147O}
}

@ARTICLE{2018PASJ...70S..20O,
       author = {{Oguri}, Masamune and {Lin}, Yen-Ting and {Lin}, Sheng-Chieh and {Nishizawa}, Atsushi J. and {More}, Anupreeta and {More}, Surhud and {Hsieh}, Bau-Ching and {Medezinski}, Elinor and {Miyatake}, Hironao and {Jian}, Hung-Yu and {Lin}, Lihwai and {Takada}, Masahiro and {Okabe}, Nobuhiro and {Speagle}, Joshua S. and {Coupon}, Jean and {Leauthaud}, Alexie and {Lupton}, Robert H. and {Miyazaki}, Satoshi and {Price}, Paul A. and {Tanaka}, Masayuki and {Chiu}, I.-Non and {Komiyama}, Yutaka and {Okura}, Yuki and {Tanaka}, Manobu M. and {Usuda}, Tomonori},
        title = "{An optically-selected cluster catalog at redshift 0.1 < z < 1.1 from the Hyper Suprime-Cam Subaru Strategic Program S16A data}",
      journal = {\pasj},
         year = 2018,
        month = jan,
       volume = {70},
          eid = {S20},
        pages = {S20},
          doi = {10.1093/pasj/psx042},
archivePrefix = {arXiv},
       eprint = {1701.00818},
 primaryClass = {astro-ph.CO},
       adsurl = {https://ui.adsabs.harvard.edu/abs/2018PASJ...70S..20O}
}

@INPROCEEDINGS{2017ASPC..512..279J,
   author = {{Juri{\'c}}, M. and {Kantor}, J. and {Lim}, K.-T. and {Lupton}, R.~H. and 
	{Dubois-Felsmann}, G. and {Jenness}, T. and {Axelrod}, T.~S. and 
	{Aleksi{\'c}}, J. and {Allsman}, R.~A. and {AlSayyad}, Y. and 
	{Alt}, J. and {Armstrong}, R. and {Basney}, J. and {Becker}, A.~C. and 
	{Becla}, J. and {Biswas}, R. and {Bosch}, J. and {Boutigny}, D. and 
	{Kind}, M.~C. and {Ciardi}, D.~R. and {Connolly}, A.~J. and 
	{Daniel}, S.~F. and {Daues}, G.~E. and {Economou}, F. and {Chiang}, H.-F. and 
	{Fausti}, A. and {Fisher-Levine}, M. and {Freemon}, D.~M. and 
	{Gris}, P. and {Hernandez}, F. and {Hoblitt}, J. and {Ivezi{\'c}}, Z. and 
	{Jammes}, F. and {Jevremovi{\'c}}, D. and {Jones}, R.~L. and 
	{Kalmbach}, J.~B. and {Kasliwal}, V.~P. and {Krughoff}, K.~S. and 
	{Lurie}, J. and {Lust}, N.~B. and {MacArthur}, L.~A. and {Melchior}, P. and 
	{Moeyens}, J. and {Nidever}, D.~L. and {Owen}, R. and {Parejko}, J.~K. and 
	{Peterson}, J.~M. and {Petravick}, D. and {Pietrowicz}, S.~R. and 
	{Price}, P.~A. and {Reiss}, D.~J. and {Shaw}, R.~A. and {Sick}, J. and 
	{Slater}, C.~T. and {Strauss}, M.~A. and {Sullivan}, I.~S. and 
	{Swinbank}, J.~D. and {Van Dyk}, S. and {Vuj{\v c}i{\'c}}, V. and 
	{Withers}, A. and {Yoachim}, P.},
    title = "{The LSST Data Management System}",
booktitle = {Astronomical Data Analysis Software and Systems XXV},
     year = 2017,
   series = {Astronomical Society of the Pacific Conference Series},
   volume = 512,
archivePrefix = "arXiv",
   eprint = {1512.07914},
 primaryClass = "astro-ph.IM",
   editor = {{Lorente}, N.~P.~F. and {Shortridge}, K. and {Wayth}, R.},
    month = dec,
    pages = {279},
   adsurl = {http://ads.nao.ac.jp/abs/2017ASPC..512..279J}
}

@ARTICLE{2012ApJ...756..158S,
   author = {{Schlafly}, E.~F. and {Finkbeiner}, D.~P. and {Juri{\'c}}, M. and 
	{Magnier}, E.~A. and {Burgett}, W.~S. and {Chambers}, K.~C. and 
	{Grav}, T. and {Hodapp}, K.~W. and {Kaiser}, N. and {Kudritzki}, R.-P. and 
	{Martin}, N.~F. and {Morgan}, J.~S. and {Price}, P.~A. and {Rix}, H.-W. and 
	{Stubbs}, C.~W. and {Tonry}, J.~L. and {Wainscoat}, R.~J.},
    title = "{Photometric Calibration of the First 1.5 Years of the Pan-STARRS1 Survey}",
  journal = {\apj},
archivePrefix = "arXiv",
   eprint = {1201.2208},
 primaryClass = "astro-ph.IM",
     year = 2012,
    month = sep,
   volume = 756,
      eid = {158},
    pages = {158},
      doi = {10.1088/0004-637X/756/2/158},
   adsurl = {http://ads.nao.ac.jp/abs/2012ApJ...756..158S}
}

@ARTICLE{2022AA...661A..11C,
       author = {{Chiu}, I.-Non and {Ghirardini}, Vittorio and {Liu}, Ang and {Grandis}, Sebastian and {Bulbul}, Esra and {Bahar}, Y. Emre and {Comparat}, Johan and {Bocquet}, Sebastian and {Clerc}, Nicolas and {Klein}, Matthias and {Liu}, Teng and {Li}, Xiangchong and {Miyatake}, Hironao and {Mohr}, Joseph and {More}, Surhud and {Oguri}, Masamune and {Okabe}, Nobuhiro and {Pacaud}, Florian and {Ramos-Ceja}, Miriam E. and {Reiprich}, Thomas H. and {Schrabback}, Tim and {Umetsu}, Keiichi},
        title = "{The eROSITA Final Equatorial-Depth Survey (eFEDS). X-ray observable-to-mass-and-redshift relations of galaxy clusters and groups with weak-lensing mass calibration from the Hyper Suprime-Cam Subaru Strategic Program survey}",
      journal = {\aap},
         year = 2022,
        month = may,
       volume = {661},
          eid = {A11},
        pages = {A11},
          doi = {10.1051/0004-6361/202141755},
archivePrefix = {arXiv},
       eprint = {2107.05652},
 primaryClass = {astro-ph.CO},
       adsurl = {https://ui.adsabs.harvard.edu/abs/2022A&A...661A..11C}
}

@ARTICLE{2016arXiv161205560C,
       author = {{Chambers}, K.~C. and {Magnier}, E.~A. and {Metcalfe}, N. and {Flewelling}, H.~A. and {Huber}, M.~E. and {Waters}, C.~Z. and {Denneau}, L. and {Draper}, P.~W. and {Farrow}, D. and {Finkbeiner}, D.~P. and {Holmberg}, C. and {Koppenhoefer}, J. and {Price}, P.~A. and {Rest}, A. and {Saglia}, R.~P. and {Schlafly}, E.~F. and {Smartt}, S.~J. and {Sweeney}, W. and {Wainscoat}, R.~J. and {Burgett}, W.~S. and {Chastel}, S. and {Grav}, T. and {Heasley}, J.~N. and {Hodapp}, K.~W. and {Jedicke}, R. and {Kaiser}, N. and {Kudritzki}, R.-P. and {Luppino}, G.~A. and {Lupton}, R.~H. and {Monet}, D.~G. and {Morgan}, J.~S. and {Onaka}, P.~M. and {Shiao}, B. and {Stubbs}, C.~W. and {Tonry}, J.~L. and {White}, R. and {Ba{\~n}ados}, E. and {Bell}, E.~F. and {Bender}, R. and {Bernard}, E.~J. and {Boegner}, M. and {Boffi}, F. and {Botticella}, M.~T. and {Calamida}, A. and {Casertano}, S. and {Chen}, W.-P. and {Chen}, X. and {Cole}, S. and {Deacon}, N. and {Frenk}, C. and {Fitzsimmons}, A. and {Gezari}, S. and {Gibbs}, V. and {Goessl}, C. and {Goggia}, T. and {Gourgue}, R. and {Goldman}, B. and {Grant}, P. and {Grebel}, E.~K. and {Hambly}, N.~C. and {Hasinger}, G. and {Heavens}, A.~F. and {Heckman}, T.~M. and {Henderson}, R. and {Henning}, T. and {Holman}, M. and {Hopp}, U. and {Ip}, W.-H. and {Isani}, S. and {Jackson}, M. and {Keyes}, C.~D. and {Koekemoer}, A.~M. and {Kotak}, R. and {Le}, D. and {Liska}, D. and {Long}, K.~S. and {Lucey}, J.~R. and {Liu}, M. and {Martin}, N.~F. and {Masci}, G. and {McLean}, B. and {Mindel}, E. and {Misra}, P. and {Morganson}, E. and {Murphy}, D.~N.~A. and {Obaika}, A. and {Narayan}, G. and {Nieto-Santisteban}, M.~A. and {Norberg}, P. and {Peacock}, J.~A. and {Pier}, E.~A. and {Postman}, M. and {Primak}, N. and {Rae}, C. and {Rai}, A. and {Riess}, A. and {Riffeser}, A. and {Rix}, H.~W. and {R{\"o}ser}, S. and {Russel}, R. and {Rutz}, L. and {Schilbach}, E. and {Schultz}, A.~S.~B. and {Scolnic}, D. and {Strolger}, L. and {Szalay}, A. and {Seitz}, S. and {Small}, E. and {Smith}, K.~W. and {Soderblom}, D.~R. and {Taylor}, P. and {Thomson}, R. and {Taylor}, A.~N. and {Thakar}, A.~R. and {Thiel}, J. and {Thilker}, D. and {Unger}, D. and {Urata}, Y. and {Valenti}, J. and {Wagner}, J. and {Walder}, T. and {Walter}, F. and {Watters}, S.~P. and {Werner}, S. and {Wood-Vasey}, W.~M. and {Wyse}, R.},
        title = "{The Pan-STARRS1 Surveys}",
      journal = {arXiv e-prints},
         year = 2016,
        month = dec,
          eid = {arXiv:1612.05560},
        pages = {arXiv:1612.05560},
          doi = {10.48550/arXiv.1612.05560},
archivePrefix = {arXiv},
       eprint = {1612.05560},
 primaryClass = {astro-ph.IM},
       adsurl = {https://ui.adsabs.harvard.edu/abs/2016arXiv161205560C}
}

@ARTICLE{2012ApJ...750...99T,
   author = {{Tonry}, J.~L. and {Stubbs}, C.~W. and {Lykke}, K.~R. and {Doherty}, P. and 
	{Shivvers}, I.~S. and {Burgett}, W.~S. and {Chambers}, K.~C. and 
	{Hodapp}, K.~W. and {Kaiser}, N. and {Kudritzki}, R.-P. and 
	{Magnier}, E.~A. and {Morgan}, J.~S. and {Price}, P.~A. and 
	{Wainscoat}, R.~J.},
    title = "{The Pan-STARRS1 Photometric System}",
  journal = {\apj},
archivePrefix = "arXiv",
   eprint = {1203.0297},
 primaryClass = "astro-ph.IM",
     year = 2012,
    month = may,
   volume = 750,
      eid = {99},
    pages = {99},
      doi = {10.1088/0004-637X/750/2/99},
   adsurl = {http://ads.nao.ac.jp/abs/2012ApJ...750...99T}
}

@ARTICLE{2013ApJS..205...20M,
   author = {{Magnier}, E.~A. and {Schlafly}, E. and {Finkbeiner}, D. and 
	{Juric}, M. and {Tonry}, J.~L. and {Burgett}, W.~S. and {Chambers}, K.~C. and 
	{Flewelling}, H.~A. and {Kaiser}, N. and {Kudritzki}, R.-P. and 
	{Morgan}, J.~S. and {Price}, P.~A. and {Sweeney}, W.~E. and 
	{Stubbs}, C.~W.},
    title = "{The Pan-STARRS 1 Photometric Reference Ladder, Release 12.01}",
  journal = {\apjs},
archivePrefix = "arXiv",
   eprint = {1303.3634},
 primaryClass = "astro-ph.IM",
     year = 2013,
    month = apr,
   volume = 205,
      eid = {20},
    pages = {20},
      doi = {10.1088/0067-0049/205/2/20},
   adsurl = {http://ads.nao.ac.jp/abs/2013ApJS..205...20M}
}

@ARTICLE{2018PASJ...70S...1M,
   author = {{Miyazaki}, S. and {Komiyama}, Y. and {Kawanomoto}, S. and {Doi}, Y. and 
	{Furusawa}, H. and {Hamana}, T. and {Hayashi}, Y. and {Ikeda}, H. and 
	{Kamata}, Y. and {Karoji}, H. and {Koike}, M. and {Kurakami}, T. and 
	{Miyama}, S. and {Morokuma}, T. and {Nakata}, F. and {Namikawa}, K. and 
	{Nakaya}, H. and {Nariai}, K. and {Obuchi}, Y. and {Oishi}, Y. and 
	{Okada}, N. and {Okura}, Y. and {Tait}, P. and {Takata}, T. and 
	{Tanaka}, Y. and {Tanaka}, M. and {Terai}, T. and {Tomono}, D. and 
	{Uraguchi}, F. and {Usuda}, T. and {Utsumi}, Y. and {Yamada}, Y. and 
	{Yamanoi}, H. and {Aihara}, H. and {Fujimori}, H. and {Mineo}, S. and 
	{Miyatake}, H. and {Oguri}, M. and {Uchida}, T. and {Tanaka}, M.~M. and 
	{Yasuda}, N. and {Takada}, M. and {Murayama}, H. and {Nishizawa}, A.~J. and 
	{Sugiyama}, N. and {Chiba}, M. and {Futamase}, T. and {Wang}, S.-Y. and 
	{Chen}, H.-Y. and {Ho}, P.~T.~P. and {Liaw}, E.~J.~Y. and {Chiu}, C.-F. and 
	{Ho}, C.-L. and {Lai}, T.-C. and {Lee}, Y.-C. and {Jeng}, D.-Z. and 
	{Iwamura}, S. and {Armstrong}, R. and {Bickerton}, S. and {Bosch}, J. and 
	{Gunn}, J.~E. and {Lupton}, R.~H. and {Loomis}, C. and {Price}, P. and 
	{Smith}, S. and {Strauss}, M.~A. and {Turner}, E.~L. and {Suzuki}, H. and 
	{Miyazaki}, Y. and {Muramatsu}, M. and {Yamamoto}, K. and {Endo}, M. and 
	{Ezaki}, Y. and {Ito}, N. and {Kawaguchi}, N. and {Sofuku}, S. and 
	{Taniike}, T. and {Akutsu}, K. and {Dojo}, N. and {Kasumi}, K. and 
	{Matsuda}, T. and {Imoto}, K. and {Miwa}, Y. and {Suzuki}, M. and 
	{Takeshi}, K. and {Yokota}, H.},
    title = "{Hyper Suprime-Cam: System design and verification of image quality}",
  journal = {\pasj},
     year = 2018,
    month = jan,
   volume = 70,
      eid = {S1},
    pages = {S1},
      doi = {10.1093/pasj/psx063},
   adsurl = {http://ads.nao.ac.jp/abs/2018PASJ...70S...1M}
}

@ARTICLE{2019ApJ...873..111I,
   author = {{Ivezi{\'c}}, {\v Z}. and {Kahn}, S.~M. and {Tyson}, J.~A. and 
	{Abel}, B. and {Acosta}, E. and {Allsman}, R. and {Alonso}, D. and 
	{AlSayyad}, Y. and {Anderson}, S.~F. and {Andrew}, J. and et al.},
    title = "{LSST: From Science Drivers to Reference Design and Anticipated Data Products}",
  journal = {\apj},
archivePrefix = "arXiv",
   eprint = {0805.2366},
     year = 2019,
    month = mar,
   volume = 873,
      eid = {111},
    pages = {111},
      doi = {10.3847/1538-4357/ab042c},
   adsurl = {http://ads.nao.ac.jp/abs/2019ApJ...873..111I}
}

@ARTICLE{2011MNRAS.414.1840M,
       author = {{Medezinski}, Elinor and {Broadhurst}, Tom and {Umetsu}, Keiichi and {Ben{\'\i}tez}, Narciso and {Taylor}, Andy},
        title = "{A weak lensing detection of the cosmological distance-redshift relation behind three massive clusters}",
      journal = {\mnras},
         year = 2011,
        month = jul,
       volume = {414},
       number = {3},
        pages = {1840-1850},
          doi = {10.1111/j.1365-2966.2011.18332.x},
archivePrefix = {arXiv},
       eprint = {1101.1955},
 primaryClass = {astro-ph.CO},
       adsurl = {https://ui.adsabs.harvard.edu/abs/2011MNRAS.414.1840M}
}

@ARTICLE{more23,
       author = {{More}, Surhud and {Sugiyama}, Sunao and {Miyatake}, Hironao and {Rau}, Markus Michael and {Shirasaki}, Masato and {Li}, Xiangchong and {Nishizawa}, Atsushi J. and {Osato}, Ken and {Zhang}, Tianqing and {Takada}, Masahiro and {Hamana}, Takashi and {Takahashi}, Ryuichi and {Dalal}, Roohi and {Mandelbaum}, Rachel and {Strauss}, Michael A. and {Kobayashi}, Yosuke and {Nishimichi}, Takahiro and {Oguri}, Masamune and {Luo}, Wentao and {Kannawadi}, Arun and {Hsieh}, Bau-Ching and {Armstrong}, Robert and {Bosch}, James and {Komiyama}, Yutaka and {Lupton}, Robert H. and {Lust}, Nate B. and {MacArthur}, Lauren A. and {Miyazaki}, Satoshi and {Murayama}, Hitoshi and {Okura}, Yuki and {Price}, Paul A. and {Tait}, Philip J. and {Tanaka}, Masayuki and {Wang}, Shiang-Yu},
        title = "{Hyper Suprime-Cam Year 3 results: Measurements of clustering of SDSS-BOSS galaxies, galaxy-galaxy lensing, and cosmic shear}",
      journal = {\prd},
         year = 2023,
        month = dec,
       volume = {108},
       number = {12},
          eid = {123520},
        pages = {123520},
          doi = {10.1103/PhysRevD.108.123520},
archivePrefix = {arXiv},
       eprint = {2304.00703},
 primaryClass = {astro-ph.CO},
       adsurl = {https://ui.adsabs.harvard.edu/abs/2023PhRvD.108l3520M}
}

@MISC{2025rubn.rept...32N,
       author = {{NSF-DOE Vera C. Rubin Observatory}},
        title = "{PSTN-019: The LSST Science Pipelines Software: Optical Survey Pipeline Reduction and Analysis Environment}",
 howpublished = {NSF-DOE Vera C. Rubin Observatory Technical Report},
         year = 2025,
        month = jan,
        pages = {32},
          doi = {10.71929/RUBIN/2570545},
       adsurl = {https://ui.adsabs.harvard.edu/abs/2025rubn.rept...32N}
}

@article{navarro1995simulations,
  title={Simulations of X-ray clusters},
  author={Navarro, Julio F and Frenk, Carlos S and White, Simon DM},
  journal={Monthly Notices of the Royal Astronomical Society},
  volume={275},
  number={3},
  pages={720--740},
  year={1995},
  publisher={Oxford University Press Oxford, UK}
}

@misc{guy2024rubin,
  title={Rubin Observatory plans for an early science program},
  author={Guy, Leanne P and Bechtol, Keith and Bellm, Eric and Blum, Bob and Graham, Melissa L and Ivezi{\'c}, {\v{Z}}eljko and Lupton, Robert H and Marshall, Phil and Strauss, Michael},
  year={2024}
}

@ARTICLE{Foreman-Mackey13,
       author = {{Foreman-Mackey}, Daniel and {Hogg}, David W. and {Lang}, Dustin and {Goodman}, Jonathan},
        title = "{emcee: The MCMC Hammer}",
      journal = {\pasp},
         year = 2013,
        month = mar,
       volume = {125},
       number = {925},
        pages = {306},
          doi = {10.1086/670067},
archivePrefix = {arXiv},
       eprint = {1202.3665},
 primaryClass = {astro-ph.IM},
       adsurl = {https://ui.adsabs.harvard.edu/abs/2013PASP..125..306F}
}

@ARTICLE{giri21,
       author = {{Giri}, Sambit K. and {Schneider}, Aurel},
        title = "{Emulation of baryonic effects on the matter power spectrum and constraints from galaxy cluster data}",
      journal = {\jcap},
         year = 2021,
        month = dec,
       volume = {2021},
       number = {12},
          eid = {046},
        pages = {046},
          doi = {10.1088/1475-7516/2021/12/046},
archivePrefix = {arXiv},
       eprint = {2108.08863},
 primaryClass = {astro-ph.CO},
       adsurl = {https://ui.adsabs.harvard.edu/abs/2021JCAP...12..046G}
}

@ARTICLE{cromer21,
       author = {{Cromer}, Dylan and {Battaglia}, Nicholas and {Miyatake}, Hironao and {Simet}, Melanie},
        title = "{Towards 1\% accurate galaxy cluster masses: including baryons in weak-lensing mass inference}",
      journal = {\jcap},
         year = 2022,
        month = oct,
       volume = {2022},
       number = {10},
          eid = {034},
        pages = {034},
          doi = {10.1088/1475-7516/2022/10/034},
archivePrefix = {arXiv},
       eprint = {2104.06925},
 primaryClass = {astro-ph.CO},
       adsurl = {https://ui.adsabs.harvard.edu/abs/2022JCAP...10..034C}
}

@ARTICLE{shin25,
       author = {{Shin}, T. and {Baxter}, E.~J. and {Lee}, E. and {Battaglia}, N. and {Alarcon}, A. and {Amon}, A. and {Becker}, M. and {Bernstein}, G. and {Bond}, J.~R. and {Campos}, A. and {Chang}, C. and {Chen}, R. and {Choi}, A. and {DeRose}, J. and {Dodelson}, S. and {Doux}, C. and {Dunkley}, J. and {Elvin-Poole}, J. and {Esteves}, J.~H. and {Everett}, S. and {Fert{\'e}}, A. and {Gatti}, M. and {Grandis}, S. and {Gruen}, D. and {Harrison}, I. and {Hill}, J.~C. and {Hilton}, M. and {Jarvis}, M. and {MacCrann}, N. and {McCullough}, J. and {Moodley}, K. and {Mroczkowski}, T. and {Myles}, J. and {Navarro Alsina}, A. and {Nicola}, A. and {Page}, L. and {Pandey}, S. and {Prat}, J. and {Raveri}, M. and {Ried Guachalla}, B. and {Rollins}, R.~P. and {Sanchez}, C. and {Secco}, L.~F. and {Sheldon}, E. and {Sif{\'o}n}, C. and {Troxel}, M. and {Tutusaus}, I. and {von der Linden}, A. and {Wollack}, E. and {Yin}, B. and {Aguena}, M. and {Allam}, S.~S. and {Alves}, O. and {Andrade-Oliveira}, F. and {Bacon}, D. and {Bocquet}, S. and {Brooks}, D. and {Camilleri}, R. and {Carnero Rosell}, A. and {Carretero}, J. and {Castander}, F.~J. and {Costanzi}, M. and {da Costa}, L. and {da Silva Pereira}, M.~E. and {Davis}, T. and {De Vicente}, J. and {Desai}, S. and {Flaugher}, B. and {Frieman}, J. and {Garcia-Bellido}, J. and {Gutierrez}, G. and {Hinton}, S. and {Hollowood}, D.~L. and {Huterer}, D. and {James}, D. and {Lee}, S. and {Marshall}, J. and {Mena-Fern{\'a}ndez}, J. and {Menanteau}, F. and {Miquel}, R. and {Mohr}, J. and {Muir}, J. and {Ogando}, R. and {Plazas Malag{\'o}n}, A. and {Porredon}, A. and {Romer}, K. and {Sanchez}, E. and {Sanchez Cid}, D. and {Sevilla}, I. and {Smith}, M. and {Soares-Santos}, M. and {Suchyta}, E. and {Swanson}, M. and {To}, C. and {Weaverdyck}, N. and {Weller}, J.},
        title = "{Weak Lensing Mass Calibration of the ACT DR5 Galaxy Clusters with the DES Year 3 Weak Lensing Data}",
      journal = {arXiv e-prints},
         year = 2025,
        month = dec,
          eid = {arXiv:2512.18935},
        pages = {arXiv:2512.18935},
          doi = {10.48550/arXiv.2512.18935},
archivePrefix = {arXiv},
       eprint = {2512.18935},
 primaryClass = {astro-ph.CO},
       adsurl = {https://ui.adsabs.harvard.edu/abs/2025arXiv251218935S}
}

@article{vonderLinden2014,
  author = {{von der Linden}, Anja and {Mantz}, Adam and {Allen}, Steven W. and {Applegate}, Douglas E. and {Morris}, R. Glenn and {Allen}, Mark T. and {Kelly}, Patrick L. and {Burke}, David L. and {Ebeling}, Harald and {Burchat}, Patricia R. and {Donovan}, David and {Schmidt}, R. W.},
  title = "{Robust weak-lensing mass calibration of Planck galaxy clusters}",
  journal = {Mon. Not. R. Astron. Soc.},
  volume = {443},
  number = {3},
  pages = {1973--1978},
  year = {2014},
  doi = {10.1093/mnras/stu1423},
  archivePrefix = {arXiv},
  eprint = {1402.2670},
  primaryClass = {astro-ph.CO}
}

@article{Hoekstra2015,
  author = {Hoekstra, H. and Herbonnet, R. and Muzzin, A. and Babul, A. and Mahdavi, A. and Viola, M. and Cacciato, M.},
  title = "{The Canadian Cluster Comparison Project: detailed study of systematics and updated weak lensing masses}",
  journal = {Mon. Not. R. Astron. Soc.},
  volume = {449},
  pages = {685--714},
  year = {2015},
  doi = {10.1093/mnras/stv275}
}

@article{Zubeldia2019,
  author = {Zubeldia, I. and Challinor, A.},
  title = "{Cosmological constraints from Planck galaxy clusters with CMB lensing mass bias calibration}",
  journal = {Mon. Not. R. Astron. Soc.},
  volume = {489},
  pages = {401--419},
  year = {2019},
  doi = {10.1093/mnras/stz2153}
}

@article{Herbonnet2020,
  author = {Herbonnet, R. and Sif{\'o}n, C. and Hoekstra, H. and Bah{\'e}, Y. and van der Burg, R.~F.~J. and Melin, J.-B. and von der Linden, A. and Sand, D. and Kay, S. and Barnes, D.},
  title = "{CCCP and MENeaCS: (updated) weak-lensing masses for 100 galaxy clusters}",
  journal = {Mon. Not. R. Astron. Soc.},
  volume = {497},
  pages = {4684--4703},
  year = {2020},
  doi = {10.1093/mnras/staa2303}
}

@article{Robertson2024,
  author = {Robertson, N.~C. and Sif{\'o}n, C. and Asgari, M. and Battaglia, N. and Bilicki, M. and Bond, J.~R. and et al.},
  title = "{ACT-DR5 Sunyaev--Zel'dovich clusters: Weak lensing mass calibration with KiDS}",
  journal = {Astron. Astrophys.},
  volume = {681},
  pages = {A87},
  year = {2024},
  doi = {10.1051/0004-6361/202346712},
  archivePrefix = {arXiv},
  eprint = {2304.10219},
  primaryClass = {astro-ph.CO}
}

@article{PennaLima2017,
  author = {Penna-Lima, M. and Bartlett, J.~G. and Rozo, E. and Melin, J.-B. and Merten, J. and Evrard, A.~E. and Postman, M. and Rykoff, E.},
  title = "{Calibrating the Planck cluster mass scale with CLASH}",
  journal = {Astron. Astrophys.},
  volume = {604},
  pages = {A89},
  year = {2017},
  doi = {10.1051/0004-6361/201629971}
}

@article{Sereno2017,
  author = {{Sereno}, Mauro and {Covone}, Giovanni and {Izzo}, Luca and {Ettori}, Stefano and {Coupon}, Jean and {Lieu}, Maggie},
  title = "{PSZ2LenS: Weak lensing analysis of the Planck clusters in the CFHTLenS and RCSLenS}",
  journal = {Mon. Not. R. Astron. Soc.},
  volume = {472},
  pages = {1946--1971},
  year = {2017},
  doi = {10.1093/mnras/stx2085}
}

@ARTICLE{Schrabback2021,
       author = {{Schrabback}, T. and {Bocquet}, S. and {Sommer}, M. and {Zohren}, H. and {van den Busch}, J.~L. and {Hern{\'a}ndez-Mart{\'\i}n}, B. and {Hoekstra}, H. and {Raihan}, S.~F. and {Schirmer}, M. and {Applegate}, D. and {Bayliss}, M. and {Benson}, B.~A. and {Bleem}, L.~E. and {Dietrich}, J.~P. and {Floyd}, B. and {Hilbert}, S. and {Hlavacek-Larrondo}, J. and {McDonald}, M. and {Saro}, A. and {Stark}, A.~A. and {Weissgerber}, N.},
        title = "{Mass calibration of distant SPT galaxy clusters through expanded weak-lensing follow-up observations with HST, VLT, \& Gemini-South}",
      journal = {\mnras},
         year = 2021,
        month = aug,
       volume = {505},
       number = {3},
        pages = {3923-3943},
          doi = {10.1093/mnras/stab1386},
archivePrefix = {arXiv},
       eprint = {2009.07591},
 primaryClass = {astro-ph.CO},
       adsurl = {https://ui.adsabs.harvard.edu/abs/2021MNRAS.505.3923S}
}

@article{Shirasaki2024,
  author = {{Shirasaki}, Masato and {Sif{\'o}n}, Crist{\'o}bal and {Miyatake}, Hironao and {Lau}, Erwin and {Zhang}, Zhuowen and {Bahcall}, Neta and {Battaglia}, Nicholas and {Devlin}, Mark and {Dunkley}, Jo and {Farahi}, Arya and {Hilton}, Matt and {Lin}, Yen-Ting and {Nagai}, Daisuke and {Staggs}, Suzanne T. and {Sunayama}, Tomomi and {Spergel}, David and {Wollack}, Edward J.},
  title = "{Masses of Sunyaev-Zel'dovich galaxy clusters detected by the Atacama Cosmology Telescope: Stacked lensing measurements with Subaru HSC year 3 data}",
  journal = {\prd},
  volume = {110},
  number = {10},
  eid = {103006},
  pages = {103006},
  year = {2024},
  doi = {10.1103/PhysRevD.110.103006},
  archivePrefix = {arXiv},
  eprint = {2407.08201},
  primaryClass = {astro-ph.CO}
}

@ARTICLE{Nishimichi2019,
       author = {{Nishimichi}, Takahiro and {Takada}, Masahiro and {Takahashi}, Ryuichi and {Osato}, Ken and {Shirasaki}, Masato and {Oogi}, Taira and {Miyatake}, Hironao and {Oguri}, Masamune and {Murata}, Ryoma and {Kobayashi}, Yosuke and {Yoshida}, Naoki},
        title = "{Dark Quest. I. Fast and Accurate Emulation of Halo Clustering Statistics and Its Application to Galaxy Clustering}",
      journal = {\apj},
         year = 2019,
        month = oct,
       volume = {884},
       number = {1},
          eid = {29},
        pages = {29},
          doi = {10.3847/1538-4357/ab3719},
archivePrefix = {arXiv},
       eprint = {1811.09504},
 primaryClass = {astro-ph.CO},
       adsurl = {https://ui.adsabs.harvard.edu/abs/2019ApJ...884...29N}
}

@article{Sunayama2020,
  author = {Sunayama, T. and Park, Y. and Takada, M. and Kobayashi, Y. and Nishimichi, T. and Kurita, T. and More, S. and Oguri, M. and Osato, K.},
  title = "{The impact of projection effects on cluster observables: stacked lensing and projected clustering}",
  journal = {Mon. Not. R. Astron. Soc.},
  volume = {496},
  number = {4},
  pages = {4468--4487},
  year = {2020},
  doi = {10.1093/mnras/staa1646},
  archivePrefix = {arXiv},
  eprint = {2002.03867},
  primaryClass = {astro-ph.CO}
}

@ARTICLE{hirata2003,
       author = {{Hirata}, Christopher and {Seljak}, Uro{\v{s}}},
        title = "{Shear calibration biases in weak-lensing surveys}",
      journal = {\mnras},
         year = 2003,
        month = aug,
       volume = {343},
       number = {2},
        pages = {459-480},
          doi = {10.1046/j.1365-8711.2003.06683.x},
archivePrefix = {arXiv},
       eprint = {astro-ph/0301054},
 primaryClass = {astro-ph},
       adsurl = {https://ui.adsabs.harvard.edu/abs/2003MNRAS.343..459H}
}

@article{tinker08,
  author = {{Tinker}, J. and {Kravtsov}, A.~V. and {Klypin}, A. and {Abazajian}, K. and {Warren}, M. and {Yepes}, G. and {Gottl{\"o}ber}, S. and {Holz}, D.~E.},
  title = "{Toward a Halo Mass Function for Precision Cosmology: The Limits of Universality}",
  journal = {ApJ},
  volume = {688},
  pages = {709-728},
  year = {2008},
  doi = {10.1086/591439}
}

@ARTICLE{2025arXiv250721459A,
       author = {{ACTDESHSC Collaboration} and {Aguena}, M. and {Aiola}, S. and {Allam}, S. and {Andrade-Oliveira}, F. and {Bacon}, D. and {Bahcall}, N. and {Battaglia}, N. and {Battistelli}, E.~S. and {Bocquet}, S. and {Bolliet}, B. and {Bond}, J.~R. and {Brooks}, D. and {Calabrese}, E. and {Carretero}, J. and {Choi}, S.~K. and {da Costa}, L.~N. and {Costanzi}, M. and {Coulton}, W. and {Davis}, T.~M. and {Desai}, S. and {Devlin}, M.~J. and {Dicker}, S. and {Doel}, P. and {Duivenvoorden}, A.~J. and {Dunkley}, J. and {Ferraro}, S. and {Flaugher}, B. and {Frieman}, J. and {Gallardo}, P.~A. and {Gatti}, M. and {Gaztanaga}, E. and {Gill}, A.~S. and {Golec}, J.~E. and {Gruen}, D. and {Gruendl}, R.~A. and {Halpern}, M. and {Hasselfield}, M. and {Hill}, J.~C. and {Hilton}, M. and {Hincks}, A.~D. and {Hinton}, S.~R. and {Hollowood}, D.~L. and {Honscheid}, K. and {Hubmayr}, J. and {Huffenberger}, K.~M. and {Hughes}, J.~P. and {James}, D.~J. and {Klein}, M. and {Knowles}, K. and {Koopman}, B.~J. and {Kosowsky}, A. and {Lahav}, O. and {Lee}, E. and {Lin}, Y. and {Lokken}, M. and {Madhavacheril}, M.~S. and {Plazas Malag{\'o}n}, A.~A. and {Marrewijk}, J. v. and {Marshall}, J.~L. and {McMahon}, J. and {Mena-Fern{\'a}ndez}, J. and {Miquel}, R. and {Miyatake}, H. and {Mohr}, J.~J. and {Moodley}, K. and {Mroczkowski}, T. and {Naess}, S. and {Nati}, F. and {Nicola}, A. and {Niemack}, M.~D. and {Ogando}, R.~L.~C. and {Oguri}, M. and {Orlowski-Scherer}, J. and {Page}, L.~A. and {Partridge}, B. and {da Silva Pereira}, M.~E. and {Porredon}, A. and {Qu}, F.~J. and {Ragavan}, D.~C. and {Ried Guachalla}, B. and {Romer}, A.~K. and {Carnero Rosell}, A. and {Rykoff}, E.~S. and {Samuroff}, S. and {Sanchez}, E. and {Sevilla-Noarbe}, I. and {Sierra}, C. and {Sif{\'o}n}, C. and {Smith}, M. and {Staggs}, S.~T. and {Suchyta}, E. and {Swanson}, M.~E.~C. and {Tucker}, D.~L. and {Vargas}, C. and {Vavagiakis}, E.~M. and {De Vicente}, J. and {Weaverdyck}, N. and {Weller}, J. and {Wollack}, E.~J. and {Zubeldia}, I.},
        title = "{The Atacama Cosmology Telescope: DR6 Sunyaev-Zel'dovich Selected Galaxy Clusters Catalog}",
      journal = {arXiv e-prints},
         year = 2025,
        month = jul,
          eid = {arXiv:2507.21459},
        pages = {arXiv:2507.21459},
          doi = {10.48550/arXiv.2507.21459},
archivePrefix = {arXiv},
       eprint = {2507.21459},
 primaryClass = {astro-ph.CO},
       adsurl = {https://ui.adsabs.harvard.edu/abs/2025arXiv250721459A}
}

@ARTICLE{evrard1990,
       author = {{Evrard}, August E.},
        title = "{Formation and Evolution of X-Ray Clusters: A Hydrodynamic Simulation of the Intracluster Medium}",
      journal = {\apj},
         year = 1990,
        month = nov,
       volume = {363},
        pages = {349},
          doi = {10.1086/169350},
       adsurl = {https://ui.adsabs.harvard.edu/abs/1990ApJ...363..349E}
}

@ARTICLE{rykoff16,
       author = {{Rykoff}, E.~S. and {Rozo}, E. and {Hollowood}, D. and {Bermeo-Hernandez}, A. and {Jeltema}, T. and {Mayers}, J. and {Romer}, A.~K. and {Rooney}, P. and {Saro}, A. and {Vergara Cervantes}, C. and {Wechsler}, R.~H. and {Wilcox}, H. and {Abbott}, T.~M.~C. and {Abdalla}, F.~B. and {Allam}, S. and {Annis}, J. and {Benoit-L{\'e}vy}, A. and {Bernstein}, G.~M. and {Bertin}, E. and {Brooks}, D. and {Burke}, D.~L. and {Capozzi}, D. and {Carnero Rosell}, A. and {Carrasco Kind}, M. and {Castander}, F.~J. and {Childress}, M. and {Collins}, C.~A. and {Cunha}, C.~E. and {D'Andrea}, C.~B. and {da Costa}, L.~N. and {Davis}, T.~M. and {Desai}, S. and {Diehl}, H.~T. and {Dietrich}, J.~P. and {Doel}, P. and {Evrard}, A.~E. and {Finley}, D.~A. and {Flaugher}, B. and {Fosalba}, P. and {Frieman}, J. and {Glazebrook}, K. and {Goldstein}, D.~A. and {Gruen}, D. and {Gruendl}, R.~A. and {Gutierrez}, G. and {Hilton}, M. and {Honscheid}, K. and {Hoyle}, B. and {James}, D.~J. and {Kay}, S.~T. and {Kuehn}, K. and {Kuropatkin}, N. and {Lahav}, O. and {Lewis}, G.~F. and {Lidman}, C. and {Lima}, M. and {Maia}, M.~A.~G. and {Mann}, R.~G. and {Marshall}, J.~L. and {Martini}, P. and {Melchior}, P. and {Miller}, C.~J. and {Miquel}, R. and {Mohr}, J.~J. and {Nichol}, R.~C. and {Nord}, B. and {Ogando}, R. and {Plazas}, A.~A. and {Reil}, K. and {Sahl{\'e}n}, M. and {Sanchez}, E. and {Santiago}, B. and {Scarpine}, V. and {Schubnell}, M. and {Sevilla-Noarbe}, I. and {Smith}, R.~C. and {Soares-Santos}, M. and {Sobreira}, F. and {Stott}, J.~P. and {Suchyta}, E. and {Swanson}, M.~E.~C. and {Tarle}, G. and {Thomas}, D. and {Tucker}, D. and {Uddin}, S. and {Viana}, P.~T.~P. and {Vikram}, V. and {Walker}, A.~R. and {Zhang}, Y. and {DES Collaboration}},
        title = "{The RedMaPPer Galaxy Cluster Catalog From DES Science Verification Data}",
      journal = {\apjs},
         year = 2016,
        month = may,
       volume = {224},
       number = {1},
          eid = {1},
        pages = {1},
          doi = {10.3847/0067-0049/224/1/1},
archivePrefix = {arXiv},
       eprint = {1601.00621},
 primaryClass = {astro-ph.CO},
       adsurl = {https://ui.adsabs.harvard.edu/abs/2016ApJS..224....1R}
}

@ARTICLE{battaglia16,
       author = {{Battaglia}, N. and {Leauthaud}, A. and {Miyatake}, H. and {Hasselfield}, M. and {Gralla}, M.~B. and {Allison}, R. and {Bond}, J.~R. and {Calabrese}, E. and {Crichton}, D. and {Devlin}, M.~J. and {Dunkley}, J. and {D{\"u}nner}, R. and {Erben}, T. and {Ferrara}, S. and {Halpern}, M. and {Hilton}, M. and {Hill}, J.~C. and {Hincks}, A.~D. and {Hlo{\v{z}}ek}, R. and {Huffenberger}, K.~M. and {Hughes}, J.~P. and {Kneib}, J.~P. and {Kosowsky}, A. and {Makler}, M. and {Marriage}, T.~A. and {Menanteau}, F. and {Miller}, L. and {Moodley}, K. and {Moraes}, B. and {Niemack}, M.~D. and {Page}, L. and {Shan}, H. and {Sehgal}, N. and {Sherwin}, B.~D. and {Sievers}, J.~L. and {Sif{\'o}n}, C. and {Spergel}, D.~N. and {Staggs}, S.~T. and {Taylor}, J.~E. and {Thornton}, R. and {van Waerbeke}, L. and {Wollack}, E.~J.},
        title = "{Weak-lensing mass calibration of the Atacama Cosmology Telescope equatorial Sunyaev-Zeldovich cluster sample with the Canada-France-Hawaii telescope stripe 82 survey}",
      journal = {\jcap},
         year = 2016,
        month = aug,
       volume = {2016},
       number = {8},
          eid = {013},
        pages = {013},
          doi = {10.1088/1475-7516/2016/08/013},
archivePrefix = {arXiv},
       eprint = {1509.08930},
 primaryClass = {astro-ph.CO},
       adsurl = {https://ui.adsabs.harvard.edu/abs/2016JCAP...08..013B}
}

@ARTICLE{miyatake19,
       author = {{Miyatake}, Hironao and {Battaglia}, Nicholas and {Hilton}, Matt and {Medezinski}, Elinor and {Nishizawa}, Atsushi J. and {More}, Surhud and {Aiola}, Simone and {Bahcall}, Neta and {Bond}, J. Richard and {Calabrese}, Erminia and {Choi}, Steve K. and {Devlin}, Mark J. and {Dunkley}, Joanna and {Dunner}, Rolando and {Fuzia}, Brittany and {Gallardo}, Patricio and {Gralla}, Megan and {Hasselfield}, Matthew and {Halpern}, Mark and {Hikage}, Chiaki and {Hill}, J. Colin and {Hincks}, Adam D. and {Hlo{\v{z}}ek}, Ren{\'e}e and {Huffenberger}, Kevin and {Hughes}, John P. and {Koopman}, Brian and {Kosowsky}, Arthur and {Louis}, Thibaut and {Madhavacheril}, Mathew S. and {McMahon}, Jeff and {Mandelbaum}, Rachel and {Marriage}, Tobias A. and {Maurin}, Lo{\"\i}c and {Miyazaki}, Satoshi and {Moodley}, Kavilan and {Murata}, Ryoma and {Naess}, Sigurd and {Newburgh}, Laura and {Niemack}, Michael D. and {Nishimichi}, Takahiro and {Okabe}, Nobuhiro and {Oguri}, Masamune and {Osato}, Ken and {Page}, Lyman and {Partridge}, Bruce and {Robertson}, Naomi and {Sehgal}, Neelima and {Sherwin}, Blake and {Shirasaki}, Masato and {Sievers}, Jonathan and {Sif{\'o}n}, Crist{\'o}bal and {Simon}, Sara and {Spergel}, David N. and {Staggs}, Suzanne T. and {Stein}, George and {Takada}, Masahiro and {Trac}, Hy and {Umetsu}, Keiichi and {van Engelen}, Alex and {Wollack}, Edward J.},
        title = "{Weak-lensing Mass Calibration of ACTPol Sunyaev-Zel{\textquoteright}dovich Clusters with the Hyper Suprime-Cam Survey}",
      journal = {\apj},
         year = 2019,
        month = apr,
       volume = {875},
       number = {1},
          eid = {63},
        pages = {63},
          doi = {10.3847/1538-4357/ab0af0},
archivePrefix = {arXiv},
       eprint = {1804.05873},
 primaryClass = {astro-ph.CO},
       adsurl = {https://ui.adsabs.harvard.edu/abs/2019ApJ...875...63M}
}

@ARTICLE{nishizawa20,
       author = {{Nishizawa}, Atsushi J. and {Hsieh}, Bau-Ching and {Tanaka}, Masayuki and {Takata}, Tadafumi},
        title = "{Photometric Redshifts for the Hyper Suprime-Cam Subaru Strategic Program Data Release 2}",
      journal = {arXiv e-prints},
         year = 2020,
        month = feb,
          eid = {arXiv:2003.01511},
        pages = {arXiv:2003.01511},
          doi = {10.48550/arXiv.2003.01511},
archivePrefix = {arXiv},
       eprint = {2003.01511},
 primaryClass = {astro-ph.GA},
       adsurl = {https://ui.adsabs.harvard.edu/abs/2020arXiv200301511N}
}

@ARTICLE{mandelbaum18,
       author = {{Mandelbaum}, Rachel and {Miyatake}, Hironao and {Hamana}, Takashi and {Oguri}, Masamune and {Simet}, Melanie and {Armstrong}, Robert and {Bosch}, James and {Murata}, Ryoma and {Lanusse}, Fran{\c{c}}ois and {Leauthaud}, Alexie and {Coupon}, Jean and {More}, Surhud and {Takada}, Masahiro and {Miyazaki}, Satoshi and {Speagle}, Joshua S. and {Shirasaki}, Masato and {Sif{\'o}n}, Crist{\'o}bal and {Huang}, Song and {Nishizawa}, Atsushi J. and {Medezinski}, Elinor and {Okura}, Yuki and {Okabe}, Nobuhiro and {Czakon}, Nicole and {Takahashi}, Ryuichi and {Coulton}, William R. and {Hikage}, Chiaki and {Komiyama}, Yutaka and {Lupton}, Robert H. and {Strauss}, Michael A. and {Tanaka}, Masayuki and {Utsumi}, Yousuke},
        title = "{The first-year shear catalog of the Subaru Hyper Suprime-Cam Subaru Strategic Program Survey}",
      journal = {\pasj},
         year = 2018,
        month = jan,
       volume = {70},
          eid = {S25},
        pages = {S25},
          doi = {10.1093/pasj/psx130},
archivePrefix = {arXiv},
       eprint = {1705.06745},
 primaryClass = {astro-ph.CO},
       adsurl = {https://ui.adsabs.harvard.edu/abs/2018PASJ...70S..25M}
}

@ARTICLE{bosch2018,
       author = {{Bosch}, James and {Armstrong}, Robert and {Bickerton}, Steven and {Furusawa}, Hisanori and {Ikeda}, Hiroyuki and {Koike}, Michitaro and {Lupton}, Robert and {Mineo}, Sogo and {Price}, Paul and {Takata}, Tadafumi and {Tanaka}, Masayuki and {Yasuda}, Naoki and {AlSayyad}, Yusra and {Becker}, Andrew C. and {Coulton}, William and {Coupon}, Jean and {Garmilla}, Jose and {Huang}, Song and {Krughoff}, K. Simon and {Lang}, Dustin and {Leauthaud}, Alexie and {Lim}, Kian-Tat and {Lust}, Nate B. and {MacArthur}, Lauren A. and {Mandelbaum}, Rachel and {Miyatake}, Hironao and {Miyazaki}, Satoshi and {Murata}, Ryoma and {More}, Surhud and {Okura}, Yuki and {Owen}, Russell and {Swinbank}, John D. and {Strauss}, Michael A. and {Yamada}, Yoshihiko and {Yamanoi}, Hitomi},
        title = "{The Hyper Suprime-Cam software pipeline}",
      journal = {\pasj},
         year = 2018,
        month = jan,
       volume = {70},
          eid = {S5},
        pages = {S5},
          doi = {10.1093/pasj/psx080},
archivePrefix = {arXiv},
       eprint = {1705.06766},
 primaryClass = {astro-ph.IM},
       adsurl = {https://ui.adsabs.harvard.edu/abs/2018PASJ...70S...5B}
}

@INPROCEEDINGS{bosch2019,
       author = {{Bosch}, James and {AlSayyad}, Yusra and {Armstrong}, Robert and {Bellm}, Eric and {Chiang}, Hsin-Fang and {Eggl}, Siegfried and {Findeisen}, Krzysztof and {Fisher-Levine}, Merlin and {Guy}, Leanne P. and {Guyonnet}, Augustin and {Ivezi{\'c}}, {\v{Z}}eljko and {Jenness}, Tim and {Kov{\'a}cs}, G{\'a}bor and {Krughoff}, K. Simon and {Lupton}, Robert H. and {Lust}, Nate B. and {MacArthur}, Lauren A. and {Meyers}, Joshua and {Moolekamp}, Fred and {Morrison}, Christopher B. and {Morton}, Timothy D. and {O'Mullane}, William and {Parejko}, John K. and {Plazas}, Andr{\'e}s A. and {Price}, Paul A. and {Rawls}, Meredith L. and {Reed}, Sophie L. and {Schellart}, Pim and {Slater}, Colin T. and {Sullivan}, Ian and {Swinbank}, John D. and {Taranu}, Dan and {Waters}, Christopher Z. and {Wood-Vasey}, W.~M.},
        title = "{An Overview of the LSST Image Processing Pipelines}",
    booktitle = {Astronomical Data Analysis Software and Systems XXVII},
         year = 2019,
       editor = {{Teuben}, Peter J. and {Pound}, Marc W. and {Thomas}, Brian A. and {Warner}, Elizabeth M.},
       series = {Astronomical Society of the Pacific Conference Series},
       volume = {523},
        month = oct,
        pages = {521},
          doi = {10.48550/arXiv.1812.03248},
archivePrefix = {arXiv},
       eprint = {1812.03248},
 primaryClass = {astro-ph.IM},
       adsurl = {https://ui.adsabs.harvard.edu/abs/2019ASPC..523..521B}
}

@ARTICLE{li22,
       author = {{Li}, Xiangchong and {Miyatake}, Hironao and {Luo}, Wentao and {More}, Surhud and {Oguri}, Masamune and {Hamana}, Takashi and {Mandelbaum}, Rachel and {Shirasaki}, Masato and {Takada}, Masahiro and {Armstrong}, Robert and {Kannawadi}, Arun and {Takita}, Satoshi and {Miyazaki}, Satoshi and {Nishizawa}, Atsushi J. and {Plazas Malagon}, Andres A. and {Strauss}, Michael A. and {Tanaka}, Masayuki and {Yoshida}, Naoki},
        title = "{The three-year shear catalog of the Subaru Hyper Suprime-Cam SSP Survey}",
      journal = {\pasj},
         year = 2022,
        month = apr,
       volume = {74},
       number = {2},
        pages = {421-459},
          doi = {10.1093/pasj/psac006},
archivePrefix = {arXiv},
       eprint = {2107.00136},
 primaryClass = {astro-ph.CO},
       adsurl = {https://ui.adsabs.harvard.edu/abs/2022PASJ...74..421L}
}

@ARTICLE{oguri14,
       author = {{Oguri}, Masamune},
        title = "{A cluster finding algorithm based on the multiband identification of red sequence galaxies}",
      journal = {\mnras},
         year = 2014,
        month = oct,
       volume = {444},
       number = {1},
        pages = {147-161},
          doi = {10.1093/mnras/stu1446},
archivePrefix = {arXiv},
       eprint = {1407.4693},
 primaryClass = {astro-ph.CO},
       adsurl = {https://ui.adsabs.harvard.edu/abs/2014MNRAS.444..147O}
}

@ARTICLE{oguri18,
       author = {{Oguri}, Masamune and {Lin}, Yen-Ting and {Lin}, Sheng-Chieh and {Nishizawa}, Atsushi J. and {More}, Anupreeta and {More}, Surhud and {Hsieh}, Bau-Ching and {Medezinski}, Elinor and {Miyatake}, Hironao and {Jian}, Hung-Yu and {Lin}, Lihwai and {Takada}, Masahiro and {Okabe}, Nobuhiro and {Speagle}, Joshua S. and {Coupon}, Jean and {Leauthaud}, Alexie and {Lupton}, Robert H. and {Miyazaki}, Satoshi and {Price}, Paul A. and {Tanaka}, Masayuki and {Chiu}, I. -Non and {Komiyama}, Yutaka and {Okura}, Yuki and {Tanaka}, Manobu M. and {Usuda}, Tomonori},
        title = "{An optically-selected cluster catalog at redshift 0.1 < z < 1.1 from the Hyper Suprime-Cam Subaru Strategic Program S16A data}",
      journal = {\pasj},
         year = 2018,
        month = jan,
       volume = {70},
          eid = {S20},
        pages = {S20},
          doi = {10.1093/pasj/psx042},
archivePrefix = {arXiv},
       eprint = {1701.00818},
 primaryClass = {astro-ph.CO},
       adsurl = {https://ui.adsabs.harvard.edu/abs/2018PASJ...70S..20O}
}

@ARTICLE{hilton21,
       author = {{Hilton}, M. and {Sif{\'o}n}, C. and {Naess}, S. and {Madhavacheril}, M. and {Oguri}, M. and {Rozo}, E. and {Rykoff}, E. and {Abbott}, T.~M.~C. and {Adhikari}, S. and {Aguena}, M. and {Aiola}, S. and {Allam}, S. and {Amodeo}, S. and {Amon}, A. and {Annis}, J. and {Ansarinejad}, B. and {Aros-Bunster}, C. and {Austermann}, J.~E. and {Avila}, S. and {Bacon}, D. and {Battaglia}, N. and {Beall}, J.~A. and {Becker}, D.~T. and {Bernstein}, G.~M. and {Bertin}, E. and {Bhandarkar}, T. and {Bhargava}, S. and {Bond}, J.~R. and {Brooks}, D. and {Burke}, D.~L. and {Calabrese}, E. and {Carrasco Kind}, M. and {Carretero}, J. and {Choi}, S.~K. and {Choi}, A. and {Conselice}, C. and {da Costa}, L.~N. and {Costanzi}, M. and {Crichton}, D. and {Crowley}, K.~T. and {D{\"u}nner}, R. and {Denison}, E.~V. and {Devlin}, M.~J. and {Dicker}, S.~R. and {Diehl}, H.~T. and {Dietrich}, J.~P. and {Doel}, P. and {Duff}, S.~M. and {Duivenvoorden}, A.~J. and {Dunkley}, J. and {Everett}, S. and {Ferraro}, S. and {Ferrero}, I. and {Fert{\'e}}, A. and {Flaugher}, B. and {Frieman}, J. and {Gallardo}, P.~A. and {Garc{\'\i}a-Bellido}, J. and {Gaztanaga}, E. and {Gerdes}, D.~W. and {Giles}, P. and {Golec}, J.~E. and {Gralla}, M.~B. and {Grandis}, S. and {Gruen}, D. and {Gruendl}, R.~A. and {Gschwend}, J. and {Gutierrez}, G. and {Han}, D. and {Hartley}, W.~G. and {Hasselfield}, M. and {Hill}, J.~C. and {Hilton}, G.~C. and {Hincks}, A.~D. and {Hinton}, S.~R. and {Ho}, S. -P.~P. and {Honscheid}, K. and {Hoyle}, B. and {Hubmayr}, J. and {Huffenberger}, K.~M. and {Hughes}, J.~P. and {Jaelani}, A.~T. and {Jain}, B. and {James}, D.~J. and {Jeltema}, T. and {Kent}, S. and {Knowles}, K. and {Koopman}, B.~J. and {Kuehn}, K. and {Lahav}, O. and {Lima}, M. and {Lin}, Y. -T. and {Lokken}, M. and {Loubser}, S.~I. and {MacCrann}, N. and {Maia}, M.~A.~G. and {Marriage}, T.~A. and {Martin}, J. and {McMahon}, J. and {Melchior}, P. and {Menanteau}, F. and {Miquel}, R. and {Miyatake}, H. and {Moodley}, K. and {Morgan}, R. and {Mroczkowski}, T. and {Nati}, F. and {Newburgh}, L.~B. and {Niemack}, M.~D. and {Nishizawa}, A.~J. and {Ogando}, R.~L.~C. and {Orlowski-Scherer}, J. and {Page}, L.~A. and {Palmese}, A. and {Partridge}, B. and {Paz-Chinch{\'o}n}, F. and {Phakathi}, P. and {Plazas}, A.~A. and {Robertson}, N.~C. and {Romer}, A.~K. and {Carnero Rosell}, A. and {Salatino}, M. and {Sanchez}, E. and {Schaan}, E. and {Schillaci}, A. and {Sehgal}, N. and {Serrano}, S. and {Shin}, T. and {Simon}, S.~M. and {Smith}, M. and {Soares-Santos}, M. and {Spergel}, D.~N. and {Staggs}, S.~T. and {Storer}, E.~R. and {Suchyta}, E. and {Swanson}, M.~E.~C. and {Tarle}, G. and {Thomas}, D. and {To}, C. and {Trac}, H. and {Ullom}, J.~N. and {Vale}, L.~R. and {Van Lanen}, J. and {Vavagiakis}, E.~M. and {De Vicente}, J. and {Wilkinson}, R.~D. and {Wollack}, E.~J. and {Xu}, Z. and {Zhang}, Y.},
        title = "{The Atacama Cosmology Telescope: A Catalog of >4000 Sunyaev-Zel{\textquoteright}dovich Galaxy Clusters}",
      journal = {\apjs},
         year = 2021,
        month = mar,
       volume = {253},
       number = {1},
          eid = {3},
        pages = {3},
          doi = {10.3847/1538-4365/abd023},
archivePrefix = {arXiv},
       eprint = {2009.11043},
 primaryClass = {astro-ph.CO},
       adsurl = {https://ui.adsabs.harvard.edu/abs/2021ApJS..253....3H}
}

@ARTICLE{bocquet19,
       author = {{Bocquet}, S. and {Dietrich}, J.~P. and {Schrabback}, T. and {Bleem}, L.~E. and {Klein}, M. and {Allen}, S.~W. and {Applegate}, D.~E. and {Ashby}, M.~L.~N. and {Bautz}, M. and {Bayliss}, M. and {Benson}, B.~A. and {Brodwin}, M. and {Bulbul}, E. and {Canning}, R.~E.~A. and {Capasso}, R. and {Carlstrom}, J.~E. and {Chang}, C.~L. and {Chiu}, I. and {Cho}, H. -M. and {Clocchiatti}, A. and {Crawford}, T.~M. and {Crites}, A.~T. and {de Haan}, T. and {Desai}, S. and {Dobbs}, M.~A. and {Foley}, R.~J. and {Forman}, W.~R. and {Garmire}, G.~P. and {George}, E.~M. and {Gladders}, M.~D. and {Gonzalez}, A.~H. and {Grandis}, S. and {Gupta}, N. and {Halverson}, N.~W. and {Hlavacek-Larrondo}, J. and {Hoekstra}, H. and {Holder}, G.~P. and {Holzapfel}, W.~L. and {Hou}, Z. and {Hrubes}, J.~D. and {Huang}, N. and {Jones}, C. and {Khullar}, G. and {Knox}, L. and {Kraft}, R. and {Lee}, A.~T. and {von der Linden}, A. and {Luong-Van}, D. and {Mantz}, A. and {Marrone}, D.~P. and {McDonald}, M. and {McMahon}, J.~J. and {Meyer}, S.~S. and {Mocanu}, L.~M. and {Mohr}, J.~J. and {Morris}, R.~G. and {Padin}, S. and {Patil}, S. and {Pryke}, C. and {Rapetti}, D. and {Reichardt}, C.~L. and {Rest}, A. and {Ruhl}, J.~E. and {Saliwanchik}, B.~R. and {Saro}, A. and {Sayre}, J.~T. and {Schaffer}, K.~K. and {Shirokoff}, E. and {Stalder}, B. and {Stanford}, S.~A. and {Staniszewski}, Z. and {Stark}, A.~A. and {Story}, K.~T. and {Strazzullo}, V. and {Stubbs}, C.~W. and {Vanderlinde}, K. and {Vieira}, J.~D. and {Vikhlinin}, A. and {Williamson}, R. and {Zenteno}, A.},
        title = "{Cluster Cosmology Constraints from the 2500 deg$^{2}$ SPT-SZ Survey: Inclusion of Weak Gravitational Lensing Data from Magellan and the Hubble Space Telescope}",
      journal = {\apj},
         year = 2019,
        month = jun,
       volume = {878},
       number = {1},
          eid = {55},
        pages = {55},
          doi = {10.3847/1538-4357/ab1f10},
archivePrefix = {arXiv},
       eprint = {1812.01679},
 primaryClass = {astro-ph.CO},
       adsurl = {https://ui.adsabs.harvard.edu/abs/2019ApJ...878...55B}
}

@ARTICLE{planck16,
       author = {{Planck Collaboration} and {Ade}, P.~A.~R. and {Aghanim}, N. and {Arnaud}, M. and {Ashdown}, M. and {Aumont}, J. and {Baccigalupi}, C. and {Banday}, A.~J. and {Barreiro}, R.~B. and {Barrena}, R. and {Bartlett}, J.~G. and {Bartolo}, N. and {Battaner}, E. and {Battye}, R. and {Benabed}, K. and {Beno{\^\i}t}, A. and {Benoit-L{\'e}vy}, A. and {Bernard}, J. -P. and {Bersanelli}, M. and {Bielewicz}, P. and {Bikmaev}, I. and {B{\"o}hringer}, H. and {Bonaldi}, A. and {Bonavera}, L. and {Bond}, J.~R. and {Borrill}, J. and {Bouchet}, F.~R. and {Bucher}, M. and {Burenin}, R. and {Burigana}, C. and {Butler}, R.~C. and {Calabrese}, E. and {Cardoso}, J. -F. and {Carvalho}, P. and {Catalano}, A. and {Challinor}, A. and {Chamballu}, A. and {Chary}, R. -R. and {Chiang}, H.~C. and {Chon}, G. and {Christensen}, P.~R. and {Clements}, D.~L. and {Colombi}, S. and {Colombo}, L.~P.~L. and {Combet}, C. and {Comis}, B. and {Couchot}, F. and {Coulais}, A. and {Crill}, B.~P. and {Curto}, A. and {Cuttaia}, F. and {Dahle}, H. and {Danese}, L. and {Davies}, R.~D. and {Davis}, R.~J. and {de Bernardis}, P. and {de Rosa}, A. and {de Zotti}, G. and {Delabrouille}, J. and {D{\'e}sert}, F. -X. and {Dickinson}, C. and {Diego}, J.~M. and {Dolag}, K. and {Dole}, H. and {Donzelli}, S. and {Dor{\'e}}, O. and {Douspis}, M. and {Ducout}, A. and {Dupac}, X. and {Efstathiou}, G. and {Eisenhardt}, P.~R.~M. and {Elsner}, F. and {En{\ss}lin}, T.~A. and {Eriksen}, H.~K. and {Falgarone}, E. and {Fergusson}, J. and {Feroz}, F. and {Ferragamo}, A. and {Finelli}, F. and {Forni}, O. and {Frailis}, M. and {Fraisse}, A.~A. and {Franceschi}, E. and {Frejsel}, A. and {Galeotta}, S. and {Galli}, S. and {Ganga}, K. and {G{\'e}nova-Santos}, R.~T. and {Giard}, M. and {Giraud-H{\'e}raud}, Y. and {Gjerl{\o}w}, E. and {Gonz{\'a}lez-Nuevo}, J. and {G{\'o}rski}, K.~M. and {Grainge}, K.~J.~B. and {Gratton}, S. and {Gregorio}, A. and {Gruppuso}, A. and {Gudmundsson}, J.~E. and {Hansen}, F.~K. and {Hanson}, D. and {Harrison}, D.~L. and {Hempel}, A. and {Henrot-Versill{\'e}}, S. and {Hern{\'a}ndez-Monteagudo}, C. and {Herranz}, D. and {Hildebrandt}, S.~R. and {Hivon}, E. and {Hobson}, M. and {Holmes}, W.~A. and {Hornstrup}, A. and {Hovest}, W. and {Huffenberger}, K.~M. and {Hurier}, G. and {Jaffe}, A.~H. and {Jaffe}, T.~R. and {Jin}, T. and {Jones}, W.~C. and {Juvela}, M. and {Keih{\"a}nen}, E. and {Keskitalo}, R. and {Khamitov}, I. and {Kisner}, T.~S. and {Kneissl}, R. and {Knoche}, J. and {Kunz}, M. and {Kurki-Suonio}, H. and {Lagache}, G. and {Lamarre}, J. -M. and {Lasenby}, A. and {Lattanzi}, M. and {Lawrence}, C.~R. and {Leonardi}, R. and {Lesgourgues}, J. and {Levrier}, F. and {Liguori}, M. and {Lilje}, P.~B. and {Linden-V{\o}rnle}, M. and {L{\'o}pez-Caniego}, M. and {Lubin}, P.~M. and {Mac{\'\i}as-P{\'e}rez}, J.~F. and {Maggio}, G. and {Maino}, D. and {Mak}, D.~S.~Y. and {Mandolesi}, N. and {Mangilli}, A. and {Martin}, P.~G. and {Mart{\'\i}nez-Gonz{\'a}lez}, E. and {Masi}, S. and {Matarrese}, S. and {Mazzotta}, P. and {McGehee}, P. and {Mei}, S. and {Melchiorri}, A. and {Melin}, J. -B. and {Mendes}, L. and {Mennella}, A. and {Migliaccio}, M. and {Mitra}, S. and {Miville-Desch{\^e}nes}, M. -A. and {Moneti}, A. and {Montier}, L. and {Morgante}, G. and {Mortlock}, D. and {Moss}, A. and {Munshi}, D. and {Murphy}, J.~A. and {Naselsky}, P. and {Nastasi}, A. and {Nati}, F. and {Natoli}, P. and {Netterfield}, C.~B. and {N{\o}rgaard-Nielsen}, H.~U. and {Noviello}, F. and {Novikov}, D. and {Novikov}, I. and {Olamaie}, M. and {Oxborrow}, C.~A. and {Paci}, F. and {Pagano}, L. and {Pajot}, F. and {Paoletti}, D. and {Pasian}, F. and {Patanchon}, G. and {Pearson}, T.~J. and {Perdereau}, O. and {Perotto}, L. and {Perrott}, Y.~C. and {Perrotta}, F. and {Pettorino}, V. and {Piacentini}, F. and {Piat}, M. and {Pierpaoli}, E. and {Pietrobon}, D. and {Plaszczynski}, S. and {Pointecouteau}, E. and {Polenta}, G. and {Pratt}, G.~W. and {Pr{\'e}zeau}, G. and {Prunet}, S. and {Puget}, J. -L. and {Rachen}, J.~P. and {Reach}, W.~T. and {Rebolo}, R. and {Reinecke}, M. and {Remazeilles}, M. and {Renault}, C. and {Renzi}, A. and {Ristorcelli}, I. and {Rocha}, G. and {Rosset}, C. and {Rossetti}, M. and {Roudier}, G. and {Rozo}, E. and {Rubi{\~n}o-Mart{\'\i}n}, J.~A. and {Rumsey}, C. and {Rusholme}, B. and {Rykoff}, E.~S. and {Sandri}, M. and {Santos}, D. and {Saunders}, R.~D.~E. and {Savelainen}, M. and {Savini}, G. and {Schammel}, M.~P. and {Scott}, D. and {Seiffert}, M.~D. and {Shellard}, E.~P.~S. and {Shimwell}, T.~W. and {Spencer}, L.~D. and {Stanford}, S.~A. and {Stern}, D. and {Stolyarov}, V. and {Stompor}, R. and {Streblyanska}, A. and {Sudiwala}, R. and {Sunyaev}, R. and {Sutton}, D. and {Suur-Uski}, A. -S. and {Sygnet}, J. -F. and {Tauber}, J.~A. and {Terenzi}, L. and {Toffolatti}, L. and {Tomasi}, M. and {Tramonte}, D. and {Tristram}, M. and {Tucci}, M. and {Tuovinen}, J. and {Umana}, G. and {Valenziano}, L. and {Valiviita}, J. and {Van Tent}, B. and {Vielva}, P. and {Villa}, F. and {Wade}, L.~A. and {Wandelt}, B.~D. and {Wehus}, I.~K. and {White}, S.~D.~M. and {Wright}, E.~L. and {Yvon}, D. and {Zacchei}, A. and {Zonca}, A.},
        title = "{Planck 2015 results. XXVII. The second Planck catalogue of Sunyaev-Zeldovich sources}",
      journal = {\aap},
         year = 2016,
        month = sep,
       volume = {594},
          eid = {A27},
        pages = {A27},
          doi = {10.1051/0004-6361/201525823},
archivePrefix = {arXiv},
       eprint = {1502.01598},
 primaryClass = {astro-ph.CO},
       adsurl = {https://ui.adsabs.harvard.edu/abs/2016A&A...594A..27P}
}

@ARTICLE{sunyaev72,
       author = {{Sunyaev}, R.~A. and {Zeldovich}, Ya. B.},
        title = "{The Observations of Relic Radiation as a Test of the Nature of X-Ray Radiation from the Clusters of Galaxies}",
      journal = {Comments on Astrophysics and Space Physics},
         year = 1972,
        month = nov,
       volume = {4},
        pages = {173},
       adsurl = {https://ui.adsabs.harvard.edu/abs/1972CoASP...4..173S}
}

@ARTICLE{rykoff14,
       author = {{Rykoff}, E.~S. and {Rozo}, E. and {Busha}, M.~T. and {Cunha}, C.~E. and {Finoguenov}, A. and {Evrard}, A. and {Hao}, J. and {Koester}, B.~P. and {Leauthaud}, A. and {Nord}, B. and {Pierre}, M. and {Reddick}, R. and {Sadibekova}, T. and {Sheldon}, E.~S. and {Wechsler}, R.~H.},
        title = "{redMaPPer. I. Algorithm and SDSS DR8 Catalog}",
      journal = {\apj},
         year = 2014,
        month = apr,
       volume = {785},
       number = {2},
          eid = {104},
        pages = {104},
          doi = {10.1088/0004-637X/785/2/104},
archivePrefix = {arXiv},
       eprint = {1303.3562},
 primaryClass = {astro-ph.CO},
       adsurl = {https://ui.adsabs.harvard.edu/abs/2014ApJ...785..104R}
}

@ARTICLE{miyazaki18,
       author = {{Miyazaki}, Satoshi and {Oguri}, Masamune and {Hamana}, Takashi and {Shirasaki}, Masato and {Koike}, Michitaro and {Komiyama}, Yutaka and {Umetsu}, Keiichi and {Utsumi}, Yousuke and {Okabe}, Nobuhiro and {More}, Surhud and {Medezinski}, Elinor and {Lin}, Yen-Ting and {Miyatake}, Hironao and {Murayama}, Hitoshi and {Ota}, Naomi and {Mitsuishi}, Ikuyuki},
        title = "{A large sample of shear-selected clusters from the Hyper Suprime-Cam Subaru Strategic Program S16A Wide field mass maps}",
      journal = {\pasj},
         year = 2018,
        month = jan,
       volume = {70},
          eid = {S27},
        pages = {S27},
          doi = {10.1093/pasj/psx120},
archivePrefix = {arXiv},
       eprint = {1802.10290},
 primaryClass = {astro-ph.CO},
       adsurl = {https://ui.adsabs.harvard.edu/abs/2018PASJ...70S..27M}
}

@ARTICLE{oguri21,
       author = {{Oguri}, Masamune and {Miyazaki}, Satoshi and {Li}, Xiangchong and {Luo}, Wentao and {Mitsuishi}, Ikuyuki and {Miyatake}, Hironao and {More}, Surhud and {Nishizawa}, Atsushi J. and {Okabe}, Nobuhiro and {Ota}, Naomi and {Plazas Malag{\'o}n}, Andr{\'e}s A. and {Utsumi}, Yousuke},
        title = "{Hundreds of weak lensing shear-selected clusters from the Hyper Suprime-Cam Subaru Strategic Program S19A data}",
      journal = {\pasj},
         year = 2021,
        month = aug,
       volume = {73},
       number = {4},
        pages = {817-829},
          doi = {10.1093/pasj/psab047},
archivePrefix = {arXiv},
       eprint = {2103.15016},
 primaryClass = {astro-ph.CO},
       adsurl = {https://ui.adsabs.harvard.edu/abs/2021PASJ...73..817O}
}

@ARTICLE{pacaud16,
       author = {{Pacaud}, F. and {Clerc}, N. and {Giles}, P.~A. and {Adami}, C. and {Sadibekova}, T. and {Pierre}, M. and {Maughan}, B.~J. and {Lieu}, M. and {Le F{\`e}vre}, J.~P. and {Alis}, S. and {Altieri}, B. and {Ardila}, F. and {Baldry}, I. and {Benoist}, C. and {Birkinshaw}, M. and {Chiappetti}, L. and {D{\'e}mocl{\`e}s}, J. and {Eckert}, D. and {Evrard}, A.~E. and {Faccioli}, L. and {Gastaldello}, F. and {Guennou}, L. and {Horellou}, C. and {Iovino}, A. and {Koulouridis}, E. and {Le Brun}, V. and {Lidman}, C. and {Liske}, J. and {Maurogordato}, S. and {Menanteau}, F. and {Owers}, M. and {Poggianti}, B. and {Pomar{\`e}de}, D. and {Pompei}, E. and {Ponman}, T.~J. and {Rapetti}, D. and {Reiprich}, T.~H. and {Smith}, G.~P. and {Tuffs}, R. and {Valageas}, P. and {Valtchanov}, I. and {Willis}, J.~P. and {Ziparo}, F.},
        title = "{The XXL Survey. II. The bright cluster sample: catalogue and luminosity function}",
      journal = {\aap},
         year = 2016,
        month = jun,
       volume = {592},
          eid = {A2},
        pages = {A2},
          doi = {10.1051/0004-6361/201526891},
archivePrefix = {arXiv},
       eprint = {1512.04264},
 primaryClass = {astro-ph.CO},
       adsurl = {https://ui.adsabs.harvard.edu/abs/2016A&A...592A...2P}
}

@ARTICLE{allen11,
       author = {{Allen}, Steven W. and {Evrard}, August E. and {Mantz}, Adam B.},
        title = "{Cosmological Parameters from Observations of Galaxy Clusters}",
      journal = {\araa},
         year = 2011,
        month = sep,
       volume = {49},
       number = {1},
        pages = {409-470},
          doi = {10.1146/annurev-astro-081710-102514},
archivePrefix = {arXiv},
       eprint = {1103.4829},
 primaryClass = {astro-ph.CO},
       adsurl = {https://ui.adsabs.harvard.edu/abs/2011ARA&A..49..409A}
}

@ARTICLE{voit05,
       author = {{Voit}, G. Mark},
        title = "{Tracing cosmic evolution with clusters of galaxies}",
      journal = {Reviews of Modern Physics},
         year = 2005,
        month = apr,
       volume = {77},
       number = {1},
        pages = {207-258},
          doi = {10.1103/RevModPhys.77.207},
archivePrefix = {arXiv},
       eprint = {astro-ph/0410173},
 primaryClass = {astro-ph},
       adsurl = {https://ui.adsabs.harvard.edu/abs/2005RvMP...77..207V},
}

@ARTICLE{bahcall98,
       author = {{Bahcall}, Neta A. and {Fan}, Xiaohui},
        title = "{The Most Massive Distant Clusters: Determining {\ensuremath{\Omega}} and {\ensuremath{\sigma}}$_{8}$}",
      journal = {\apj},
         year = 1998,
        month = sep,
       volume = {504},
       number = {1},
        pages = {1-6},
          doi = {10.1086/306088},
archivePrefix = {arXiv},
       eprint = {astro-ph/9803277},
 primaryClass = {astro-ph},
       adsurl = {https://ui.adsabs.harvard.edu/abs/1998ApJ...504....1B}
}

@ARTICLE{reiprich02,
       author = {{Reiprich}, Thomas H. and {B{\"o}hringer}, Hans},
        title = "{The Mass Function of an X-Ray Flux-limited Sample of Galaxy Clusters}",
      journal = {\apj},
         year = 2002,
        month = mar,
       volume = {567},
       number = {2},
        pages = {716-740},
          doi = {10.1086/338753},
archivePrefix = {arXiv},
       eprint = {astro-ph/0111285},
 primaryClass = {astro-ph},
       adsurl = {https://ui.adsabs.harvard.edu/abs/2002ApJ...567..716R}
}

@ARTICLE{medezinski18,
       author = {{Medezinski}, Elinor and {Battaglia}, Nicholas and {Umetsu}, Keiichi and {Oguri}, Masamune and {Miyatake}, Hironao and {Nishizawa}, Atsushi J. and {Sif{\'o}n}, Crist{\'o}bal and {Spergel}, David N. and {Chiu}, I. -Non and {Lin}, Yen-Ting and {Bahcall}, Neta and {Komiyama}, Yutaka},
        title = "{Planck Sunyaev-Zel'dovich cluster mass calibration using Hyper Suprime-Cam weak lensing}",
      journal = {\pasj},
         year = 2018,
        month = jan,
       volume = {70},
          eid = {S28},
        pages = {S28},
          doi = {10.1093/pasj/psx128},
archivePrefix = {arXiv},
       eprint = {1706.00434},
 primaryClass = {astro-ph.CO},
       adsurl = {https://ui.adsabs.harvard.edu/abs/2018PASJ...70S..28M}
}

@ARTICLE{arnaud10,
       author = {{Arnaud}, M. and {Pratt}, G.~W. and {Piffaretti}, R. and {B{\"o}hringer}, H. and {Croston}, J.~H. and {Pointecouteau}, E.},
        title = "{The universal galaxy cluster pressure profile from a representative sample of nearby systems (REXCESS) and the Y$_{SZ}$ - M$_{500}$ relation}",
      journal = {\aap},
         year = 2010,
        month = jul,
       volume = {517},
          eid = {A92},
        pages = {A92},
          doi = {10.1051/0004-6361/200913416},
archivePrefix = {arXiv},
       eprint = {0910.1234},
 primaryClass = {astro-ph.CO},
       adsurl = {https://ui.adsabs.harvard.edu/abs/2010A&A...517A..92A}
}

@ARTICLE{diemer18,
       author = {{Diemer}, Benedikt},
        title = "{COLOSSUS: A Python Toolkit for Cosmology, Large-scale Structure, and Dark Matter Halos}",
      journal = {\apjs},
         year = 2018,
        month = dec,
       volume = {239},
       number = {2},
          eid = {35},
        pages = {35},
          doi = {10.3847/1538-4365/aaee8c},
archivePrefix = {arXiv},
       eprint = {1712.04512},
 primaryClass = {astro-ph.CO},
       adsurl = {https://ui.adsabs.harvard.edu/abs/2018ApJS..239...35D}
}

@article{Aihara2017b,
  author = {Aihara, H. and Arimoto, N. and Armstrong, R. and Arnouts, S. and Bahcall, N.~A. and Bickerton, S. and Bosch, J. and Bundy, K. and Capak, P.~L. and Chan, J.~H.~H. and Chiba, M. and Coupon, J. and Egami, E. and Enoki, M. and Finet, F. and Fujimori, H. and Fujimoto, S. and Furusawa, H. and Furusawa, J. and Goto, T. and Greco, J. and Greene, J.~E. and Gunn, J.~E. and Hamana, T. and Harikane, Y. and Hashimoto, Y. and Hattori, T. and Hayashi, M. and Hayashi, Y. and He{\l}miniak, K.~G. and Higuchi, R. and Hikage, C. and Ho, P.~T.~P. and Hsieh, B.-C. and Huang, K. and Huang, S. and Ikeda, H. and Imanishi, M. and Inoue, A.~K. and Iwasawa, K. and Iwata, I. and Jaelani, A.~T. and Jian, H.-Y. and Kamata, Y. and Karoji, H. and Kashikawa, N. and Katayama, N. and Kawanomoto, S. and Kayo, I. and Koda, J. and Koike, M. and Kojima, T. and Komiyama, Y. and Konno, A. and Koshida, S. and Koyama, Y. and Kusakabe, H. and Leauthaud, A. and Lee, C.-H. and Lin, L. and Lin, Y.-T. and Lupton, R.~H. and Mandelbaum, R. and Matsuoka, Y. and Medezinski, E. and Mineo, S. and Miyama, S. and Miyatake, H. and Miyazaki, S. and Momose, R. and More, A. and More, S. and Moritani, Y. and Moriya, T.~J. and Morokuma, T. and Mukae, S. and Murayama, H. and Nagao, T. and Nakata, F. and Niikura, H. and Nishizawa, A.~J. and Obuchi, Y. and Oguri, M. and Oishi, Y. and Okabe, N. and Okamoto, S. and Okura, Y. and Ono, Y. and Onodera, M. and Onoue, M. and Ouchi, M. and Pan, T.-S. and Pyo, T.-S. and Shibuya, T. and Shimasaku, K. and Shimono, A. and Shirasaki, M. and Silverman, J.~D. and Simet, M. and Speagle, J. and Spergel, D.~N. and Strauss, M.~A. and Sugahara, Y. and Sugiyama, N. and Suto, Y. and Suyu, S.~H. and Suzuki, N. and Tait, P.~J. and Takada, M. and Takata, T. and Tamura, N. and Tanaka, M.~M. and Tanaka, M. and Tanaka, M. and Tanaka, Y. and Terai, T. and Terashima, Y. and Toba, Y. and Tominaga, N. and Toshikawa, J. and Turner, E.~L. and Uchida, T. and Uchiyama, H. and Umetsu, K. and Uraguchi, F. and Urata, Y. and Usuda, T. and Utsumi, Y. and Wang, S.-Y. and Watanabe, M. and Wong, K.~C. and Yabe, K. and Yamada, Y. and Yamanoi, H. and Yasuda, N. and Yeh, S. and Yonehara, A. and Yuma, S.},
  title = "{The Hyper Suprime-Cam SSP Survey: Overview and survey design}",
  journal = {Publications of the Astronomical Society of Japan},
  year = {2018},
  volume = {70},
  eid = {S4},
  pages = {S4},
  doi = {10.1093/pasj/psx066},
  archivePrefix = {arXiv},
  eprint = {1704.05858},
  primaryClass = {astro-ph.IM},
  adsurl = {https://ui.adsabs.harvard.edu/abs/2018PASJ...70S...4A}
}

@article{Carlstrom2011,
  author = {Carlstrom, J.E. and Ade, P.A.R. and Aird, K.A. et al.},
  title = "{The 10 Meter South Pole Telescope}",
  journal = {PASP},
  volume = {123},
  pages = {568},
  year = {2011}
}

@article{Corless2007,
  author = {Corless, V.~L. and King, L.~J.},
  title = "{A statistical study of weak lensing by triaxial dark matter haloes: consequences for parameter estimation}",
  journal = {MNRAS},
  volume = {380},
  number = {1},
  pages = {149--161},
  year = {2007},
  doi = {10.1111/j.1365-2966.2007.12018.x},
  archivePrefix = {arXiv},
  eprint = {astro-ph/0611913},
  primaryClass = {astro-ph}
}

@article{Gruen2015,
  author = {Gruen, D. and Seitz, S. and Becker, M.~R. and Friedrich, O. and Mana, A.},
  title = "{Cosmic variance of the galaxy cluster weak lensing signal}",
  journal = {MNRAS},
  volume = {449},
  number = {4},
  pages = {4264--4276},
  year = {2015},
  doi = {10.1093/mnras/stv532},
  archivePrefix = {arXiv},
  eprint = {1501.01632},
  primaryClass = {astro-ph.CO}
}

@article{Smith2016,
  author = {Smith, G.P. and Mazzotta, P. and Okabe, N. et al.},
  title = "{LoCuSS: Testing hydrostatic equilibrium in galaxy clusters}",
  journal = {MNRAS},
  volume = {456},
  pages = {L74},
  year = {2016}
}

@ARTICLE{Thornton2016_ACT,
       author = {{Thornton}, R.~J. and {Ade}, P.~A.~R. and {Aiola}, S. and {Angil{\`e}}, F.~E. and {Amiri}, M. and {Beall}, J.~A. and {Becker}, D.~T. and {Cho}, H.-M. and {Choi}, S.~K. and {Corlies}, P. and {Coughlin}, K.~P. and {Datta}, R. and {Devlin}, M.~J. and {Dicker}, S.~R. and {D{\"u}nner}, R. and {Fowler}, J.~W. and {Fox}, A.~E. and {Gallardo}, P.~A. and {Gao}, J. and {Grace}, E. and {Halpern}, M. and {Hasselfield}, M. and {Henderson}, S.~W. and {Hilton}, G.~C. and {Hincks}, A.~D. and {Ho}, S.~P. and {Hubmayr}, J. and {Irwin}, K.~D. and {Klein}, J. and {Koopman}, B. and {Li}, Dale and {Louis}, T. and {Lungu}, M. and {Maurin}, L. and {McMahon}, J. and {Munson}, C.~D. and {Naess}, S. and {Nati}, F. and {Newburgh}, L. and {Nibarger}, J. and {Niemack}, M.~D. and {Niraula}, P. and {Nolta}, M.~R. and {Page}, L.~A. and {Pappas}, C.~G. and {Schillaci}, A. and {Schmitt}, B.~L. and {Sehgal}, N. and {Sievers}, J.~L. and {Simon}, S.~M. and {Staggs}, S.~T. and {Tucker}, C. and {Uehara}, M. and {van Lanen}, J. and {Ward}, J.~T. and {Wollack}, E.~J.},
        title = "{The Atacama Cosmology Telescope: The Polarization-sensitive ACTPol Instrument}",
      journal = {\apjs},
         year = 2016,
        month = dec,
       volume = {227},
       number = {2},
          eid = {21},
        pages = {21},
          doi = {10.3847/1538-4365/227/2/21},
archivePrefix = {arXiv},
       eprint = {1605.06569},
 primaryClass = {astro-ph.IM},
       adsurl = {https://ui.adsabs.harvard.edu/abs/2016ApJS..227...21T}
}

@ARTICLE{Tanaka2015,
  author = {{Tanaka}, Masayuki},
  title = "{Photometric Redshift with Bayesian Priors on Physical Properties of Galaxies}",
  journal = {\apj},
  year = {2015},
  volume = {801},
  number = {1},
  eid = {20},
  pages = {20},
  doi = {10.1088/0004-637X/801/1/20},
  archivePrefix = {arXiv},
  eprint = {1501.02047},
  primaryClass = {astro-ph.GA},
  adsurl = {https://ui.adsabs.harvard.edu/abs/2015ApJ...801...20T}
}

@ARTICLE{HSC_PZ,
  author = {Tanaka, M. and Coupon, J. and Hsieh, B.-C. and others},
  title = {Photometric Redshifts for Hyper Suprime-Cam Subaru Strategic Program Data Release 1},
  journal = {Publications of the Astronomical Society of Japan},
  year = {2018},
  volume = {70},
  eid = {S9},
  pages = {S9},
  doi = {10.1093/pasj/psx077},
  archivePrefix = {arXiv},
  eprint = {1704.05988},
  primaryClass = {astro-ph.GA},
  adsurl = {https://ui.adsabs.harvard.edu/abs/2018PASJ...70S...9T}
}

@ARTICLE{mandelbaum2006,
  author = {Mandelbaum, R. and Seljak, U. and Kauffmann, G. and Hirata, C. M. and Brinkmann, J.},
  title = {Galaxy halo masses and satellite fractions from galaxy-galaxy lensing in the Sloan Digital Sky Survey},
  journal = {MNRAS},
  year = {2006},
  volume = {368},
  pages = {715--731},
  doi = {10.1111/j.1365-2966.2006.10156.x}
}

@ARTICLE{hirata2004,
  author = {Hirata, C. and Seljak, U.},
  title = {Intrinsic alignment-lensing interference as a contaminant of cosmic shear},
  journal = {Phys. Rev. D},
  year = {2004},
  volume = {70},
  pages = {063526},
  doi = {10.1103/PhysRevD.70.063526}
}

@ARTICLE{sheldon2004,
  author = {Sheldon, E.~S. and Johnston, D.~E. and Frieman, J.~A. and Scranton, R. and McKay, T.~A. and Connolly, A.~J. and Budav{\'a}ri, T. and Zehavi, I. and Bahcall, N.~A. and Brinkmann, J. and Fukugita, M.},
  title = {The Galaxy-Mass Correlation Function Measured from Weak Lensing in the Sloan Digital Sky Survey},
  journal = {AJ},
  year = {2004},
  volume = {127},
  number = {5},
  pages = {2544--2564},
  doi = {10.1086/383293},
  archivePrefix = {arXiv},
  eprint = {astro-ph/0312036},
  primaryClass = {astro-ph}
}

@ARTICLE{2016AA...594A..13P,
       author = {{Planck Collaboration} and {Ade}, P.~A.~R. and {Aghanim}, N. and {Arnaud}, M. and {Ashdown}, M. and {Aumont}, J. and {Baccigalupi}, C. and {Banday}, A.~J. and {Barreiro}, R.~B. and {Bartlett}, J.~G. and {Bartolo}, N. and {Battaner}, E. and {Battye}, R. and {Benabed}, K. and {Beno{\^\i}t}, A. and {Benoit-L{\'e}vy}, A. and {Bernard}, J.-P. and {Bersanelli}, M. and {Bielewicz}, P. and {Bock}, J.~J. and {Bonaldi}, A. and {Bonavera}, L. and {Bond}, J.~R. and {Borrill}, J. and {Bouchet}, F.~R. and {Boulanger}, F. and {Bucher}, M. and {Burigana}, C. and {Butler}, R.~C. and {Calabrese}, E. and {Cardoso}, J.-F. and {Catalano}, A. and {Challinor}, A. and {Chamballu}, A. and {Chary}, R.-R. and {Chiang}, H.~C. and {Chluba}, J. and {Christensen}, P.~R. and {Church}, S. and {Clements}, D.~L. and {Colombi}, S. and {Colombo}, L.~P.~L. and {Combet}, C. and {Coulais}, A. and {Crill}, B.~P. and {Curto}, A. and {Cuttaia}, F. and {Danese}, L. and {Davies}, R.~D. and {Davis}, R.~J. and {de Bernardis}, P. and {de Rosa}, A. and {de Zotti}, G. and {Delabrouille}, J. and {D{\'e}sert}, F.-X. and {Di Valentino}, E. and {Dickinson}, C. and {Diego}, J.~M. and {Dolag}, K. and {Dole}, H. and {Donzelli}, S. and {Dor{\'e}}, O. and {Douspis}, M. and {Ducout}, A. and {Dunkley}, J. and {Dupac}, X. and {Efstathiou}, G. and {Elsner}, F. and {En{\ss}lin}, T.~A. and {Eriksen}, H.~K. and {Farhang}, M. and {Fergusson}, J. and {Finelli}, F. and {Forni}, O. and {Frailis}, M. and {Fraisse}, A.~A. and {Franceschi}, E. and {Frejsel}, A. and {Galeotta}, S. and {Galli}, S. and {Ganga}, K. and {Gauthier}, C. and {Gerbino}, M. and {Ghosh}, T. and {Giard}, M. and {Giraud-H{\'e}raud}, Y. and {Giusarma}, E. and {Gjerl{\o}w}, E. and {Gonz{\'a}lez-Nuevo}, J. and {G{\'o}rski}, K.~M. and {Gratton}, S. and {Gregorio}, A. and {Gruppuso}, A. and {Gudmundsson}, J.~E. and {Hamann}, J. and {Hansen}, F.~K. and {Hanson}, D. and {Harrison}, D.~L. and {Helou}, G. and {Henrot-Versill{\'e}}, S. and {Hern{\'a}ndez-Monteagudo}, C. and {Herranz}, D. and {Hildebrandt}, S.~R. and {Hivon}, E. and {Hobson}, M. and {Holmes}, W.~A. and {Hornstrup}, A. and {Hovest}, W. and {Huang}, Z. and {Huffenberger}, K.~M. and {Hurier}, G. and {Jaffe}, A.~H. and {Jaffe}, T.~R. and {Jones}, W.~C. and {Juvela}, M. and {Keih{\"a}nen}, E. and {Keskitalo}, R. and {Kisner}, T.~S. and {Kneissl}, R. and {Knoche}, J. and {Knox}, L. and {Kunz}, M. and {Kurki-Suonio}, H. and {Lagache}, G. and {L{\"a}hteenm{\"a}ki}, A. and {Lamarre}, J.-M. and {Lasenby}, A. and {Lattanzi}, M. and {Lawrence}, C.~R. and {Leahy}, J.~P. and {Leonardi}, R. and {Lesgourgues}, J. and {Levrier}, F. and {Lewis}, A. and {Liguori}, M. and {Lilje}, P.~B. and {Linden-V{\o}rnle}, M. and {L{\'o}pez-Caniego}, M. and {Lubin}, P.~M. and {Mac{\'\i}as-P{\'e}rez}, J.~F. and {Maggio}, G. and {Maino}, D. and {Mandolesi}, N. and {Mangilli}, A. and {Marchini}, A. and {Maris}, M. and {Martin}, P.~G. and {Martinelli}, M. and {Mart{\'\i}nez-Gonz{\'a}lez}, E. and {Masi}, S. and {Matarrese}, S. and {McGehee}, P. and {Meinhold}, P.~R. and {Melchiorri}, A. and {Melin}, J.-B. and {Mendes}, L. and {Mennella}, A. and {Migliaccio}, M. and {Millea}, M. and {Mitra}, S. and {Miville-Desch{\^e}nes}, M.-A. and {Moneti}, A. and {Montier}, L. and {Morgante}, G. and {Mortlock}, D. and {Moss}, A. and {Munshi}, D. and {Murphy}, J.~A. and {Naselsky}, P. and {Nati}, F. and {Natoli}, P. and {Netterfield}, C.~B. and {N{\o}rgaard-Nielsen}, H.~U. and {Noviello}, F. and {Novikov}, D. and {Novikov}, I. and {Oxborrow}, C.~A. and {Paci}, F. and {Pagano}, L. and {Pajot}, F. and {Paladini}, R. and {Paoletti}, D. and {Partridge}, B. and {Pasian}, F. and {Patanchon}, G. and {Pearson}, T.~J. and {Perdereau}, O. and {Perotto}, L. and {Perrotta}, F. and {Pettorino}, V. and {Piacentini}, F. and {Piat}, M. and {Pierpaoli}, E. and {Pietrobon}, D. and {Plaszczynski}, S. and {Pointecouteau}, E. and {Polenta}, G. and {Popa}, L. and {Pratt}, G.~W. and {Pr{\'e}zeau}, G.},
        title = "{Planck 2015 results. XIII. Cosmological parameters}",
      journal = {\aap},
         year = 2016,
        month = sep,
       volume = {594},
          eid = {A13},
        pages = {A13},
          doi = {10.1051/0004-6361/201525830},
archivePrefix = {arXiv},
       eprint = {1502.01589},
 primaryClass = {astro-ph.CO},
       adsurl = {https://ui.adsabs.harvard.edu/abs/2016A&A...594A..13P}
}

@ARTICLE{2016AA...594A..24P,
       author = {{Planck Collaboration} and {Ade}, P.~A.~R. and {Aghanim}, N. and {Arnaud}, M. and {Ashdown}, M. and {Aumont}, J. and {Baccigalupi}, C. and {Banday}, A.~J. and {Barreiro}, R.~B. and {Bartlett}, J.~G. and {Bartolo}, N. and {Battaner}, E. and {Battye}, R. and {Benabed}, K. and {Beno{\^\i}t}, A. and {Benoit-L{\'e}vy}, A. and {Bernard}, J.-P. and {Bersanelli}, M. and {Bielewicz}, P. and {Bock}, J.~J. and {Bonaldi}, A. and {Bonavera}, L. and {Bond}, J.~R. and {Borrill}, J. and {Bouchet}, F.~R. and {Bucher}, M. and {Burigana}, C. and {Butler}, R.~C. and {Calabrese}, E. and {Cardoso}, J.-F. and {Catalano}, A. and {Challinor}, A. and {Chamballu}, A. and {Chary}, R.-R. and {Chiang}, H.~C. and {Christensen}, P.~R. and {Church}, S. and {Clements}, D.~L. and {Colombi}, S. and {Colombo}, L.~P.~L. and {Combet}, C. and {Comis}, B. and {Couchot}, F. and {Coulais}, A. and {Crill}, B.~P. and {Curto}, A. and {Cuttaia}, F. and {Danese}, L. and {Davies}, R.~D. and {Davis}, R.~J. and {de Bernardis}, P. and {de Rosa}, A. and {de Zotti}, G. and {Delabrouille}, J. and {D{\'e}sert}, F.-X. and {Diego}, J.~M. and {Dolag}, K. and {Dole}, H. and {Donzelli}, S. and {Dor{\'e}}, O. and {Douspis}, M. and {Ducout}, A. and {Dupac}, X. and {Efstathiou}, G. and {Elsner}, F. and {En{\ss}lin}, T.~A. and {Eriksen}, H.~K. and {Falgarone}, E. and {Fergusson}, J. and {Finelli}, F. and {Forni}, O. and {Frailis}, M. and {Fraisse}, A.~A. and {Franceschi}, E. and {Frejsel}, A. and {Galeotta}, S. and {Galli}, S. and {Ganga}, K. and {Giard}, M. and {Giraud-H{\'e}raud}, Y. and {Gjerl{\o}w}, E. and {Gonz{\'a}lez-Nuevo}, J. and {G{\'o}rski}, K.~M. and {Gratton}, S. and {Gregorio}, A. and {Gruppuso}, A. and {Gudmundsson}, J.~E. and {Hansen}, F.~K. and {Hanson}, D. and {Harrison}, D.~L. and {Henrot-Versill{\'e}}, S. and {Hern{\'a}ndez-Monteagudo}, C. and {Herranz}, D. and {Hildebrandt}, S.~R. and {Hivon}, E. and {Hobson}, M. and {Holmes}, W.~A. and {Hornstrup}, A. and {Hovest}, W. and {Huffenberger}, K.~M. and {Hurier}, G. and {Jaffe}, A.~H. and {Jaffe}, T.~R. and {Jones}, W.~C. and {Juvela}, M. and {Keih{\"a}nen}, E. and {Keskitalo}, R. and {Kisner}, T.~S. and {Kneissl}, R. and {Knoche}, J. and {Kunz}, M. and {Kurki-Suonio}, H. and {Lagache}, G. and {L{\"a}hteenm{\"a}ki}, A. and {Lamarre}, J.-M. and {Lasenby}, A. and {Lattanzi}, M. and {Lawrence}, C.~R. and {Leonardi}, R. and {Lesgourgues}, J. and {Levrier}, F. and {Liguori}, M. and {Lilje}, P.~B. and {Linden-V{\o}rnle}, M. and {L{\'o}pez-Caniego}, M. and {Lubin}, P.~M. and {Mac{\'\i}as-P{\'e}rez}, J.~F. and {Maggio}, G. and {Maino}, D. and {Mandolesi}, N. and {Mangilli}, A. and {Maris}, M. and {Martin}, P.~G. and {Mart{\'\i}nez-Gonz{\'a}lez}, E. and {Masi}, S. and {Matarrese}, S. and {McGehee}, P. and {Meinhold}, P.~R. and {Melchiorri}, A. and {Melin}, J.-B. and {Mendes}, L. and {Mennella}, A. and {Migliaccio}, M. and {Mitra}, S. and {Miville-Desch{\^e}nes}, M.-A. and {Moneti}, A. and {Montier}, L. and {Morgante}, G. and {Mortlock}, D. and {Moss}, A. and {Munshi}, D. and {Murphy}, J.~A. and {Naselsky}, P. and {Nati}, F. and {Natoli}, P. and {Netterfield}, C.~B. and {N{\o}rgaard-Nielsen}, H.~U. and {Noviello}, F. and {Novikov}, D. and {Novikov}, I. and {Oxborrow}, C.~A. and {Paci}, F. and {Pagano}, L. and {Pajot}, F. and {Paoletti}, D. and {Partridge}, B. and {Pasian}, F. and {Patanchon}, G. and {Pearson}, T.~J. and {Perdereau}, O. and {Perotto}, L. and {Perrotta}, F. and {Pettorino}, V. and {Piacentini}, F. and {Piat}, M. and {Pierpaoli}, E. and {Pietrobon}, D. and {Plaszczynski}, S. and {Pointecouteau}, E. and {Polenta}, G. and {Popa}, L. and {Pratt}, G.~W. and {Pr{\'e}zeau}, G. and {Prunet}, S. and {Puget}, J.-L. and {Rachen}, J.~P. and {Rebolo}, R. and {Reinecke}, M. and {Remazeilles}, M. and {Renault}, C. and {Renzi}, A. and {Ristorcelli}, I. and {Rocha}, G. and {Roman}, M. and {Rosset}, C. and {Rossetti}, M. and {Roudier}, G. and {Rubi{\~n}o-Mart{\'\i}n}, J.~A. and {Rusholme}, B. and {Sandri}, M.},
        title = "{Planck 2015 results. XXIV. Cosmology from Sunyaev-Zeldovich cluster counts}",
      journal = {\aap},
         year = 2016,
        month = sep,
       volume = {594},
          eid = {A24},
        pages = {A24},
          doi = {10.1051/0004-6361/201525833},
archivePrefix = {arXiv},
       eprint = {1502.01597},
 primaryClass = {astro-ph.CO},
       adsurl = {https://ui.adsabs.harvard.edu/abs/2016A&A...594A..24P}
}

@ARTICLE{2018MNRAS.474.2635S,
       author = {{Schrabback}, T. and {Applegate}, D. and {Dietrich}, J.~P. and {Hoekstra}, H. and {Bocquet}, S. and {Gonzalez}, A.~H. and {von der Linden}, A. and {McDonald}, M. and {Morrison}, C.~B. and {Raihan}, S.~F. and {Allen}, S.~W. and {Bayliss}, M. and {Benson}, B.~A. and {Bleem}, L.~E. and {Chiu}, I. and {Desai}, S. and {Foley}, R.~J. and {de Haan}, T. and {High}, F.~W. and {Hilbert}, S. and {Mantz}, A.~B. and {Massey}, R. and {Mohr}, J. and {Reichardt}, C.~L. and {Saro}, A. and {Simon}, P. and {Stern}, C. and {Stubbs}, C.~W. and {Zenteno}, A.},
        title = "{Cluster mass calibration at high redshift: HST weak lensing analysis of 13 distant galaxy clusters from the South Pole Telescope Sunyaev-Zel'dovich Survey}",
      journal = {\mnras},
         year = 2018,
        month = feb,
       volume = {474},
       number = {2},
        pages = {2635-2678},
          doi = {10.1093/mnras/stx2666},
archivePrefix = {arXiv},
       eprint = {1611.03866},
 primaryClass = {astro-ph.CO},
       adsurl = {https://ui.adsabs.harvard.edu/abs/2018MNRAS.474.2635S}
}

@ARTICLE{2019MNRAS.485...69S,
       author = {{Stern}, C. and {Dietrich}, J.~P. and {Bocquet}, S. and {Applegate}, D. and {Mohr}, J.~J. and {Bridle}, S.~L. and {Carrasco Kind}, M. and {Gruen}, D. and {Jarvis}, M. and {Kacprzak}, T. and {Saro}, A. and {Sheldon}, E. and {Troxel}, M.~A. and {Zuntz}, J. and {Benson}, B.~A. and {Capasso}, R. and {Chiu}, I. and {Desai}, S. and {Rapetti}, D. and {Reichardt}, C.~L. and {Saliwanchik}, B. and {Schrabback}, T. and {Gupta}, N. and {Abbott}, T.~M.~C. and {Abdalla}, F.~B. and {Avila}, S. and {Bertin}, E. and {Brooks}, D. and {Burke}, D.~L. and {Carnero Rosell}, A. and {Carretero}, J. and {Castander}, F.~J. and {D'Andrea}, C.~B. and {da Costa}, L.~N. and {Davis}, C. and {De Vicente}, J. and {Diehl}, H.~T. and {Doel}, P. and {Estrada}, J. and {Evrard}, A.~E. and {Flaugher}, B. and {Fosalba}, P. and {Frieman}, J. and {Garc{\'\i}a-Bellido}, J. and {Gaztanaga}, E. and {Gruendl}, R.~A. and {Gschwend}, J. and {Gutierrez}, G. and {Hollowood}, D. and {Jeltema}, T. and {Kirk}, D. and {Kuehn}, K. and {Kuropatkin}, N. and {Lahav}, O. and {Lima}, M. and {Maia}, M.~A.~G. and {March}, M. and {Melchior}, P. and {Menanteau}, F. and {Miquel}, R. and {Plazas}, A.~A. and {Romer}, A.~K. and {Sanchez}, E. and {Schindler}, R. and {Schubnell}, M. and {Sevilla-Noarbe}, I. and {Smith}, M. and {Smith}, R.~C. and {Sobreira}, F. and {Suchyta}, E. and {Swanson}, M.~E.~C. and {Tarle}, G. and {Walker}, A.~R. and {DES Collaboration} and {SPT Collaboration}},
        title = "{Weak-lensing analysis of SPT-selected galaxy clusters using Dark Energy Survey Science Verification data}",
      journal = {\mnras},
         year = 2019,
        month = may,
       volume = {485},
       number = {1},
        pages = {69-87},
          doi = {10.1093/mnras/stz234},
archivePrefix = {arXiv},
       eprint = {1802.04533},
 primaryClass = {astro-ph.CO},
       adsurl = {https://ui.adsabs.harvard.edu/abs/2019MNRAS.485...69S}
}

@ARTICLE{2016JCAP...08..058B,
       author = {{Battaglia}, N.},
        title = "{The tau of galaxy clusters}",
      journal = {\jcap},
         year = 2016,
        month = aug,
       volume = {2016},
       number = {8},
          eid = {058},
        pages = {058},
          doi = {10.1088/1475-7516/2016/08/058},
archivePrefix = {arXiv},
       eprint = {1607.02442},
 primaryClass = {astro-ph.CO},
       adsurl = {https://ui.adsabs.harvard.edu/abs/2016JCAP...08..058B}
}

@ARTICLE{2013MNRAS.429.3627M,
       author = {{Miyatake}, Hironao and {Nishizawa}, Atsushi J. and {Takada}, Masahiro and {Mandelbaum}, Rachel and {Mineo}, Sogo and {Aihara}, Hiroaki and {Spergel}, David N. and {Bickerton}, Steven J. and {Bond}, J. Richard and {Gralla}, Megan and {Hajian}, Amir and {Hilton}, Matt and {Hincks}, Adam D. and {Hughes}, John P. and {Infante}, Leopoldo and {Lin}, Yen-Ting and {Lupton}, Robert H. and {Marriage}, Tobias A. and {Marsden}, Danica and {Menanteau}, Felipe and {Miyazaki}, Satoshi and {Moodley}, Kavilan and {Niemack}, Michael D. and {Oguri}, Masamune and {Price}, Paul A. and {Reese}, Erik D. and {Sif{\'o}n}, Crist{\'o}bal and {Wollack}, Edward J. and {Yasuda}, Naoki},
        title = "{Subaru weak lensing measurement of a z = 0.81 cluster discovered by the Atacama Cosmology Telescope Survey}",
      journal = {\mnras},
         year = 2013,
        month = mar,
       volume = {429},
       number = {4},
        pages = {3627-3644},
          doi = {10.1093/mnras/sts643},
archivePrefix = {arXiv},
       eprint = {1209.4643},
 primaryClass = {astro-ph.CO},
       adsurl = {https://ui.adsabs.harvard.edu/abs/2013MNRAS.429.3627M}
}

@ARTICLE{Aymerich2025,
       author = {{Aymerich}, G. and {Grandis}, S. and {Douspis}, M. and {Pratt}, G.~W. and {Salvati}, L. and {Andrade-Santos}, F. and {Bocquet}, S. and {Costanzi}, M. and {Forman}, W.~R. and {Jones}, C. and {Aguena}, M. and {Andrade-Oliveira}, F. and {Bacon}, D. and {Brooks}, D. and {Burke}, D.~L. and {Carretero}, J. and {da Costa}, L.~N. and {da Silva Pereira}, M.~E. and {Davis}, T.~M. and {De Vicente}, J. and {Desai}, S. and {Diehl}, H.~T. and {Doel}, P. and {Everett}, S. and {Flaugher}, B. and {Frieman}, J. and {Gaztanaga}, E. and {Gruen}, D. and {Gutierrez}, G. and {Hinton}, S.~R. and {Hollowood}, D.~L. and {Honscheid}, K. and {James}, D.~J. and {Lee}, S. and {Marshall}, J.~L. and {Mena-Fern{\'a}ndez}, J. and {Miquel}, R. and {Mohr}, J.~J. and {Ogando}, R.~L.~C. and {Plazas Malag{\'o}n}, A.~A. and {Porredon}, A. and {Prat}, J. and {Romer}, A.~K. and {Samuroff}, S. and {Sanchez}, E. and {Sanchez Cid}, D. and {Smith}, M. and {Suchyta}, E. and {Swanson}, M.~E.~C. and {Tucker}, D.~L. and {Weaverdyck}, N. and {Weller}, J. and {Yamamoto}, M.},
        title = "{Cosmological constraints from the Planck cluster catalogue with DES shear profiles and Chandra observations}",
      journal = {arXiv e-prints},
         year = 2025,
        month = sep,
          eid = {arXiv:2509.02068},
        pages = {arXiv:2509.02068},
          doi = {10.48550/arXiv.2509.02068},
archivePrefix = {arXiv},
       eprint = {2509.02068},
 primaryClass = {astro-ph.CO},
       adsurl = {https://ui.adsabs.harvard.edu/abs/2025arXiv250902068A}
}

@ARTICLE{Battaglia2012,
       author = {{Battaglia}, N. and {Bond}, J.~R. and {Pfrommer}, C. and {Sievers}, J.~L.},
        title = "{On the Cluster Physics of Sunyaev-Zel'dovich and X-Ray Surveys. I. The Influence of Feedback, Non-thermal Pressure, and Cluster Shapes on Y-M Scaling Relations}",
      journal = {\apj},
         year = 2012,
        month = oct,
       volume = {758},
       number = {2},
          eid = {74},
        pages = {74},
          doi = {10.1088/0004-637X/758/2/74},
archivePrefix = {arXiv},
       eprint = {1109.3709},
 primaryClass = {astro-ph.CO},
       adsurl = {https://ui.adsabs.harvard.edu/abs/2012ApJ...758...74B}
}

@ARTICLE{Wicker2023,
       author = {{Wicker}, R. and {Douspis}, M. and {Salvati}, L. and {Aghanim}, N.},
        title = "{Constraining the mass and redshift evolution of the hydrostatic mass bias using the gas mass fraction in galaxy clusters}",
      journal = {\aap},
         year = 2023,
        month = jun,
       volume = {674},
          eid = {A48},
        pages = {A48},
          doi = {10.1051/0004-6361/202243922},
archivePrefix = {arXiv},
       eprint = {2204.12823},
 primaryClass = {astro-ph.CO},
       adsurl = {https://ui.adsabs.harvard.edu/abs/2023A&A...674A..48W}
}

@ARTICLE{CCL2019,
       author = {{Chisari}, Nora Elisa and {Alonso}, David and {Krause}, Elisabeth and {Leonard}, C. Danielle and {Bull}, Philip and {Nev{\'e}u}, J{\'e}r{\'e}my and others},
        title = "{Core Cosmology Library: Precision Cosmological Predictions for LSST}",
      journal = {\apjs},
         year = 2019,
        month = may,
       volume = {242},
       number = {1},
          eid = {2},
        pages = {2},
          doi = {10.3847/1538-4365/ab1658},
archivePrefix = {arXiv},
       eprint = {1812.05995},
 primaryClass = {astro-ph.CO},
       adsurl = {https://ui.adsabs.harvard.edu/abs/2019ApJS..242....2C}
}

@ARTICLE{Hoekstra2003,
       author = {{Hoekstra}, Henk},
        title = "{How well can we determine cluster mass profiles from weak lensing?}",
      journal = {\mnras},
         year = 2003,
        month = mar,
       volume = {339},
       number = {4},
        pages = {1155-1162},
          doi = {10.1046/j.1365-8711.2003.06264.x},
archivePrefix = {arXiv},
       eprint = {astro-ph/0208351},
 primaryClass = {astro-ph},
       adsurl = {https://ui.adsabs.harvard.edu/abs/2003MNRAS.339.1155H}
}

@ARTICLE{Schneider1998,
       author = {{Schneider}, Peter and {van Waerbeke}, Ludovic and {Jain}, Bhuvnesh and {Kruse}, Guido},
        title = "{A new measure for cosmic shear}",
      journal = {\mnras},
         year = 1998,
        month = jun,
       volume = {296},
       number = {4},
        pages = {873-892},
          doi = {10.1046/j.1365-8711.1998.01422.x},
archivePrefix = {arXiv},
       eprint = {astro-ph/9708143},
 primaryClass = {astro-ph},
       adsurl = {https://ui.adsabs.harvard.edu/abs/1998MNRAS.296..873S}
}

@ARTICLE{Diemer2015,
       author = {{Diemer}, Benedikt and {Kravtsov}, Andrey V.},
        title = "{A Universal Model for Halo Concentrations}",
      journal = {\apj},
         year = 2015,
        month = jan,
       volume = {799},
       number = {1},
          eid = {108},
        pages = {108},
          doi = {10.1088/0004-637X/799/1/108},
archivePrefix = {arXiv},
       eprint = {1407.4730},
 primaryClass = {astro-ph.CO}
}

@ARTICLE{2012NJPh...14e5018R,
       author = {{Rasia}, E. and {Meneghetti}, M. and {Martino}, R. and {Borgani}, S. and {Bonafede}, A. and {Dolag}, K. and {Ettori}, S. and {Fabjan}, D. and {Giocoli}, C. and {Mazzotta}, P. and {Merten}, J. and {Radovich}, M. and {Tornatore}, L.},
        title = "{Lensing and x-ray mass estimates of clusters (simulations)}",
      journal = {New Journal of Physics},
         year = 2012,
        month = may,
       volume = {14},
       number = {5},
          eid = {055018},
        pages = {055018},
          doi = {10.1088/1367-2630/14/5/055018},
archivePrefix = {arXiv},
       eprint = {1201.1569},
 primaryClass = {astro-ph.CO},
       adsurl = {https://ui.adsabs.harvard.edu/abs/2012NJPh...14e5018R}
}

@ARTICLE{wu19,
       author = {{Wu}, Hao-Yi and {Weinberg}, David H. and {Salcedo}, Andr{\'e}s N. and
         {Wibking}, Benjamin D. and {Zu}, Ying},
        title = "{Covariance of the galaxy cluster weak lensing profiles}",
      journal = {\mnras},
         year = 2019,
        month = dec,
       volume = {490},
       number = {2},
        pages = {2606-2626},
          doi = {10.1093/mnras/stz2617},
archivePrefix = {arXiv},
       eprint = {1812.07732},
 primaryClass = {astro-ph.CO}
}

@ARTICLE{salvati18,
       author = {{Salvati}, L. and {Douspis}, M. and {Aghanim}, N.},
        title = "{Constraints from thermal Sunyaev-Zel'dovich cluster counts and power spectrum combined with CMB}",
      journal = {\aap},
         year = 2018,
        month = jun,
       volume = {614},
          eid = {A13},
        pages = {A13},
          doi = {10.1051/0004-6361/201731990},
archivePrefix = {arXiv},
       eprint = {1708.00697},
 primaryClass = {astro-ph.CO}
}

@ARTICLE{shirasaki_lau17,
       author = {{Shirasaki}, Masato and {Lau}, Erwin T. and {Nagai}, Daisuke},
        title = "{Modelling baryonic effects on galaxy cluster mass profiles}",
      journal = {\mnras},
         year = 2018,
        month = jun,
       volume = {477},
       number = {2},
        pages = {2804-2814},
          doi = {10.1093/mnras/sty763},
archivePrefix = {arXiv},
       eprint = {1711.06366},
 primaryClass = {astro-ph.CO},
       adsurl = {https://ui.adsabs.harvard.edu/abs/2018MNRAS.477.2804S}
}

@ARTICLE{SimonsObservatory2019,
       author = {{The Simons Observatory Collaboration} and {Ade}, Peter and {Aguirre}, James and {Ahmed}, Zeeshan and {Aiola}, Simone and {Ali}, Aamir and {Alonso}, David and {Alvarez}, Marcelo A. and {Arnold}, Kam and {Ashton}, Peter and others},
        title = "{The Simons Observatory: science goals and forecasts}",
      journal = {\jcap},
         year = 2019,
        month = feb,
       volume = {2019},
       number = {2},
          eid = {056},
        pages = {056},
          doi = {10.1088/1475-7516/2019/02/056},
archivePrefix = {arXiv},
       eprint = {1808.07445},
 primaryClass = {astro-ph.CO}
}

@ARTICLE{CMBS42016,
       author = {{CMB-S4 Collaboration} and {Abazajian}, Kevork N. and {Adshead}, Peter and {Ahmed}, Zeeshan and {Allen}, Steven W. and {Alonso}, David and {Arnold}, Kam S. and {Baccigalupi}, Carlo and {Bartlett}, James G. and {Battaglia}, Nicholas and others},
        title = "{CMB-S4 Science Book, First Edition}",
      journal = {arXiv e-prints},
         year = 2016,
        month = oct,
          eid = {arXiv:1610.02743},
        pages = {arXiv:1610.02743},
          doi = {10.48550/arXiv.1610.02743},
archivePrefix = {arXiv},
       eprint = {1610.02743},
 primaryClass = {astro-ph.CO}
}

@article{Giglio2023,
  author    = {Giglio, A. D. and Costa, M. U. P. D.},
  title     = {The use of artificial intelligence to improve the scientific writing of non-native English speakers},
  journal   = {Revista da Associacao Medica Brasileira (1992)},
  year      = {2023},
  volume    = {69},
  number    = {9},
  pages     = {e20230560},
  doi       = {10.1590/1806-9282.20230560}
}

@INPROCEEDINGS{alonso_pavon2025,
       author = {{Alonso Pav{\'o}n}, Jos{\'e} Antonio and {Plazas Malag{\'o}n}, Andr{\'e}s},
        title = "{Recommendations to overcome language barriers in the Vera C. Rubin Observatory Research Ecosystem}",
    booktitle = {Bulletin of the American Astronomical Society},
         year = 2025,
       volume = {57},
        month = jul,
          eid = {2025i015},
        pages = {2025i015},
          doi = {10.3847/25c2cfeb.983bab04},
archivePrefix = {arXiv},
       eprint = {2507.18682},
 primaryClass = {astro-ph.IM},
       adsurl = {https://ui.adsabs.harvard.edu/abs/2025BAAS...57a.015A}
}

@ARTICLE{BeckerKravtsov2011,
       author = {{Becker}, Matthew R. and {Kravtsov}, Andrey V.},
        title = "{On the Accuracy of Weak-lensing Cluster Mass Reconstructions}",
      journal = {\apj},
         year = 2011,
        month = oct,
       volume = {740},
       number = {1},
          eid = {25},
        pages = {25},
          doi = {10.1088/0004-637X/740/1/25},
archivePrefix = {arXiv},
       eprint = {1011.1681},
 primaryClass = {astro-ph.CO}
}

@ARTICLE{Grandis2021,
       author = {{Grandis}, S. and {Bocquet}, S. and {Dietrich}, J.~P. and {Schrabback}, T. and {Hyeonghan}, K. and {Allen}, S.~W.},
        title = "{Calibration of bias and scatter involved in cluster mass measurements using optical weak gravitational lensing}",
      journal = {\mnras},
         year = 2021,
        month = nov,
       volume = {507},
       number = {4},
        pages = {5671-5689},
          doi = {10.1093/mnras/stab2414},
archivePrefix = {arXiv},
       eprint = {2106.13783},
 primaryClass = {astro-ph.CO},
       adsurl = {https://ui.adsabs.harvard.edu/abs/2021MNRAS.507.5671G}
}

@ARTICLE{Grandis2024,
       author = {{Grandis}, S. and {Ghirardini}, V. and {Bocquet}, S. and {Garrel}, C. and {Mohr}, J.~J. and {Liu}, A. and {Kluge}, M. and {Kimmig}, L. and {Reiprich}, T.~H. and {Alarcon}, A. and {Amon}, A. and {Artis}, E. and {Bahar}, Y.~E. and {Balzer}, F. and {Bechtol}, K. and {Becker}, M.~R. and {Bernstein}, G. and {Bulbul}, E. and {Campos}, A. and {Carnero Rosell}, A. and {Carrasco Kind}, M. and {Cawthon}, R. and {Chang}, C. and {Chen}, R. and {Chiu}, I. and {Choi}, A. and {Clerc}, N. and {Comparat}, J. and {Cordero}, J. and {Davis}, C. and {Derose}, J. and {Diehl}, H.~T. and {Dodelson}, S. and {Doux}, C. and {Drlica-Wagner}, A. and {Eckert}, K. and {Elvin-Poole}, J. and {Everett}, S. and {Ferte}, A. and {Gatti}, M. and {Giannini}, G. and {Giles}, P. and {Gruen}, D. and {Gruendl}, R.~A. and {Harrison}, I. and {Hartley}, W.~G. and {Herner}, K. and {Huff}, E.~M. and {Kleinebreil}, F. and {Kuropatkin}, N. and {Leget}, P.~F. and {Maccrann}, N. and {Mccullough}, J. and {Merloni}, A. and {Myles}, J. and {Nandra}, K. and {Navarro-Alsina}, A. and {Okabe}, N. and {Pacaud}, F. and {Pandey}, S. and {Prat}, J. and {Predehl}, P. and {Ramos}, M. and {Raveri}, M. and {Rollins}, R.~P. and {Roodman}, A. and {Ross}, A.~J. and {Rykoff}, E.~S. and {Sanchez}, C. and {Sanders}, J. and {Schrabback}, T. and {Secco}, L.~F. and {Seppi}, R. and {Sevilla-Noarbe}, I. and {Sheldon}, E. and {Shin}, T. and {Troxel}, M. and {Tutusaus}, I. and {Varga}, T.~N. and {Wu}, H. and {Yanny}, B. and {Yin}, B. and {Zhang}, X. and {Zhang}, Y. and {Alves}, O. and {Bhargava}, S. and {Brooks}, D. and {Burke}, D.~L. and {Carretero}, J. and {Costanzi}, M. and {da Costa}, L.~N. and {Pereira}, M.~E.~S. and {De Vicente}, J. and {Desai}, S. and {Doel}, P. and {Ferrero}, I. and {Flaugher}, B. and {Friedel}, D. and {Frieman}, J. and {Garc{\'\i}a-Bellido}, J. and {Gutierrez}, G. and {Hinton}, S.~R. and {Hollowood}, D.~L. and {Honscheid}, K. and {James}, D.~J. and {Jeffrey}, N. and {Lahav}, O. and {Lee}, S. and {Marshall}, J.~L. and {Menanteau}, F. and {Ogando}, R.~L.~C. and {Pieres}, A. and {Plazas Malag{\'o}n}, A.~A. and {Romer}, A.~K. and {Sanchez}, E. and {Schubnell}, M. and {Smith}, M. and {Suchyta}, E. and {Swanson}, M.~E.~C. and {Tarle}, G. and {Weaverdyck}, N. and {Weller}, J.},
        title = "{The SRG/eROSITA All-Sky Survey: Dark Energy Survey year 3 weak gravitational lensing by eRASS1 selected galaxy clusters}",
      journal = {\aap},
         year = 2024,
        month = jul,
       volume = {687},
          eid = {A178},
        pages = {A178},
          doi = {10.1051/0004-6361/202348615},
archivePrefix = {arXiv},
       eprint = {2402.08455},
 primaryClass = {astro-ph.CO},
       adsurl = {https://ui.adsabs.harvard.edu/abs/2024A&A...687A.178G}
}

@ARTICLE{2504.01076,
       author = {{Chiu}, I.-Non and {Ghirardini}, Vittorio and {Grandis}, Sebastian and {Okabe}, Nobuhiro and {Artis}, Emmanuel and {Bulbul}, Esra and {Emre Bahar}, Y. and {Balzer}, Fabian and {Clerc}, Nicolas and {Comparat}, Johan and {Hsieh}, Bau-Ching and {Kleinebreil}, Florian and {Kluge}, Matthias and {Liu}, Ang and {Monteiro-Oliveira}, Rog{\'e}rio and {Oguri}, Masamune and {Pacaud}, Florian and {Ramos Ceja}, Miriam and {Reiprich}, Thomas H. and {Sanders}, Jeremy and {Schrabback}, Tim and {Seppi}, Riccardo and {Sommer}, Martin and {Tam}, Sut-Ieng and {Umetsu}, Keiichi and {Zhang}, Xiaoyuan},
        title = "{The SRG/eROSITA All-Sky Survey: The weak-lensing mass calibration and the stellar mass-to-halo mass relation from the Hyper Suprime-Cam Subaru Strategic Program}",
      journal = {\aap},
         year = 2025,
       volume = {704},
          eid = {A110},
        pages = {A110},
          doi = {10.1051/0004-6361/202554942},
archivePrefix = {arXiv},
       eprint = {2504.01076},
 primaryClass = {astro-ph.CO}
}

@ARTICLE{Ghirardini2024,
       author = {{Ghirardini}, V. and {Bulbul}, E. and {Artis}, E. and {Clerc}, N. and {Garrel}, C. and {Grandis}, S. and {Kluge}, M. and {Liu}, A. and {Bahar}, Y.~E. and {Balzer}, F. and {Chiu}, I. and {Comparat}, J. and {Gruen}, D. and {Kleinebreil}, F. and {Krippendorf}, S. and {Merloni}, A. and {Nandra}, K. and {Okabe}, N. and {Pacaud}, F. and {Predehl}, P. and {Ramos-Ceja}, M.~E. and {Reiprich}, T.~H. and {Sanders}, J.~S. and {Schrabback}, T. and {Seppi}, R. and {Zelmer}, S. and {Zhang}, X.},
        title = "{The SRG/eROSITA all-sky survey: Cosmology constraints from cluster abundances in the western Galactic hemisphere}",
      journal = {\aap},
         year = 2024,
       volume = {689},
          eid = {A298},
        pages = {A298},
          doi = {10.1051/0004-6361/202348852},
archivePrefix = {arXiv},
       eprint = {2402.08458},
 primaryClass = {astro-ph.CO},
       adsurl = {https://ui.adsabs.harvard.edu/abs/2024A&A...689A.298G}
}

@ARTICLE{2602.12238,
       author = {{Pantos}, Ioannis and {Perivolaropoulos}, Leandros},
        title = "{Status of the $S_8$ Tension: A 2026 Review of Probe Discrepancies}",
      journal = {arXiv e-prints},
         year = 2026,
          eid = {arXiv:2602.12238},
        pages = {arXiv:2602.12238},
          doi = {10.48550/arXiv.2602.12238},
archivePrefix = {arXiv},
       eprint = {2602.12238},
 primaryClass = {astro-ph.CO}
}

\end{document}